\documentclass[
fleqn,
12pt,
a4paper,
tightenlines,
notitlepage
]{revtex4-1}

\usepackage{bm,bbm}
\usepackage[colorlinks=true,linkcolor={magenta},citecolor={cyan}, urlcolor={magenta}]{hyperref}
\usepackage{xcolor}
\usepackage{graphicx}
\usepackage{amsmath}
\usepackage{multirow}
\usepackage{comment}
\usepackage{soul}
\usepackage[normalem]{ulem}

\newcommand{\msd}[1]{#1}
\newcommand{\fin}[1]{#1}

\newcommand{\note}[1]{#1}
\newcommand{\seg}[1]{{#1}}
\newcommand{\krish}[1]{#1}

\renewcommand{\sout}[1]{{}}

\renewcommand\thesubsection{SUPPLEMENTARY NOTE \arabic{subsection}}
\renewcommand\thesubsubsection{\arabic{subsection}.\arabic{subsubsection}}
\makeatletter\renewcommand\p@subsubsection{SUPPLEMENTARY NOTE }

\begin{document}

\title{Programming Mechanics in Knitted Materials, Stitch by Stitch}

\author{Krishma Singal\textsuperscript{1,+}}
\author{Michael S.~Dimitriyev\textsuperscript{2,3,+}}
\author{Sarah E.~Gonzalez\textsuperscript{1,+}}
\author{Alexander P.~Cachine\textsuperscript{1}}
\author{Sam Quinn\textsuperscript{1}}
\author{Elisabetta A.~Matsumoto\textsuperscript{1,4,*}}
\affiliation{\textsuperscript{1}School of Physics, Georgia Institute of Technology, Atlanta, Georgia 30332, USA}
\affiliation{\textsuperscript{2}Department of Polymer Science and Engineering, University of Massachusetts, Amherst, Massachusetts 01003, USA}
\affiliation{\textsuperscript{3}Department of Materials Science and Engineering, Texas A\&M University, College Station, Texas 77843, USA}
\affiliation{\textsuperscript{4} International Institute for Sustainability with Knotted Chiral Meta Matter (WPI-SKCM), Hiroshima University, Boulder, Higashihiroshima 739-8526, Japan}

\affiliation{\textsuperscript{*}\href{mailto:sabetta@gatech.edu}{sabetta@gatech.edu}; \textsuperscript{+}these authors contributed equally to this work}


\begin{abstract}
Knitting turns yarn, a 1D material, into a 2D fabric that is flexible, durable \cite{Warren2018}, and can be patterned to adopt a wide range of 3D geometries\cite{Narayanan2018}. 
Like other mechanical metamaterials\cite{Bertoldi2017}, the elasticity of knitted fabrics is an emergent property of the local stitch topology and pattern that cannot solely be attributed to the yarn itself.
Thus, knitting can be viewed as an additive manufacturing technique that allows for stitch-by-stitch programming of elastic properties and has applications in many fields ranging from soft robotics\cite{Abel2012,Albaugh2019,Sanchez2021} and wearable electronics\cite{Zeng2014,Cherenack2012} to engineered tissue\cite{Magnan2020} and architected materials\cite{Thomsen2016,Scott2018}.
However, predicting these mechanical properties based on the stitch type remains elusive. 
Here we untangle the relationship between changes in stitch topology and emergent elasticity in several types of knitted fabrics. 
We combine experiment and simulation to construct a constitutive model for the nonlinear bulk response of these fabrics.
This model serves as a basis for composite fabrics with bespoke mechanical properties, which crucially do not depend on the constituent yarn.
\end{abstract}

\flushbottom
\maketitle

\thispagestyle{empty}
 
Knitting has long been regarded as an art that turns natural fibers into garments.
Recently, engineers have begun to use knitting as an additive manufacturing technique to construct textiles with bespoke mechanical properties and geometries from `yarns' made from a myriad of materials.
\seg{Textiles research has traditionally been housed in both textile engineering and computer graphics; however, the growing interest of textiles as metamaterials in other fields creates the need for cross-disciplinary pollination.}
\seg{From that viewpoint, k}nitted textiles are mechanical metamaterials whose properties are imbued by the pattern of stitches, which exists irrespective of the choice of particular yarn.
By choosing the appropriate stitches and their ordering, \seg{one} can sculpt the local mechanical response of a textile using a yarn of their choice.
Tunable compliance and tensile strength of knitted and braided structures made from bio-compatible yarns are used for medical bandages\cite{Magnan2020}, surgical grafts\cite{Freeman2009,Goyal2019}, and mesh implants\cite{Mikolajczyk2016,Shuang2019,Liu2019,Yu2019}.
\krish{The mechanical properties of knitted textiles make them ideal for strain\cite{Mattmann2008,Seyedin2019} and pressure\cite{Vu2020,Yan2022,mcdonald_knitted_2020} sensors used in medical monitoring and therapeutics\cite{Fan2020,Tian2021,Chen2021} as well as soft actuators\cite{Scott2013,Abel2013,Albaugh2019,Han2017,Rivera2020}.}
Likewise, knitted textiles can harvest energy from human movement\cite{Wang2016,Kwak2017,Choi2017} and even store energy as wearable supercapacitors\cite{Bao2012,Jost2013}.
By spatially varying the pattern of stitches, we can generate textiles with high or low stiffnesses (Fig.~1). 
With the aid of computerized knitting machines, we can on-the-fly program regions of variable stiffness into a larger textile.
Unlike other composites, the entangled microstructure that gives rise to a knitted fabric's variable rigidity also holds it together along seamless interfaces.
Continuously modifying the in-plane rigidity of a textile across a region can mitigate the damage often associated with large stresses at interfaces\cite{Suresh2001}.

To facilitate the rational design of textiles, we need to understand the fundamental mechanics of knitted materials.
\krish{Here\seg{, inspired by the design of hand-knit garments,} we study how the mechanical behavior of weft knitted fabrics is encoded by the topology of their stitches as a first step towards creating a design tool for programmable textile metamaterials.}

The stitch pattern and mechanical properties of the constituent yarn are quasi-independent knobs we can fine tune.
A consequence of our model is that knitting can be used to program mechanics at any lengthscale, from polymeric and colloidal assemblies\cite{Goodrich2017} to light-weight tensile support in building construction\cite{Thomsen2016}. 

\krish{The computer graphics community has made great strides in creating knit fabric simulations with visual fidelity\cite{kaldor8, kaldor_efficient_2010, cirio_yarn-level_2017}, often with the goal of modeling entire sheets\cite{sperl_estimation_2022} of fabric and garments\cite{liu_knitting_2021}. There has not yet been a systematic study of how changes in stitch topology affect the fabric elasticity\cite{tekerek_experimental_2020} -- even modeling stockinette (sometimes called jersey or plain-knit) fabric is quite complex \cite{Choi2006,Postle2002,Poincloux2018PRX}.} 
\seg{In this work, our goal is to study knit fabrics from three different types of models: a minimal model of yarn-level simulation at the microscopic level, a constitutive model at the textile level, and our ``Reduced-Symmetry'' model at the intermediate level to unite these two points of view.}
\seg{\sout{Such a model would enable knitwear designers, soft roboticists \cite{Sanchez2021}, and other engineers to fine tune the elastic response of their fabrics irrespective of the yarn used.}}
\krish{Traditionally elastic response in knitted textiles is achieved by modifying the properties of the yarn often using blends of natural (wool and cotton) and synthetic fibers (polyester\seg{, nylon, or other plastics}) which contribute to microplastic pollution\cite{grandview}.}
\seg{To maximise extensibility, manufacturers reduce the amount of natural fibers used in the fabric and increase the amount of elastane and/or other elastomeric fibers. 
Our goal is to use stitch type as a way of modulating the bulk elasticity of fabrics made of inelastic yarn, irregardless of fiber composition, so that the desired elastic response of a textile can be achieved with natural and/or biodegradable fibers and without synthetic materials. Recent research has shown that a broad range of synthetic materials can degrade when in contact with skin secretions, which increases the potential for dermal absorption of compounds within those fibers\cite{Abafe2023}. 
}

\begin{figure*}[t]
    \includegraphics[width=\textwidth]{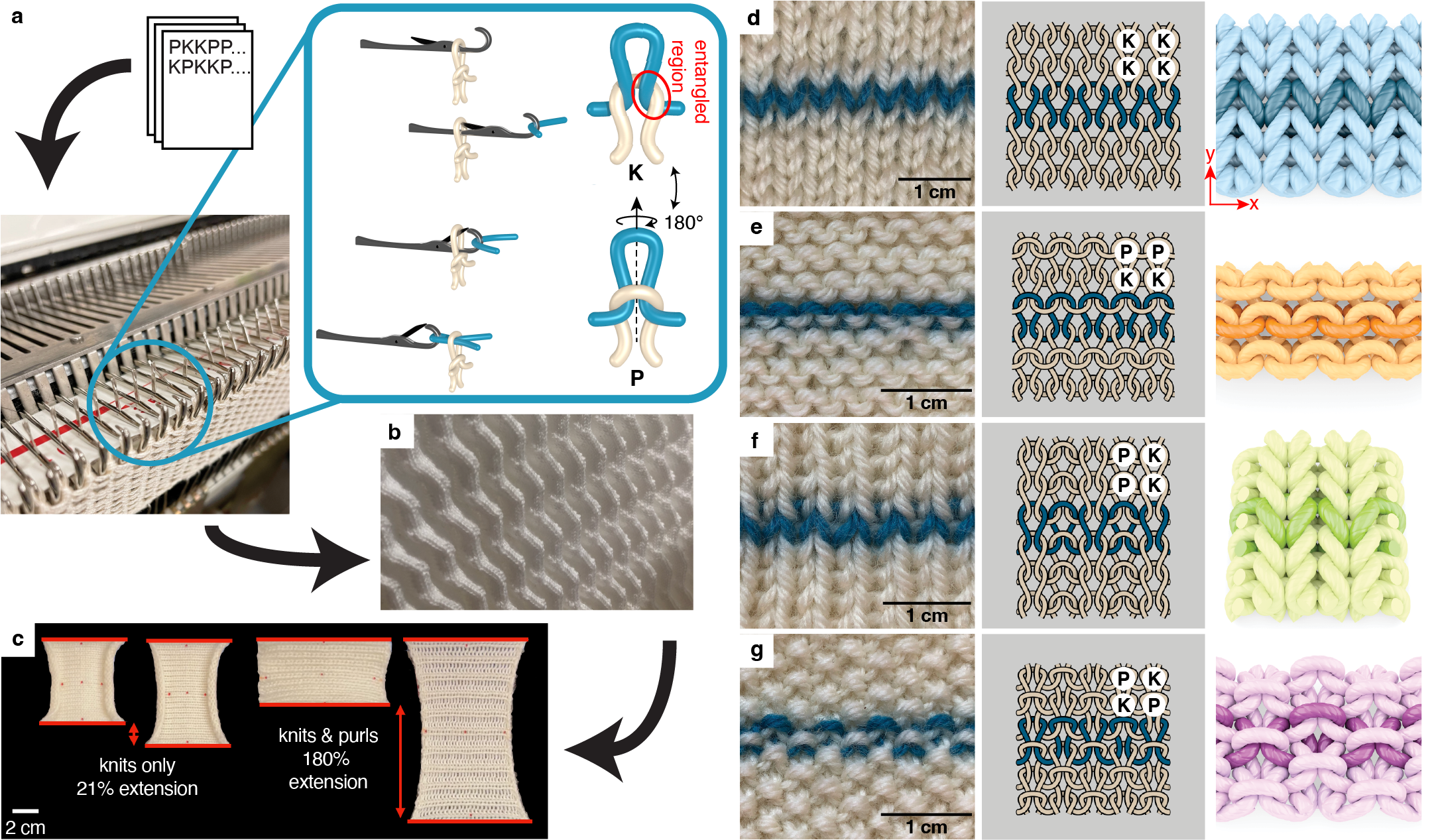}
    \centering
    \label{fig:fig1}
    \caption{{\bf Knitted materials have elastic responses that can be programmed by the pattern of {\bf K}s and {\bf P}s.} (\textbf{a}) A schematic of the knitting process where a knitting machine converts a code of {\bf K}s and {\bf P}s into a textile such as the Issey Miyake \cite{IsseyMiyake} sweater shown in (\textbf{b}). The knitting machine manipulates a bed of latch needles that pull new loops of yarn through existing loops to build the knitted fabric. An entangled region of the stitch is identified by the red circle in the inset of (\textbf{a}). (\textbf{c}) Knitted fabrics with a mix of both {\bf K}s and {\bf P}s are markedly more extensible (under the same applied stress) than ones with only a single type of stitch. (\textbf{d}-\textbf{g}) Close up images (left), line diagrams (center), and simulation results (right) of four fabrics: (\textbf{d}) stockinette, (\textbf{e}) garter, (\textbf{f}) rib, and (\textbf{g}) seed.}
\end{figure*}

\textbf{Topology and Elasticity}

Knitted textiles are composed from a rectangular lattice of slip knots.
The two foundational stitches in knitting are the \emph{knit stitch} (denoted \textbf{K}\seg{, also known as a front stitch}) and the \emph{purl stitch} (denoted \textbf{P}\seg{, also known as a back stitch}).
These two stitches form the bulk of a textile's structure, although many more complicated stitches exist \cite{Markande2020}.
The knit stitch is formed by passing a loop of yarn from the back to the front of the textile through an existing loop, while the purl stitch pulls the new loop from the front to the back.
Therefore, knits and purls are fundamentally the same object, just related by a $180^\circ$ rotation about the $y$-direction of the fabric (Fig.~1a).
A schematic of the knitting process is shown in Fig.~1a,b.
Combining \textbf{K}s and \textbf{P}s in different patterns generates textiles with markedly different linear elastic responses (Fig.~1c). 
Our goal is to untangle this relationship between stitch pattern and mechanical response using four common knitted fabrics: stockinette (Fig.~1d), garter (Fig.~1e\seg{, also known as links-links}), rib (Fig.~1f), and seed (Fig.~1g).

The combination of entangled elastic segments and confinement makes knitted fabrics different from many mechanical metamaterials.
The microstructure of a knitted fabric has \emph{entangled regions} whose contact interactions dictate the stiffness and unconstrained regions that enable extensibility.
Changing the ordering of yarn in an entangled region changes the topology of the fabric. 
Therefore, the topological method of knot theory is used to study textiles \cite{Markande2020,Grishanov2009}.
Previous studies have shown that the ordering of crossings within a knot can have a major impact on its strength\cite{Patil2020}, indicating a strong relationship between topology and mechanics.

We measured the elastic response of each of the four common knitted fabrics (Fig.~1d-g) in a series of uniaxial stretching experiments\cite{quaglini_experimental_2008} and simulations (see Methods\seg{;} \seg{Supplementary Fig.~1;} \seg{Supplementary Fig.~3;} \seg{Supplementary Tables 7, 9, and 11;} and \seg{Supplementary Notes 1-7}).
We fabricated and characterized samples made from two types of yarn, an acrylic yarn (Fig.~2) and a pearlized-cotton (\seg{Supplementary Fig.~2}), which have different mechanical properties (see Methods). 
With the fabric under fixed uniaxial loading, we measured the bulk fabric deformation using computer vision\cite{Ershov2022,Schindelin2012} (see Methods and \seg{Supplementary Fig.~1}). 
The maximal longitudinal components of the average stress $\sigma$ versus strain $\varepsilon$ measurements are shown in Fig.~2, where the $x$- and $y$-directions are along the rows and columns of the fabric (Fig.~1d).

\begin{figure}[t]
    \includegraphics[width=16cm]{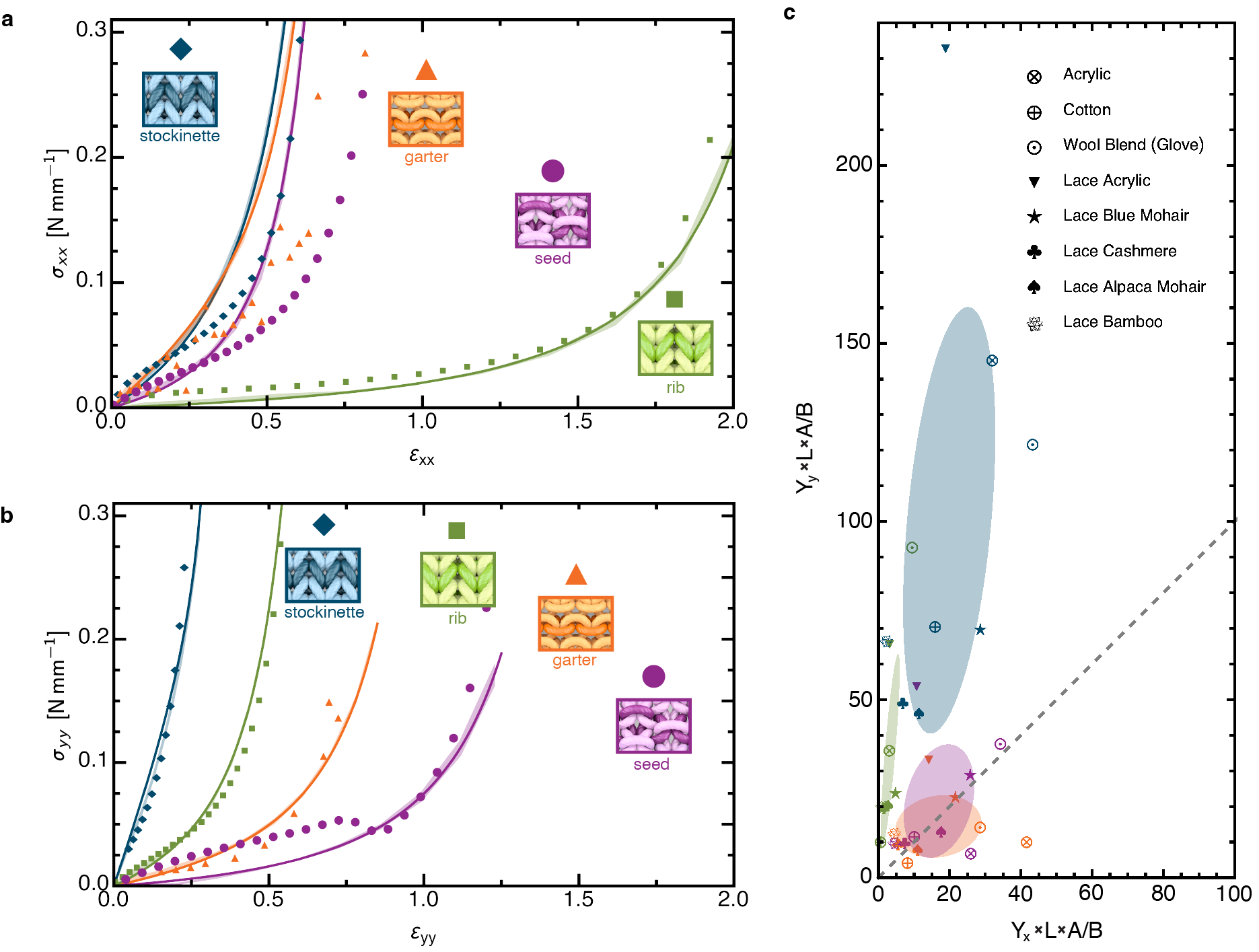}
    \centering
    \label{fig:fig2}
    \caption{
    {\bf Experimental and simulated results of uniaxial stretching.} 
The stress-versus-strain relations for the four fabrics made from the acrylic yarn in the (\textbf{a}) $x$- and (\textbf{b}) $y$-directions. 
All of the data for each type of fabric is displayed by a different color: stockinette in blue, garter in orange, rib in green, and seed in purple.
The experimental data is shown in the translucent regions where the width of the region is one standard deviation of the four experiment runs. 
The simulation data is shown with solid symbols.
The solid curves are fits to the constitutive relations. 
\seg{\sout{Dashed lines depict the linear response at zero stress.}}
This is a system where the linear response for each fabric is significantly different despite only small differences in the stitch configuration, whereas the nonlinear parts are quite similar.
Experiments applying force in the $x$-direction show the extreme extensibility of the rib pattern compared with the other three. 
Garter and seed dominate in the $y$-direction. 
Note, the experimental measurements for seed fabric differ from that of simulations due to a compression-related buckling instability 
in the computation, investigated in Supplementary Note 4 and \seg{Supplementary Fig.~8}. \seg{(\textbf{c}) Normalized rigidity plot of all fabric samples, where $Y_i$ is the Young's modulus in the $i$\textsuperscript{th} direction in N/mm (Supplementary Tables 10, 12, 14, and 20), $L$ is the length of yarn per stitch in mm (Supplementary Tables 2, 3, and 6), $A$ is the area of one stitch in mm$^2$ (Supplementary Tables 5 and 6), and $B$ is the bending modulus in N mm$^2$ (Supplementary Table 1). The colored ellipses represent one standard deviation for each of the four fabric types and are oriented along the principal axes. The gray dashed line represents an isotropic mechanical response. The same analysis was conducted on the un-normalized rigidities, shown in Supplementary Fig.~11.}
    }
\end{figure}

Under small stresses, the responses of all the fabrics are linear, and the Young's moduli are given by the slopes of the stress-vs-strain curves (Fig.~2, \seg{Supplementary Tables 10 and 12}).
Under high stresses, their responses become nonlinear, displaying strain-stiffening behavior as the yarn within the stitch becomes taut. 
Of the four fabrics, rib is by far the softest in the $x$-direction while stockinette is the stiffest (Fig.~2a).
Similarly, the garter and seed fabrics are softer in the $y$-direction (Fig.~2b).
\seg{In Figure~2c, we have plotted the normalized Young's modulus in the $x$-direction by the normalized Young's modulus in the $y$-direction for samples made from eight different types of yarn of varying sizes and constituent fibers.
The clusters of data confirm that the relative anisotropy is fairly consistent across each type of fabric, regardless of the constituent fiber (Supplementary Fig.~11).}

\textbf{Numerical Model}

\krish{Simulations help us unravel the effect that stitch topology and microstructure have on the macroscopic elasticity of the fabric.}
\seg{Stitch-level simulations (also known as loop modeling) have been of interest to a variety of fields, including textile engineering and metamaterials. 
Current simulations typically have at least one of three primary limiting constraints: they do not consider compressible yarn \cite{abghary1026,Poincloux2018PRX,Duhovic2006}, they only consider one fabric type\cite{Duhovic2006, Abel2012, abghary1026, htoo_3-dimension_2017, Poincloux2018PRX}, or they only compare simulation to experimental results for visual fidelity and not mechanical response\cite{kaldor8,ru_modeling_2023}. 
Our simulation method considers all three of these factors to investigate the role of stitch topology on the mechanical behavior of knit fabrics.
}

\krish{Yarn is an inherently hierarchical material with short staple fibers spun into indefinitely long yarn.
To model the complex mechanics of yarn, we use yarn characterization experiments to measure the dominant energetic contributions: bending and compression (Supplementary Notes 2 and 3, \seg{Supplementary Tables 1 and 4}).
The torsional rigidity of a balanced, spun yarn is comparatively negligible, and the stretching energy is so large that we consider the yarn to be inextensible.
We simulate the yarn as a space curve $\bm{\gamma}(s)$ subject to a bending energy that is quadratic in the curvature, $E_{\rm bend} = (B/2)\int_0^L {\rm d}s\,|\partial_s\hat{\mathbf{t}}|^2$, where $s$ is the arclength parameter, $\hat{\mathbf{t}} \equiv \partial_s\bm{\gamma}$ is the unit tangent vector of the curve at each point, and the yarn parameters (the yarn length per stitch $L$\seg{, also known as loop length,} and the bending modulus $B$) are measured experimentally (see Methods\seg{; Supplementary Note 2;} \seg{Supplementary Fig.~4;} \seg{Supplementary Fig.~9;} \seg{and Supplementary Tables 1, 2, and 3}). 
\seg{\sout{Existing models that take a bottom-up approach to fabric elasticity typically consider incompressible yarns\cite{duhovic_simulating_2006} however} To capture yarn-yarn interactions,} we use an elastic core-shell model informed by experiments (see Methods, \seg{Supplementary Fig.~5}, and Supplementary Note 3)\seg{.} \seg{\sout{in our simulations to capture the elastic interaction between yarn and to} This also} prevent\seg{s} yarn segments from passing through one another.
\seg{By implementing a minimal model in simulations, we can determine the key ingredients that contribute to the different mechanical behavior of different fabric types so that our results can be efficiently utilized in the fields of mechanical metamaterials and extreme mechanics.}}

The periodic nature of knitted textiles enables us to reduce the system to a closed segment of yarn in a box with boundaries identified (\seg{Supplementary Fig.~6}).
We numerically minimize the total yarn energy, while varying simulation box dimensions (see Methods and Supplementary Note 4).
Through our model, we effectively capture not only the geometry of knitted fabrics (Fig.~1d-g) \cite{Choi2006,Knittel2020,Wadekar2020} but the emergent elastic response as well (Fig.~2).
The simulations reproduce the key features of the experiments: (i) the differences between the extensional rigidities of each fabric resulting from their unique topologies in the low-tension regime and (ii) the divergent strain-stiffening behavior corresponding to the maximum extensibility of each stitch in the high-tension regime.
\krish{The simulations enable us to disentangle the ways in which contact energy and bending energy individually contribute to the local deformations of the yarn.
In the low stress regime, bending energy is the dominant contributor to elastic response.
In the high stress regime, compression energy shows a marked increase, as shown in \seg{Supplementary Fig.~7} and \seg{Supplementary Table~8}.}

\textbf{Microstructure and Modulus}

Knit stitches and purl stitches have fundamentally the same mechanical behavior. 
However, if we encode them -- like binary bits -- into a full textile, we see additional emergent behavior. 
In this way, we can view knit fabrics as a composite where each stitch has a fundamental elasticity and the yarn that connects each pair of stitches modifies the behavior based on its local symmetry.
When two knit or two purl stitches are next to each other, they are joined by a connecting yarn segment which has even symmetry (Fig.~3a).
When a knit stitch is joined to a purl stitch, however, the connecting yarn segment has odd symmetry (Fig.~3b). 
In the linear regime, the even and odd segments act as springs with different stiffnesses, as diagrammed in \seg{Supplementary Fig.~10c-f}.
We approximate the effective stiffness of the connecting yarn segments by taking its shape (from a fabric that has no forced applied to it) and calculate the work required to deform it infinitesimally (Supplementary Note 8, \seg{Supplementary Tables 15 and 16}).
When we do this to linear order, we find that the symmetric region has a stiffness that approximately scales as $Y_{\rm even} \sim 180B/\left[\lambda^3(1-\delta_{\rm even})\right]$, where $\lambda$ is the length of the segment (shown in \seg{Supplementary Fig.~10a,b}) and $\delta_{\rm even}$ is a geometry-dependent factor. 
The odd connecting yarn segment effectively acts as a moment arm where the two neighboring stitches apply a torque that causes it to rotate.
To linear order, the stiffness is approximately $Y_{\rm odd} \sim 12B/\left[\lambda^3(1-\delta_{\rm odd})\right]$.
Therefore, odd connecting yarn segments can be of order ten times softer compared to even connecting yarn segments (see Supplementary Note 8).
It is consequently harder to extend fabrics with identical neighboring stitches (\textbf{K}-\textbf{K} or \textbf{P}-\textbf{P}) than alternating neighboring stitches (\textbf{K}-\textbf{P}). 
This explains the relative stiffness of stockinette fabric, consisting only of even connecting yarn segments (Fig.~3c,g), compared with seed fabric, consisting only of odd connecting yarn segments (Fig.~3f,j).
Garter (Fig.~3d,h) and rib (Fig.~3e,i) fabrics each contain a mixture of segments but are much easier to stretch along the directions containing odd connecting yarn segments.
A pair of similar stitches (\textbf{K}-\textbf{K} or \textbf{P}-\textbf{P}) joined in the $y$-direction are in general stiffer than a pair of equivalent stitches joined in the $x$-direction because the $y$-direction has two connecting yarn segments in parallel between every pair of stitches (\seg{Supplementary Fig.~10c-f}). 

\begin{figure}[h]
    \includegraphics[width=\textwidth]{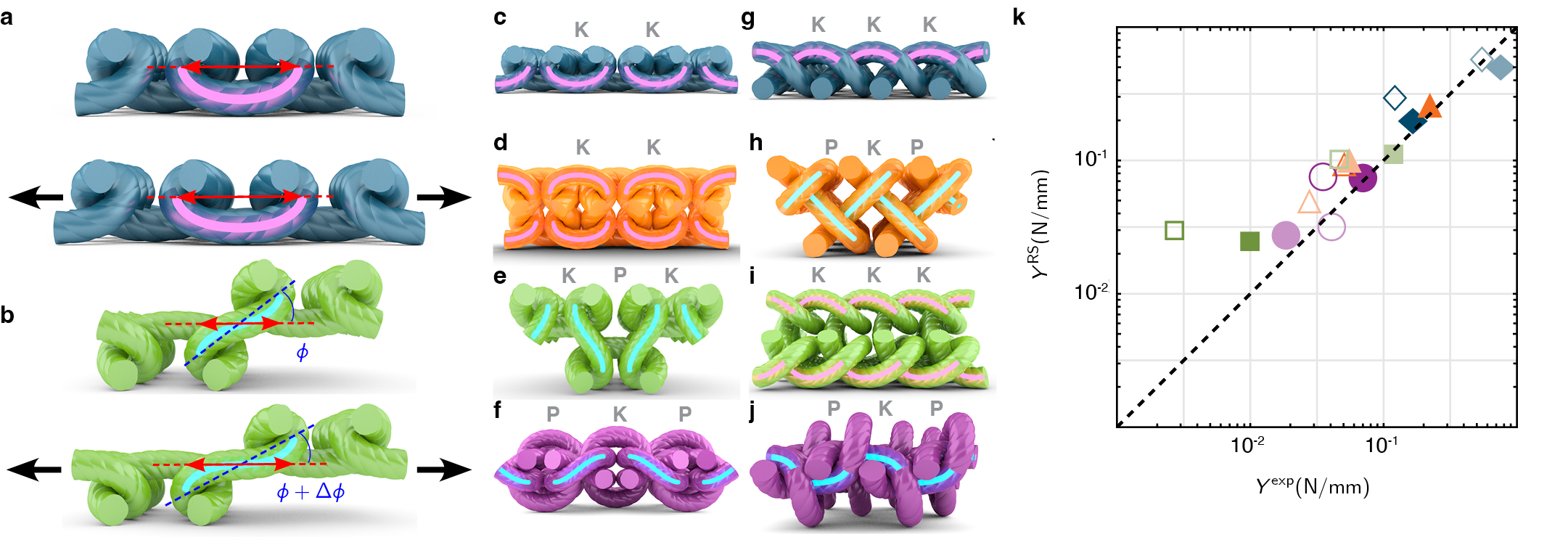}
    \centering
    \label{fig:fig3}
    \caption{
    {\bf Symmetry in the yarn segments between stitches.} 
Two similar stitches (\textbf{K}-\textbf{K} or \textbf{P}-\textbf{P}) are joined by a yarn segment with even symmetry, highlighted in pink (\textbf{a} top). Extensional deformations cause curvature deformations of the yarn segment (\textbf{a} bottom). Alternating stitches (\textbf{K}-\textbf{P}) are joined by a yarn segment with odd symmetry, highlighted in cyan (\textbf{b} top). These segments are able to rotate to accommodate extensional deformation (\textbf{b} bottom).
Symmetries of stitches are shown in the $x$-direction (\textbf{c}-\textbf{f}) and the $y$-direction (\textbf{g}-\textbf{j}).
(\textbf{c},\textbf{g}) Stockinette fabric has only even connecting yarn segments in both $x$- (\textbf{c}) and $y$-directions (\textbf{g}). 
(\textbf{d},\textbf{h}) Garter fabric has even connecting yarn segments in $x$-direction (\textbf{d}) and odd connecting yarn segments in the $y$-direction (\textbf{h}). 
(\textbf{e},\textbf{i}) Rib fabric has odd connecting yarn segments in the $x$-direction (\textbf{e}) and even connecting yarn segments in the $y$-direction (\textbf{i}). 
(\textbf{f},\textbf{j}) Since seed fabric is based on a checkerboard pattern, it only has odd connecting yarn segments. 
\seg{The renderings in (\textbf{a}-\textbf{j}) are repeated unit cells of sample stitch-level simulation outputs.}
A comparison of Young's moduli measured in experimental samples $Y^{\rm exp}$ with those computed in the reduced-symmetry (RS) model $Y^{\rm RS}$ \seg{(Supplementary Tables 17 and 18)} is shown in (\textbf{k}).
Dark and light symbols indicate extensional rigidity in the $x$-direction and $y$-direction, respectively, filled symbols indicate acrylic yarn, and open symbols indicate cotton yarn. This demonstrates that our simple composite model has both qualitative and quantitative agreement with our experimental measurements. 
}
\end{figure}

\krish{Using the ``rule of mixtures'' from the theory of fiber composites \cite{Hill1964}, we build an effective elastic model for fabrics consisting of knit or purl stitches alternating with connecting yarn segments of the appropriate symmetry.
We call this the Reduced Symmetry (RS) model.
In the low stress regime, we are treating the fabrics as a composite of geometries, rather than a composite of materials.
This allows for a direct estimate of the linear elastic rigidity using yarn geometry informed by simulations and bending modulus alone.}
To establish the dependence of the fabrics' anisotropic elastic response on stitch symmetry, we compare RS model estimates (using geometric parameters shown in Supplementary Tables 8 and 9) of the Young's moduli to those measured in experiments while varying stitch pattern, direction of extension, and type of yarn (Fig.~3k) (Supplementary Note 8).
Young's moduli estimated from our RS model closely agree with those measured in experiments, yet are systematically slightly stiffer.

In the high-tension limit, all yarn segments between neighboring entangled regions straighten along their mid-lengths and are forced to curve sharply as they enter the entangled regions due to contact confinement.
This localization of curvature to entangled regions under increasing stress represents a transition from the low-stress, linear elasticity dictated by stitch topology, $\sigma_{\rm low}(\varepsilon) \sim Y\varepsilon$, to high-stress, strain-stiffening elasticity, $\sigma_{\rm high}(\varepsilon) \sim \beta(1 - \alpha \varepsilon)^{-2}$, where $Y$ is a Young's modulus and $\beta$ and $\alpha$ are parameters characterizing the non-linear response. Each of these three parameters depend on the direction of extension.
With this reasoning, we arrive at a stress-strain constitutive relationship $\sigma(\varepsilon) = \sigma_{\rm low}(\varepsilon) + \sigma_{\rm high}(\varepsilon)$ (Supplementary Note 6).
Figs.~2a,b \seg{and Supplementary Figures 2 and 12} show self-consistent fits of this model to our data. \seg{This model is able to describe all knitted fabrics made from inextensible spun fibers (Supplementary Tables 9, 11, 13, and 19).}
This form of constitutive model resembles \seg{\sout{models} the force-extension relationship for stiff,} DNA-like polymers \cite{Marko1995} as well as amorphous fiber networks \cite{Broedersz2011}.

\textbf{Applications}

\krish{While our measurements and models capture the bulk constitutive properties of knitted fabric, the presence of boundaries can give rise to significant inhomogeneous response.
The bulk constitutive model can nonetheless well-approximate the full deformation of a finite swatch of knitted fabric, as illustrated in Fig.~4a,b, where we compare the $x$-component of the displacement field of a sample of garter fabric stretched in the $y$-direction (measured using digital image correlation, DIC) against a finite element analysis (FEA) that applies our constitutive model to a \seg{\sout{finite piece of fabric} two-dimensional sheet} with more realistic boundary conditions \seg{without directly considering the local microstructure} (Supplementary Note 9). \krish{We used garter experiments to directly obtain fits to our constitutive model for use in the FEA, without homogenizing the yarn level simulations \cite{liu_multiscale_2019,sperl_estimation_2022}.}
Notably, our constitutive model -- derived from microscopic fabric properties -- accurately captures the non-affine deformation of the fabric near its corners (where the principal stretch directions are no longer purely along the $x$- and $y$-axes) and reproduces the shape of the free boundary.}

Emergent elasticity sets knitting apart from other additive manufacturing techniques, because merely dictating the local topology by interchanging knits and purls (not changing the constituent yarn) programs the fabric's local elastic response. 
We can take advantage of the local anisotropic response of each different type of fabric by combining them into a seamless garment, in this example a prototype for a therapeutic glove (Supplementary Note 10 and \seg{Supplementary Fig.~13}).
The goal of our prototype is to direct the stiff elastic response to support the wrist joint in cases of repetitive stress injury, while enabling natural motion for the rest of the hand (Fig.~4c-e).
In Fig.~4d, the local extensibility field is represented with rectangles oriented along the principal directions with side lengths given by the extensibility in the $x$-direction, $1/Y_x$, and $y$-direction, $1/Y_y$ (see Methods, \seg{Supplementary Fig.~12}, and \seg{Supplementary Tables 19 and 20}). This shows that the stiffest region (stockinette fabric in dark blue) is designed to support the radiocarpal joints and to help keep the carpal and metacarpal bones aligned. Isotropic material (seed fabric in pink) still allows the carpometacarpal joint connecting the thumb to the wrist to move freely. Rib (green) and garter (orange) fabrics enable the fingers to extend and contract for natural motion (Fig.~4e and Supplementary Video 1). Importantly, knitted textiles can easily be crafted to fit any anatomy.

\begin{figure*}[t]
    \includegraphics[width=\textwidth]{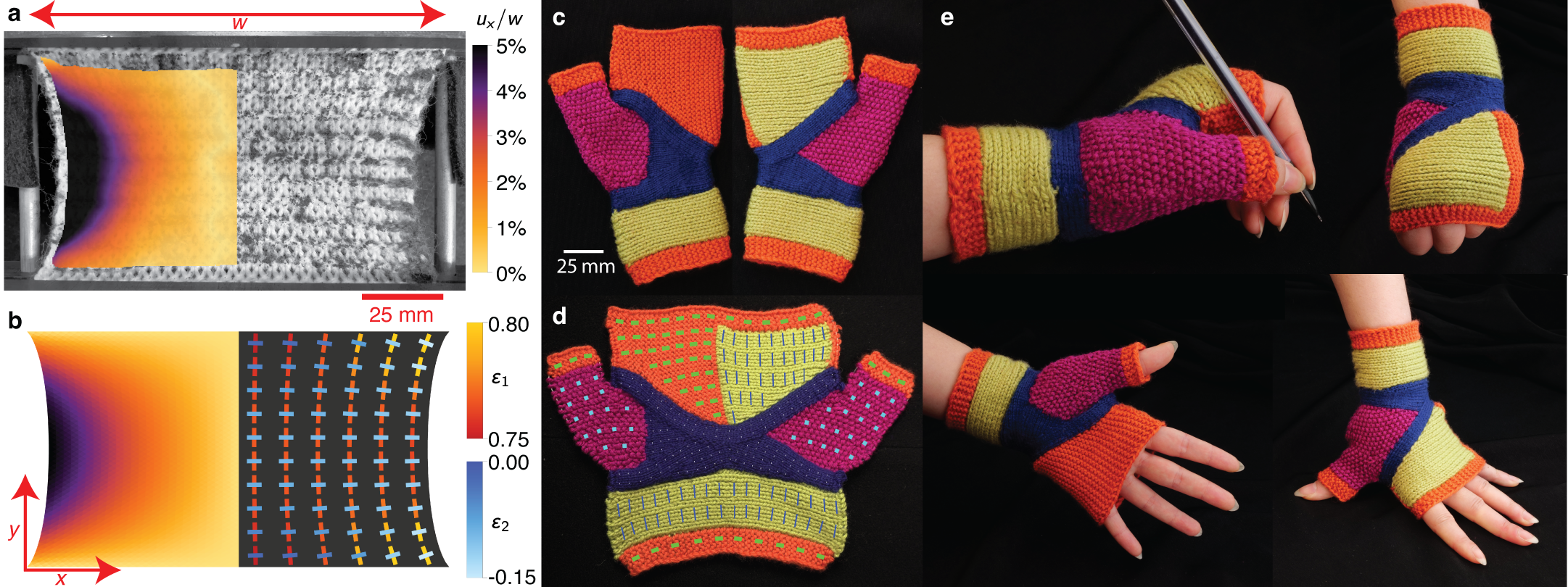}
    \centering
    \label{fig:fig4}
    \caption{
    {\bf The anisotropic and nonaffine global response of knitted textiles and an application of them.} 
    (\textbf{a,b}) Large applied stresses result in nonaffine deformations to a knitted fabric. (\textbf{a}) The $x$-component of the displacement field ($u_x$), obtained from DIC measurements, is shown overlaid on an image of garter fabric. 
    The color represents the magnitude of $u_x$, in units of fabric width $w$.
    (\textbf{b}) Finite element analysis (FEA) of our constitutive model reproduce the (left) displacement field seen in experiments and (right) the crosses show the principal directions and magnitudes of the local strain tensor. The values of local principal strains (scale bars in orange for $\varepsilon_1$ and blue for $\varepsilon_2$) show the degree of local extension and transverse compression.
    (\textbf{c}) Therapeutic glove prototype uses all four types of fabrics to generate anisotropic elastic response to motion of the hand. (\textbf{d}) The extensibility field of each type of fabric, shown as an overlay of rectangles, are oriented along the principal stiffness directions. The edge lengths are given by $1/Y_x$ and $1/Y_y$ respectively. (\textbf{e}) The stiffest stitch pattern, stockinette (blue), supports the wrist joint, while the isotropic seed (pink) grants mobility to the thumb. Highly anisotropic rib (green) and garter (orange) enable the wrist and fingers to flex along their easy direction. 
    }
\end{figure*}

\textbf{Discussion}

We present a picture of knitted fabric mechanics that is based on a micromechanical model of yarn. 
\krish{Drawing from composite theory}, we have developed a mesoscale model for the relationship between bulk elastic response and local topology, entanglement, and symmetry.
Our experiments and simulations demonstrate that changing the topology of stitches in a knitted fabric leads to remarkably different elastic responses, as seen in four standard types of knitted fabric.
The stitch micromechanics forms the basis of a nonlinear constitutive relation that models the behavior of textiles as 2D continuous materials.
The non-affine deformation of fabrics measured using digital image correlation (Fig.~4a) matches qualitatively and quantitatively with finite element simulations using our constitutive model (Fig.~4b) (see Methods and Supplementary Note 9).
Our long-term goal is to automate textile metamaterial production via a pipeline that takes desired mechanical performance and, using a computational model, generates a textile with compatible local properties.
\krish{This work can advance creation of non-proprietary software for designing fabric, as well as using mechanics to inform design, enabling textile engineers to tailor bespoke materials for a wide range of applications from performance sportswear \cite{Chen2021,Kanakaraj2015} to biomedical devices \cite{Zeng2014}.
With new developments in cost-effective methods to automate \cite{Narayanan2018,Kaspar2019} and program \cite{Hofmann2019} industrial knitting machines, 
we can build towards an open-source computational design platform that combines aspects of aesthetic, functional, and mechanical design.}

\section*{Acknowledgments}
We thank Ali Dahaj, Daria Atkinson, James McCord, Michael Czajkowski, \seg{Paul Loveman}, Peter Yunker, Robin Selinger, and Timothy Atherton for useful conversations. {\bf Funding:}  KS was supported in part by the Research Corporation for the Advancement of Science Cottrell Scholar Award. MSD, SG and EAM were supported by National Science Foundation Grant No.~DMR-1847172. This work was supported in part by the National Science Foundation Grant No.~DMS-1439786 and the Alfred P. Sloan Foundation award G-2019-11406 while the authors were in residence attending ICERM's Illustrating Mathematics program.

\section*{Author contributions statement}

EAM designed the study, KS performed the \seg{uniaxial stretching and yarn compression} experiments, \seg{KS and APC performed bending modulus experiments}, KS, SQ, MSD, and SEG analyzed the data, MSD and SEG performed the simulations, and KS, MSD, SEG, and EAM wrote the manuscript.

\section*{Additional information}

{\bf Data and materials availability:} Will be made available. {\bf Competing interests:} The authors declare no competing interests.

%

\clearpage

\section*{Methods}

\subsection*{Materials and Fabrication}
\seg{\sout{We used 2 types of yarn in our experiments.}} 
\seg{
We performed experiments on eight types of yarn that are classified in three categories: (1) two are large-gauge yarns (9-12 wraps per inch, WPI), (2) five are fine-gauge yarn (30-40 WPI) , and (3) the yarn used for the therapeutic glove prototype (14-18 WPI) (see the Knitted Glove Prototype section).}

We used Brava worsted yarn (28455-White) from KnitPicks\texttrademark{}, which is 100\% acrylic yarn, hereafter referred to as the ``acrylic yarn'' and 082L Pearl cotton 3/2 (color 1800-13 sapphire) from Halcyon Yarn\texttrademark{}, which is 100\% cotton yarn, hereafter referred to as the ``cotton yarn.'' 
For each of the types of fabrics, we recorded the average yarn diameter within the fabric stitches as well as the average yarn lengths per stitch. \seg{Supplementary Tables 2 and 3} display the measurements for the acrylic and cotton yarn, respectively.
We measured the bending rigidity, an approximate interaction potential, and the stress versus strain relationship for both types of yarn. \krish{We perform \seg{\sout{4} four} uniaxial experiment runs on the samples to obtain the stress versus strain relationship.}

We used a Taitexma\texttrademark{} Industrial Knitting Machine to create \seg{\sout{4} four} types of fabrics with both the acrylic and cotton yarn: stockinette, garter, \mbox{$1\times 1$} rib, and seed. Each fabric sample consisted of 31 rows and columns and were made at equal tensions and stitch size settings on the machine (Supplementary Note 11).

For an accurate model development, we obtained finer details of the fabric stitches. We created smaller copies of the experimental samples. We used a caliper to measure the average diameter of the yarn \textit{in situ}. We then dissected them to obtain average yarn lengths per stitch for the four types of fabric.

\seg{We additionally fabricated five sets of samples (where each set contained the four types of fabrics) made from different lace weight yarns. Of the five, three yarns were from ColourMart\texttrademark{}: heavy lace weight alpaca mohair silk mokka 811 ecru, heavy lace weight kid mohair and silk special celeste, and 2/28NM lace weight cashmere 8l brume (beige) each referred to as ``lace-weight alpaca mohair", ``lace-weight blue mohair", and ``lace-weight cashmere" respectively. The other two lace weights were Bambu 12 Gauge 100\% Bamboo in the color 010 Rice from Silk City Fibers\texttrademark{}, hereafter referred to as ``lace-weight bamboo'', and Tamm Petit 2/30 T4201 White 100\% acrylic yarn from The Knit Knack Shop\texttrademark{}, hereafter referred to as ``lace-weight acrylic.'' These samples were fabricated on a STOLL CMS 530 HP Industrial Knitting Machine and each contained 32 rows and 32 columns. Stockinette and garter were made with a stitch size setting of 12 while rib and seed were made at size 11 (Supplementary Note 11 and \seg{Supplementary Fig.~14}). All other machine parameters were kept the same. Each sample was fabricated twice with buffer regions either along its vertical or horizontal axis to aid with the uniaxial stretching experiments.

Similar to the acrylic and cotton yarns mentioned above, we measured the bending rigidity for each of these yarns and extract the stress versus strain relationship via uniaxial experiments. \seg{\sout{5} Five} experiment runs were performed on each sample. 

To obtain the length of yarn per stitch for the lace weight samples, each sample was weighed and, using the mass density for the yarn types, the average length of yarn per stitch was estimated.}

\subsection*{Uniaxial Stretching Experiments}

To perform the uniaxial stretching experiments, we designed a setup such that fabric samples had external forces uniformly applied to the boundary. We 3D printed clamps to use on both ends of the fabric samples and then had a dynamometer hooked on to one of the clamps that could be moved with a threaded rod. All components of the experiment were designed to move on guiding rails to keep everything level and prevent lateral and torsional motion. We designed the clamps with several teeth to effectively hold down both ends of the fabric sample and prevent slipping. 

For each sample, we clamped the fabric on opposite ends. During the experiments, we positioned and leveled a camera above the sample. Colored pins were placed in the fabric and red points were painted on the clamps to aid with tracking during the analysis. The dynamometer was zeroed before the experiment and then incrementally moved by turning the threaded rod, applying the external force $F_x$ (or $F_y$) to the sample boundary until reaching its maximum force (30 N). Experiments were performed slowly, to approximate a quasistatic regime, stretching from a relaxed configuration to maximum extension over 1-3 minutes. \krish{An initialization is done for each sample where they are run through the entire experiment. This run is not included in the presented data as it is meant to break apart initial fiber connections and handling bias.} Between subsequent experiment runs, the fabrics were reset to their initial resting length and briefly \seg{\sout{``fluffed''} stretched} in their transverse direction. We then waited five minutes before the next experiment run. We performed experiments along both axes of the fabrics (along its $x$- and $y$-direction). 

We looked at the uniaxial response by tracking the length and waist dimensions as the external force is exerted on the boundary.
For the overall bulk response, we used Fiji (\url{https://imagej.net/Fiji}) image processing software with the TrackMate plugin to track the pins and clamps on each of the sample videos and analyzed the position change of the coordinates (see \seg{Supplementary Fig.~1}). 
The dynamometer reading was recorded using optical character recognition (OCR) on its seven segment display, ensuring stress and strain data were synchronized.
\krish{Raw experimental data for the \seg{\sout{four}} experimental runs on \seg{\sout{each fabric}} \seg{the acrylic, cotton, and therapeutic glove fabric samples} can be found in \seg{Supplementary Fig.~3}.}

\seg{For the uniaxial experiments on the lace weight samples and the therapeutic glove swatches, we use an Instron Universal Testing Machine (UTM) Model 68SC-1. We 3D printed unique clamps with teeth to fit into the machine grips and ensure no slip boundary conditions during testing. A camera is focused on the sample while stretching and two pins are placed along the transverse direction. The clamp separation is measured and the displacement is tracked by the Instron software. The samples are stretched at a rate of 0.5 mm/sec and the lace weights are stretched to 25 N while the glove samples are stretched to 30 N. The remaining experimental procedure and analysis is the same as detailed previously except that the force data was synced by matching time steps from the tracked transverse data (acquired with Fiji) and the force data (acquired with the UTM).

}

For one uniaxial experiment, we captured the nonaffine displacement fields throughout the entire sample under stress. We clamped the acrylic garter sample and dusted graphite powder to create a speckle pattern for tracking aid. The camera was again leveled above the sample but positioned closer to capture more detailed deformation. To analyze and track the displacement fields we use the 2D digital image correlation (DIC) MATLAB software, Ncorr (\url{https://www.ncorr.com/}). 

\subsection*{Yarn bending modulus measurement}

Yarn has a hierarchical filamentous structure with internal stresses and friction arising from the manufacturing process that complicates determining a bending modulus through cantilever experiments.
Since probe-based measurements, such as the three-point flexural test, inevitably lead to compression of the yarn's cross-section, we find that cantilever experiments yield the most consistent results, using simple approximations to the yarn shape.
A schematic of the setup is shown in \seg{Supplementary Fig.~4}.

Looking at four increments of yarn length ranging from 10 cm to 25 cm, we cut out \seg{\sout{5} five} samples at each length and perform bending experiments on the yarn. \seg{For the lace weight yarns, the lengths cut were 6 cm, 9 cm, 10.5 cm, 12 cm, and 15 cm.} Each sample is cut with an additional 10 cm of yarn that is adhered on a flat surface with double-sided tape. The yarn is hung off the edge of the platform and bends under its own weight due to gravity, adopting an approximately parabolic shape. A camera is positioned level to the setup and images the yarns' behavior. We apply a blur and binarize filter to the images to isolate the yarn. Taking the points that compose the yarn shape, we fit a 4\textsuperscript{th} degree polynomial to find the approximate centerline of the yarn (see Supplementary Note 2). 

\subsection*{Yarn compressibility measurement}

We used a Zwick/Roell Z010 Universal Testing Machine (UTM) to perform compression experiments on the yarn. Three yarn samples of length 20 mm (for the acrylic yarn) and 30 mm (for the cotton yarn) were compressed between a probe tip of 5 mm in diameter and a custom acrylic stage also 5 mm in diameter. The UTM probe tip was slowly lowered onto the yarn, resulting in quasistatic measurements of the restoring force as a function of probe height.
A schematic of the setup is shown in \seg{Supplementary Fig.~5}.

\subsection*{Elastica-model simulations}

To simulate the equilibrium configurations of knitted stitches, we modeled yarn as inextensible elastica with bending modulus $B$ and fixed total length $L$ per stitch.
Interactions between overlapping yarn were treated with a hard-core, soft-shell model with a functional form derived from experimental measurements.
Equilibrium configurations were determined by numerically minimizing the total yarn energy, given by the sum of the bending energy $E_{\rm bend} = (B/2)\int_0^L{\rm d}s\,\left|\partial_s\hat{\mathbf{t}}\right|^2$ and the core-shell interaction energy $V_{\rm int}$, with a fixed total length constraint.
To perform this numerical minimization, we represented yarn configurations as degree-5 B\'ezier spline curves with degrees of freedom encoded by a collection of B\'ezier curve control points.
The resulting space curves are twice continuously differentiable with respect to its arclength parameter $s$, and thus have continuous curvature.
For more details, see Supplementary Note 4.

\subsection*{Knitted Glove Prototype}

To craft the knitted glove prototype, we used Rowan\texttrademark{}  Baby Cashsoft Merino which is composed of 57\% wool, 33\% acrylic, and 10\% cashmere. We used four different colors to knit the four types of fabric in the glove: blue for stockinette, orange for garter, green for rib, and pink for seed. 

The fabric was knit by hand (see Supplementary Note 11) on \seg{US} size 2 needles (2.75 mm in diameter) except for the stockinette regions which were knit on \seg{US} size 0 needles (2 mm in diameter). 

Miniature test swatches were made of each type of fabric to assist with glove design and to perform uniaxial stretching experiments on. \seg{A stockinette sample was knitted on US size 2 needles for comparison.} \seg{\sout{One experimental run for each sample was completed.}} \seg{Each sample underwent \sout{5} five experimental runs on the UTM.} Each sample has 25 columns and \seg{\sout{32} 34} rows.

\section*{Supplementary Text}

\subsection{\label{sec:uniaxial} Uniaxial stretching experiment}

From the uniaxial stretching experiments, we extracted the average fabric response by expressing the applied force $F_x$ (or $F_y$) in terms of stress components $\sigma_{xx} = F_x/W_y$ (or $\sigma_{yy} = F_y/W_x$), where $W_x$ and $W_y$ are fabric widths measured at the clamped edges.

\begin{figure}[ht!]
\centering
\includegraphics[width=\textwidth]{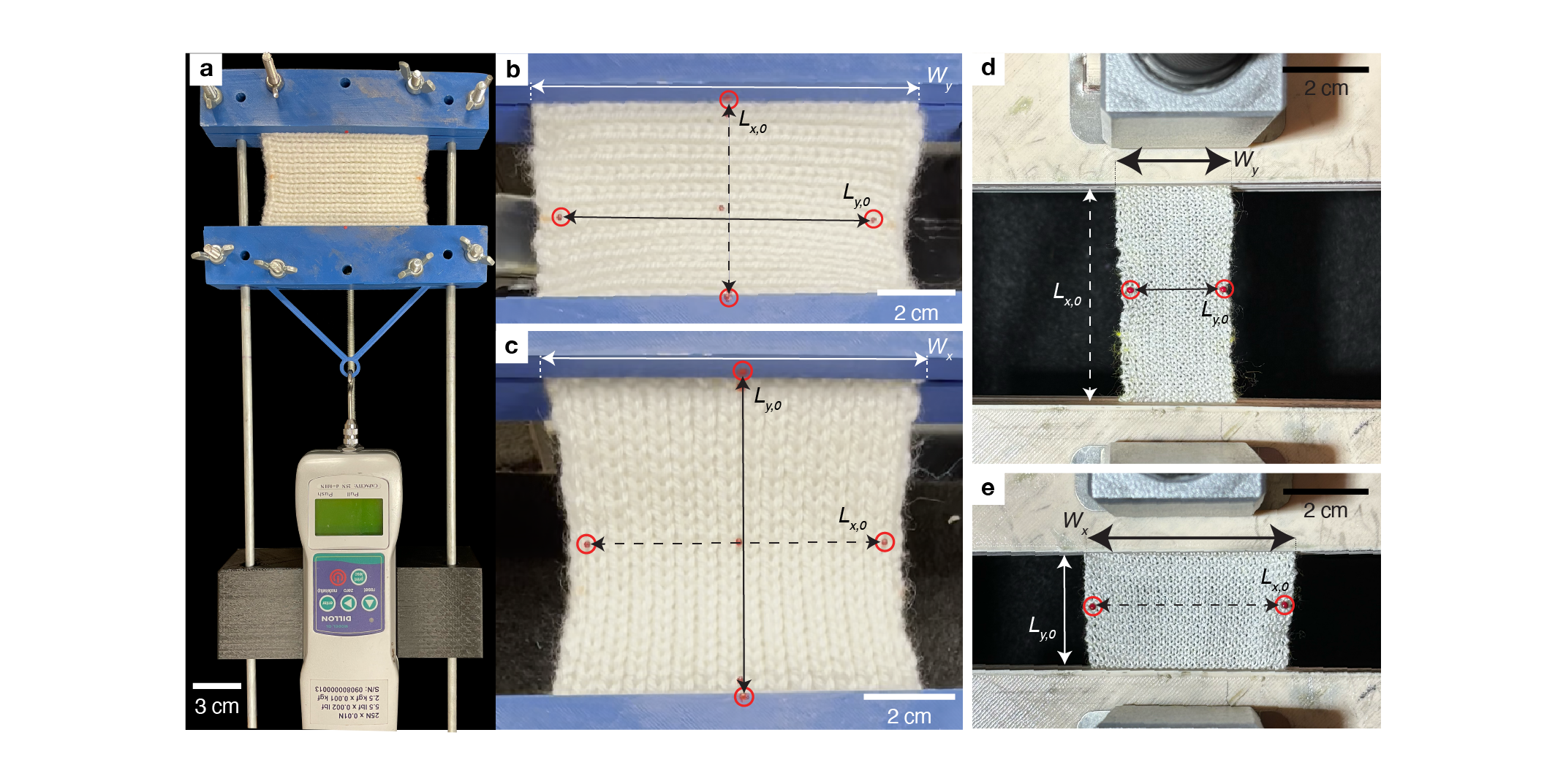}
\caption{\label{fig:pin_tracking} (\textbf{a}) Top view of the experimental setup. During experiments we control displacement while measuring force. We perform uniaxial stress-strain experiments with forces applied along the rows (\textbf{b}) and columns (\textbf{c}). A rib fabric being stretched along its (\textbf{b}) $x$- and (\textbf{c}) $y$-directions. The four points we tracked to characterize the bulk response of the fabric are circled in red. The initial pin separation along the $x$-axis is $L_{x,0}$ and along the $y$-axis is $L_{y,0}$. The values $W_x$ and $W_y$ are the widths of the fabric held down at the clamps. \seg{The lace weight and the therapeutic glove prototype samples underwent uniaxial stretching experiments on an Instron UTM. A bamboo garter sample stretched along its (\textbf{c}) $x$- and (\textbf{d}) $y$-axis. These tests only require two red pins to be tracked while the displacement between the clamps is tracked by the UTM software.}}
\end{figure}

To obtain the average strain response of the fabric, we focused on the displacement of four points placed on the axes of symmetry of the fabric. Two points were \msd{marked with} pins located on the transverse axis and the other two points \msd{were marked by painted dots} on the clamps along the direction of stretching.
In the case where the fabric is stretched along the $y$-direction, two red points painted on the clamps are aligned such that the line connecting them lies along the y-axis and has length $L_y$. 
The other two pins were positioned close to the waist of the fabric such that the line connecting them lies along the $x$-axis and has length $L_x$. 
\seg{For the lace weight and the glove prototype samples, there are only two pins along the transverse axis. The clamp displacement is tracked via the UTM.}
The principal components of the strain tensor are then $\varepsilon_{xx} = (L_x - L_{x,0})/L_{x,0}$ and $\varepsilon_{yy} = (L_y - L_{y,0})/L_{y,0}$, where $L_{y,0}$ and $L_{y,0}$ are the respective pin separations of the un-stretched fabric (see \ref{fig:pin_tracking}).

\begin{figure}[ht!]
\centering
\includegraphics[width=\textwidth]{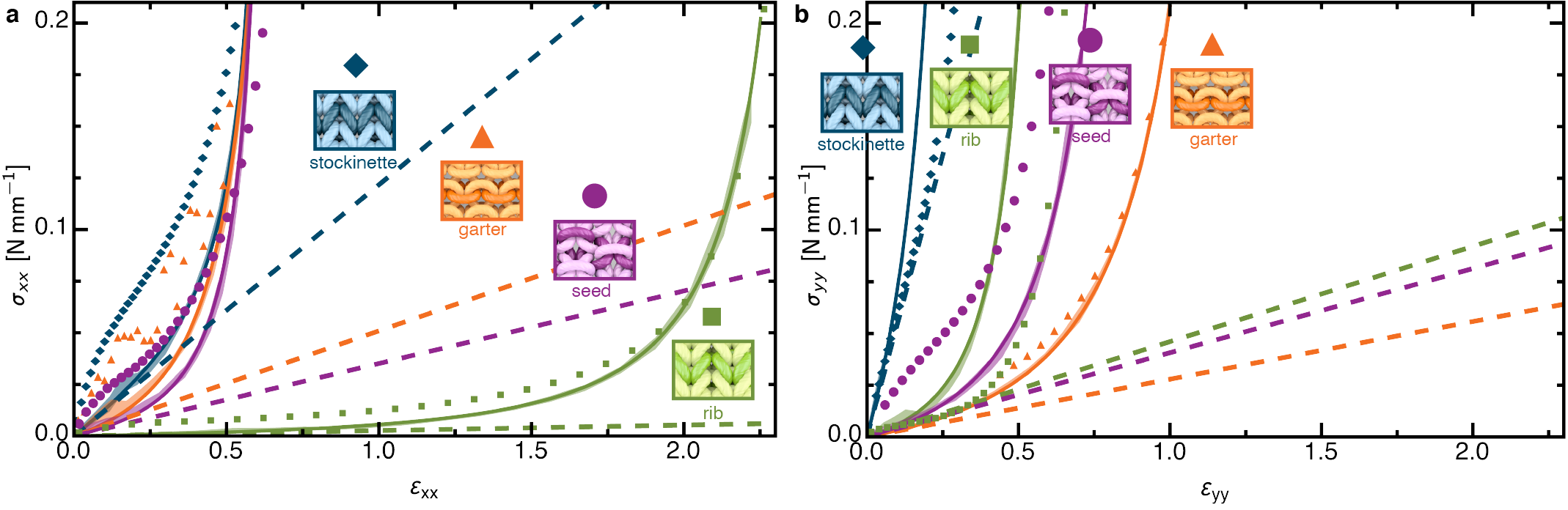}
\caption{\label{fig:cotton} The stress-versus-strain relations for the four fabrics made from the cotton yarn in the (\textbf{a}) $x$- and (\textbf{b}) $y$-directions. 
All of the data for each type of fabric is displayed by a different color: stockinette in blue, garter in orange, rib in green, and seed in purple.
The experimental data is shown in the translucent regions where the width of the region is one standard deviation of the \note{four experiment runs}. 
The simulation data is shown with solid symbols.
The solid curves are fits to the constitutive relations. 
Dashed lines depict the linear response at zero stress. 
Experiments applying force in the $x$-direction show the extreme extensibility of the rib pattern compared with the other three. 
Garter and seed dominate in the $y$-direction. 
}
\end{figure}

Figure \ref{fig:cotton} shows the results of the stress-strain analysis for the cotton yarn.

\begin{figure}[ht!]
\centering
\includegraphics[width=\textwidth]{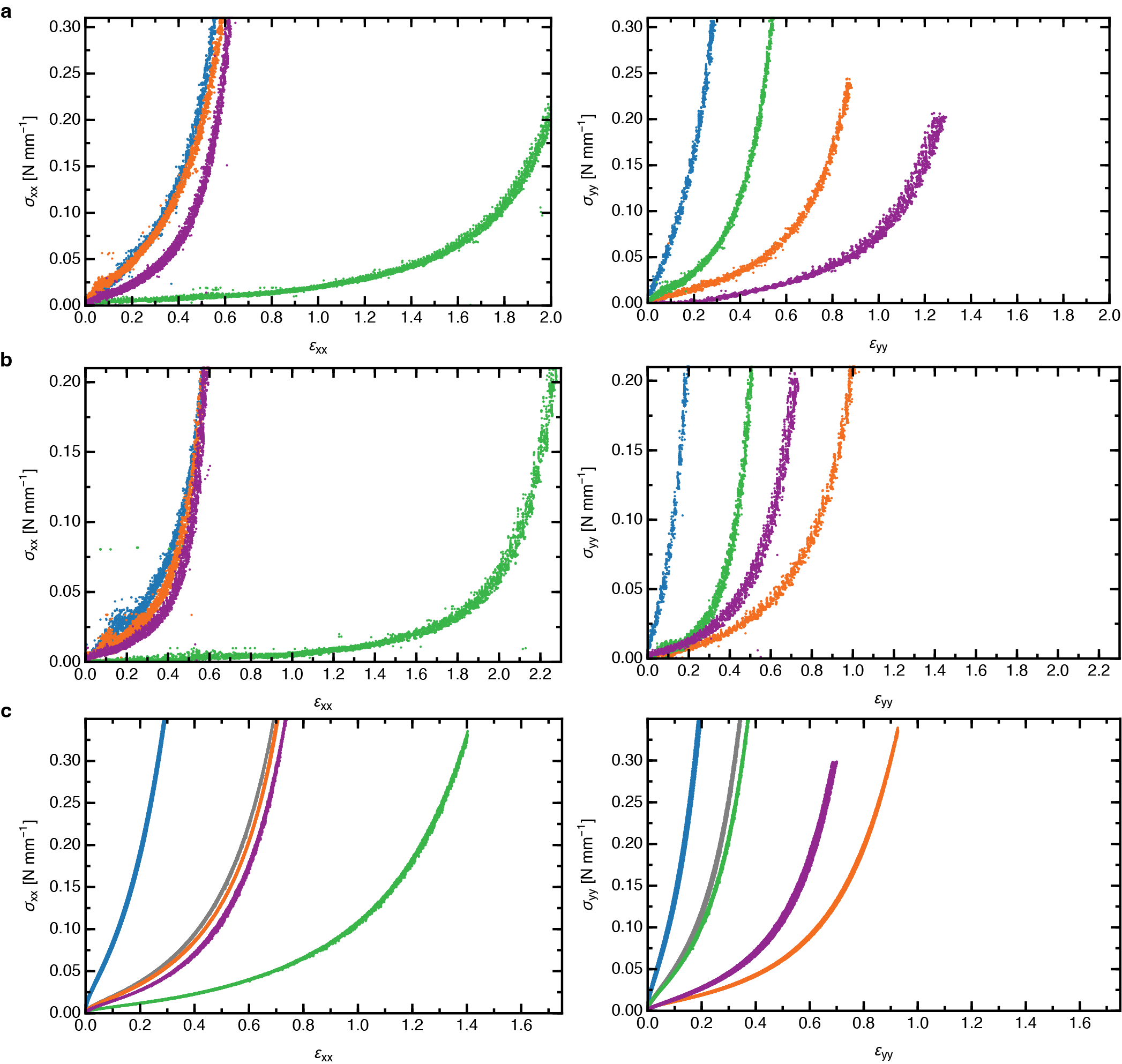}
\caption{\label{fig:rawData} The raw stress-versus-strain experimental data for the (\textbf{a}) acrylic, (\textbf{b}) cotton, and (\textbf{c}), therapeutic glove samples. All of the data for each type of fabric is displayed by a different color: stockinette in blue, garter in orange, rib in green, and seed in purple. \seg{For the therapeutic glove, there is an additional experimental data set shown in gray for a stockinette sample made with 2.75mm knitting needles.}
}
\end{figure}

\subsection{Measuring the bending modulus}

\begin{figure}[ht!]
\centering
\includegraphics[width=\textwidth]{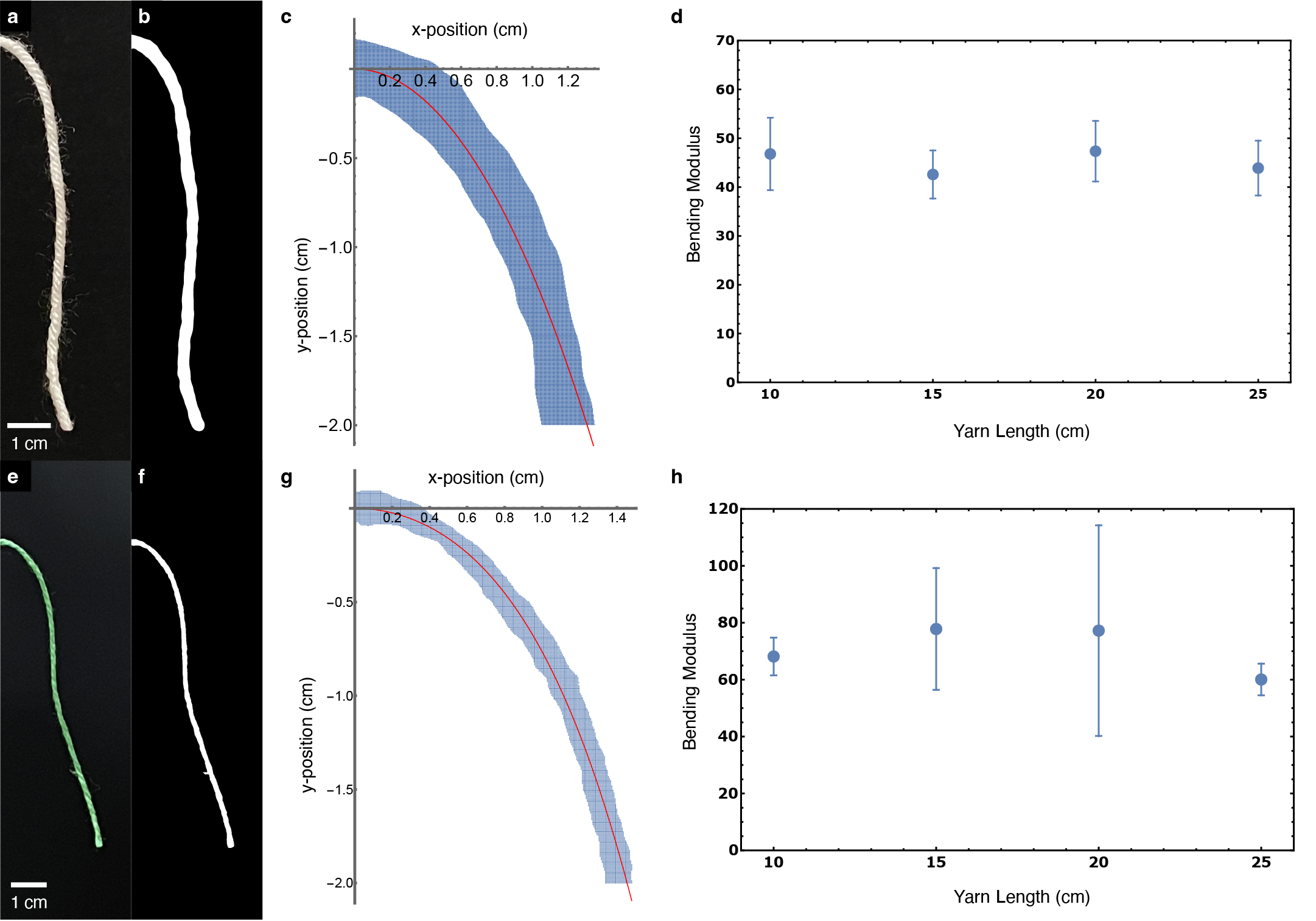}
\caption{\label{fig:bending} Depiction of the method used to determine yarn bending rigidity. The acrylic yarn bending procedure and results are shown in (\textbf{a-d}) while the cotton yarn in (\textbf{e-h}). (\textbf{a,e}) A segment of yarn is allowed to hang off of a table, with the left end held in place by double sided tape (not shown). The image of the hanging yarn's shape is then binarized with a chosen cutoff intensity to yield the region shown in (\textbf{b,f}). The curvature near the taped end is found by fitting a quartic polynomial to the binarized image, with (\textbf{c,g}) showing a close-up view of the clamped end. (\textbf{d,h}) shows the mean and standard deviation of the measured bending modulus for yarn samples of lengths 10 mm - 25 mm. }
\end{figure}

The yarn of linear mass density $\lambda$ and length $L$ is adhered to the edge of a flat surface at $x = 0$ and $y = 0$ and the free end is allowed to drape under gravity.
Equilibrium distributions of the internal moment $\mathbf{M}(s)$ and force $\mathbf{T}(s)$ are governed by the Kirchhoff rod equations~\cite{LandauLifshitz1986}, 
\begin{subequations}
\begin{align}
    \partial_s \mathbf{T}(s) - \lambda g \hat{\mathbf{y}} &= 0 \, , \label{eq:bending_1}\\
    \partial_s \mathbf{M}(s) + \hat{\mathbf{t}}(s) \times \mathbf{T}(s) &= 0 \, , \label{eq:bending_2}
\end{align}
\end{subequations}
where $s \in [0,L]$ is the arclength coordinate with $s=0$ at the fixed end and $s=L$ at the free end.
Integrating Eqs.~\ref{eq:bending_1}, \ref{eq:bending_2} and using the free end boundary conditions $\mathbf{T}(L) = \mathbf{M}(L) = 0$, we have
\begin{subequations}
\begin{align}
    \mathbf{T}(s) &= \lambda g (s - L)\hat{\mathbf{y}} \, , \\
    \mathbf{M}(s) &= \lambda g \hat{\mathbf{y}} \cdot \int_s^L {\rm d}s'\,\hat{\mathbf{t}}(s')(s' - L) \, . \label{eq:internal_moment_s}
\end{align}
\end{subequations}
Evaluating \ref{eq:internal_moment_s} at the clamping point $s = 0$ and integrating by parts, we find
\begin{equation}
    \mathbf{M}(0) = -\lambda L g x^* \hat{\mathbf{y}} \, ,
\end{equation}
where $x^* = L^{-1}\int_0^L {\rm d}s\, x(s)$ is the $x$-component of the center of mass of the hanging yarn, recovering the basic result that the total gravitational torque applied to the yarn at the fixed boundary is simply the total gravitational force of the yarn, $-\lambda L g \hat{\mathbf{y}}$, times its lever arm $x^*$.
Finally, assuming the linear constitutive relationship $M(s) = B \kappa(s)$, we solve for the bending modulus $B$ of the yarn, viz.
\begin{equation}
    B = \frac{\lambda L g x^*}{\kappa(0)} \label{eq:bending_mod_result} \, ,
\end{equation}
where $\kappa(0)$ is the curvature discontinuity of the yarn at the clamping point. We perform experiments on yarn samples of varying length (see \ref{fig:bending}).

\seg{In order to extract the curvature discontinuity at the suspension point, we must determine the shape of the yarn in space. After an image is taken and cropped to the suspended yarn, the exposure is adjusted to maximize contrast between the yarn and the background.}
\krish{We apply a blur and binarize filter to the yarn images and fit the white pixels in the domain $y > -y_{\rm max}/5$ to the 4th-order polynomial curve $y(x) = (a/2) x^2 + (b/4) x^4$. This is reminiscent of how Cornelissen and Akkerman \cite{corn} used a polynomial fit to study yarn deflection during cantilever experiments.}
Images are blurred based on how many pixels out we can see stray fibers.
\seg{For yarns containing certain fibers, stray filaments will inhomogeneously extend many yarn-radii from the spun center of the yarn. For these exceptionally fuzzy yarn types (alpaca mohair and blue mohair), the blur required to erase stray fibers is so great that a reliable fit to the core of the yarn is not feasible. In these cases, a simple blur filter is not sufficient in isolating the core and, after adjusting the exposure, the outer halo of the yarn was manually painted out before filtering the image.}
Using \seg{\sout{this} the polynomial} fit, we extract the curvature discontinuity $\kappa(0) = a/(1 + a^2)^{3/2}$.
The $x$-component of the center of mass, $x^*$, is approximated by the average over all $x$-coordinates of the binarized image of the yarn.

\seg{\sout{For acrylic, we find $B \approx 45.15 \pm 5.96$ mN mm\textsuperscript{2}. 
For cotton, we find $B \approx 70.8 \pm 21.4$ mN mm\textsuperscript{2}.
The results of these fits are also reported in \ref{table:yarn_params}.}}
\seg{The results of these fits for all yarn types used in this study are reported in \ref{table:lacebending}.}

\begin{table}[h!]
\centering
{
\setlength{\tabcolsep}{2mm}
\begin{tabular}{|c|c|}
    \hline
     \textbf{Yarn Type} & $B$ (mN mm\textsuperscript{2})  \\
    \hline\hline 
    \seg{Acrylic Yarn} &\seg{$45.15 \pm 5.96$} \\
    \hline
    \seg{Cotton Yarn} &\seg{$70.8 \pm 21.4$} \\
    \hline
    \seg{Wool blend (glove)} & \seg{$32.80 \pm 4.24$} \\
    \hline
    \seg{Lace-Weight Alpaca Mohair} &\seg{$6.44 \pm 2.04$} \\
    \hline
    \seg{Lace-Weight Blue Mohair} & \seg{$4.89 \pm 1.33$} \\
    \hline
    \seg{Lace-Weight Cashmere} & \seg{$4.34 \pm 1.07$} \\
    \hline
    \seg{Lace-Weight Bamboo} & \seg{$3.48 \pm 0.78$} \\
    \hline
    \seg{Lace-Weight Acrylic} & \seg{$3.72 \pm 0.81$} \\
    \hline
    
\end{tabular}
}
\caption{\label{table:lacebending} \seg{List of yarn bending moduli obtained from cantilever experiments for all yarns used in this study.} }
\end{table}

\subsection{Measuring the yarn compressibility and effective interaction potential}

\begin{figure}[ht!]
\centering
\includegraphics[]{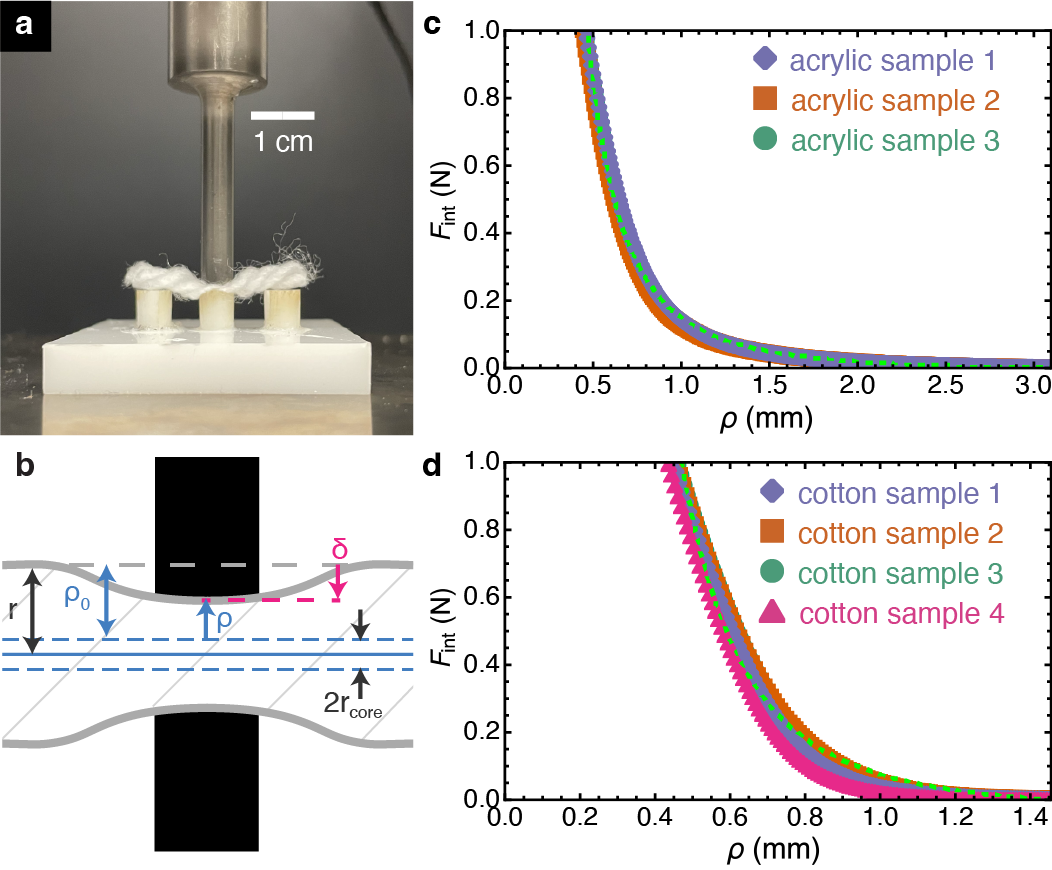}
\caption{\label{fig:compression} Depiction of the method used to determine the restoring force of the yarn under compression. (\textbf{a}) shows the experimental setup with the UTM probe pressing down on yarn that is supported by three rigid pillars. The center pillar has width equal to the probe diameter to best approximate the symmetric deformation illustrated in (\textbf{b}). (\textbf{c}) and (\textbf{d}) show the measured restoring force as a function of probe-to-midline distance $\rho$ for acrylic and cotton samples, respectively. The dashed curve is a fit of the data to the assumed function form given in \ref{eq:compress_fit}.}
\end{figure}

Using force vs.~probe height compression measurements from the yarn compression experiments (see \ref{fig:compression}), we find an effective yarn interaction potential energy for use in the simulations.
The yarn compression data show nonlinear behavior of the yarn's resisting force as a function of probe height, exhibiting a soft regime for low compression that stiffens as the constituent fibers are forced to pack into a small volume for high compression.
Noting that the stress versus strain measurements of the fabric attain maximum stresses of approximately 1 N/mm, we argue that it is sufficient to find an approximate force versus compression depth that follows the yarn compression data up to 1 N.
We fit the compression data to a model force law given by
\begin{equation} \label{eq:compress_fit}
    F_{\rm probe}(\rho) = A_{\rm comp}k\frac{\rho_0}{p}\left[\left(\frac{\rho_0}{\rho}\right)^p - 1\right]
\end{equation}
for forces between $10^{-3}$ N and 1 N, where the lower bound was chosen to cut out fluctuations on the measured force presumably due to the corona of wispy fibers sticking out of the yarn.
\msd{Note that this form is similar to the contact interaction assumed by \note{Kaldor 2008} \cite{kaldor8}, except that the exponent $p$ is left as a fitting parameter.}
Here, $\rho = \delta_{\rm c} - \delta$ is the thickness of the yarn when the probe is at depth $\delta$, where $\delta_{\rm c}$ is a cutoff depth, modeling an effective incompressible ``core'' of the yarn.
We use the probe depth at 3 N as the cutoff height.
This model was chosen because it captures the compression-stiffening behavior of the yarn for low to moderate compression, where $p > 1$ is a fitting parameter that encodes this nonlinear behavior.
Furthermore, $F_{\rm probe} \rightarrow 0$ as $\rho \rightarrow \rho_0$, where $\rho_0 = \delta_c - \delta_0$ is the uncompressed thickness of the yarn, where $\delta_0$ is the probe depth at the edge of the yarn (here taken to be when $F_{\rm probe} \approx 10^{-3}$ N).
The fitting parameter $k$ sets the scale of the yarn's compressional rigidity per area, with $A_{\rm comp}$ representing the compressed area of the yarn, which we approximate as the diameter of the UTM tool (5 mm) times the diameter of the yarn being compressed.
For small compressions, where $\rho - \rho_0 = \Delta\rho \ll \rho_0$, the compression force is approximately $F_{\rm probe} \simeq -A_{\rm comp}k\Delta\rho$ and $A_{\rm comp}k$ measures an effective spring constant.
The results of these fits are shown in \ref{table:yarn_params}.

In the simulations, two yarn segments in contact mutually compress each other.
We approximate the compression in terms of the centerline-to-centerline distance of two yarn segments, $R(s,s') \equiv |\bm{\gamma}(s) - \bm{\gamma}(s')|$, where $\bm{\gamma}(s)$ and $\bm{\gamma}(s')$ are two centerline points.
For fixed centerline points, the compressed thickness is taken to be $\rho = R - 2r_{\rm core}$, where $r_{\rm core}$ is an effective ``core radius'' of the yarn, representing the hard core cutoff radius, and $\rho_0 = 2r - 2r_{\rm core}$, where $r$ is the outer radius of the yarn.
\msd{This type of soft-shell, hard-core model has been used previously in the simulation method of \note{Sperl 2022} \cite{sperl_estimation_2022}.}
The interaction potential energy density is given by
\begin{equation}
    \mathcal{V}_{\rm int}(\zeta) = \left\{\begin{array}{cc}
    k\frac{(2r - 2r_{\rm core})^2}{p(p-1)}\left[\zeta^{1-p} - 1 - (p-1)(1 - \zeta)\right] & \,\,\text{for}\,\, \zeta < 1 \\
    0 & \,\,\text{for}\,\, \zeta \geq 1 \end{array}\right.\, ,
\end{equation}
where $\zeta \equiv (R - 2r_{\rm core})/(2r - 2r_{\rm core})$ is a non-dimensional measure of compressed thickness.

\krish{For a more physically accurate model of compression in the future, we would like to study compression in the method that \note{Park and Oh \cite{park_bending_2006}} developed for bending, which takes into account the hierarchical structure of spun yarn.}

\begin{table}[h!]
\centering
{
\setlength{\tabcolsep}{2mm}
\begin{tabular}{|c|c|c|c|c|c|}
    \hline
    \textbf{Method} & \textbf{Measurement} & \textbf{Stockinette} & \textbf{Garter} & \textbf{Rib} & \textbf{Seed} \\
    \hline\hline
    \multirow{2}{*}{Machine Knit} & \begin{tabular}{@{}c@{}}Yarn per \\ stitch (mm)\end{tabular} & $11.28 \pm 0.62$ & $10.55 \pm 0.27$ & $16.10 \pm 1.58$ & $15.67 \pm 0.76$ \\
    \cline{2-6}
    & \begin{tabular}{@{}c@{}} Yarn diameter \\ (mm)\end{tabular}  & $1.47 \pm 0.11$ & $2.07 \pm 0.10$ & $2.35 \pm 0.15$ & $2.39 \pm 0.12$ \\
    \hline
    \multirow{2}{*}{Hand Knit} & \begin{tabular}{@{}c@{}}Yarn per \\ stitch (mm)\end{tabular} & $17.85 \pm 0.95$ & $18.07 \pm 0.81$ & $18.32 \pm 0.95$ & $18.33 \pm 1.15$ \\
    \cline{2-6}
    & \begin{tabular}{@{}c@{}} Yarn diameter \\ (mm)\end{tabular} & $2.16 \pm 0.15$ & $2.42 \pm 0.15$ & $2.49 \pm 0.25$ & $2.78 \pm 0.32$ \\
    \hline
\end{tabular}
}
\caption{\label{table:acrylic_dims} The average yarn per stitch and yarn diameter within the stitches for the four types of fabrics made with acrylic yarn. The diameters were measured while the fabrics were in their relaxed (force-free) state. We created samples both by hand and with the knitting machine and note the significant changes in the range of values between the two methods.}
\end{table}

\begin{table}[h!]
\centering
{
\setlength{\tabcolsep}{2mm}
\begin{tabular}{|c|c|c|c|c|}
    \hline
    \textbf{Measurement} & \textbf{Stockinette} & \textbf{Garter} & \textbf{Rib} & \textbf{Seed} \\
    \hline\hline \begin{tabular}{@{}c@{}}Yarn per \\ stitch (mm)\end{tabular} & $12.28 \pm 0.59$ & $16.05 \pm 0.54$ & $17.94 \pm 0.40$ & $17.32 \pm 1.10$ \\
    \hline
    \begin{tabular}{@{}c@{}} Yarn diameter \\ (mm)\end{tabular}  & $1.31 \pm 0.11$ & $1.59 \pm 0.22$ & $1.38 \pm 0.08$ & $1.49 \pm 0.14$  \\
    \hline
\end{tabular}
}
\caption{\label{table:cotton_dims} The average yarn per stitch and relaxed yarn diameter within the stitches for the four types of fabrics made with the cotton yarn. The diameters were measured while the fabrics were in their relaxed (force-free) state. All of the samples were made using the knitting machine.}
\end{table}

\begin{table}[h!]
\centering
{
\setlength{\tabcolsep}{2mm}
\begin{tabular}{|c|c|c|c|}
    \hline
     & $B$ (mN mm\textsuperscript{2}) & $k$ (mN mm\textsuperscript{-2}) & $p$  \\
    \hline\hline 
    Acrylic yarn (3 samples) &\msd{$45.15 \pm 5.96$} & \fin{$0.62 \pm 0.10$} & \fin{$2.42 \pm 0.02$} \\
    \hline
    Cotton yarn (4 samples) & \fin{$70.8 \pm 21.4$} & \fin{$11.49 \pm 3.00$} & \fin{$2.94 \pm 0.11$} \\
    \hline
\end{tabular}
}
\caption{\label{table:yarn_params} List of yarn bending moduli ($B$), obtained from cantilever experiments, and compression model parameters ($k$ and $p$), obtained by fitting to UTM data. }
\end{table}

\begin{table}[h!]
\centering
{
\setlength{\tabcolsep}{2mm}
\begin{tabular}{|c|c|c|c|c|}
    \hline
    \textbf{Yarn Type} & \textbf{Stockinette} & \textbf{Garter} & \textbf{Rib} & \textbf{Seed} \\
    \hline\hline 
    \begin{tabular}{@{}c@{}} \seg{Acrylic Yarn} \end{tabular} & \seg{$ 7.74 $} & \seg{$ 7.98 $} & \seg{$ 8.50 $} & \seg{$ 10.70 $} \\
    \hline
    \begin{tabular}{@{}c@{}} \seg{Cotton Yarn} \end{tabular}  & \seg{$ 7.49 $} & \seg{$ 7.00 $} & \seg{$ 8.81 $} & \seg{$ 11.81 $}  \\
    \hline
\end{tabular}
}
\caption{\label{table:laceweight_dims} \seg{The average area per stitch (in mm\textsuperscript{2}) for the four types of fabrics made with acrylic and cotton yarn. The caliper used to measure the stitch areas had a measurement precision of 0.01 mm.}}
\end{table}

\begin{table}[h!]
\centering
{
\setlength{\tabcolsep}{2mm}

{\color{black}\begin{tabular}{|c|c|c|c|c|c|}
    \hline
    \textbf{Yarn Type} & \textbf{Measurement} & \textbf{Stockinette} & \textbf{Garter} & \textbf{Rib} & \textbf{Seed} \\
    \hline\hline
    \multirow{2}{*}{\begin{tabular}{@{}c@{}}Lace-Weight \\ Alpaca Mohair\end{tabular}} & \begin{tabular}{@{}c@{}}Yarn per \\ stitch (mm)\end{tabular} & $6.93 \pm  0.35 $ & $6.47 \pm 0.33 $ & $ 7.40 \pm 0.38 $ & $ 7.40 \pm 0.38 $ \\
    \cline{2-6}
    & \begin{tabular}{@{}c@{}} Stitch Area \\ (mm\textsuperscript{2})\end{tabular}  & $ 1.78 $ & $ 1.68 $ & $ 1.61 $ & $ 2.15 $ \\
    \hline
    \multirow{2}{*}{\begin{tabular}{@{}c@{}}Lace-Weight \\ Blue Mohair\end{tabular}} & \begin{tabular}{@{}c@{}}Yarn per \\ stitch (mm)\end{tabular} & $ 6.73 \pm 0.09 $ & $ 6.28 \pm 0.08 $ & $ 7.40 \pm 0.10 $ & $ 7.62 \pm 0.10 $ \\
    \cline{2-6}
    & \begin{tabular}{@{}c@{}} Stitch Area \\ (mm\textsuperscript{2})\end{tabular} & $1.87$ & $1.77$ & $1.60$ & $2.15$ \\
    \hline
    \multirow{2}{*}{\begin{tabular}{@{}c@{}}Lace-Weight \\ Cashmere \end{tabular}} & \begin{tabular}{@{}c@{}}Yarn per \\ stitch (mm)\end{tabular} & $6.03 \pm 0.32 $ & $ 6.03 \pm 0.32 $ & $ 6.82 \pm 0.36 $ & $ 7.08 \pm 0.38 $ \\
    \cline{2-6}
    & \begin{tabular}{@{}c@{}} Stitch Area \\ (mm\textsuperscript{2})\end{tabular} & $ 1.83 $ & $ 1.63 $ & $ 1.72 $ & $ 2.17 $ \\
    \hline
    \multirow{2}{*}{\begin{tabular}{@{}c@{}}Lace-Weight \\ Bamboo \end{tabular}} & \begin{tabular}{@{}c@{}}Yarn per \\ stitch (mm)\end{tabular} & $ 5.49 \pm 0.65 $ & $ 5.99 \pm 0.71 $ & $ 6.73 \pm 0.80 $ & $ 6.48 \pm 0.77 $ \\
    \cline{2-6}
    & \begin{tabular}{@{}c@{}} Stitch Area \\ (mm\textsuperscript{2})\end{tabular} & $ 1.76 $ & $ 1.70 $ & $ 1.41 $ & $ 1.85 $ \\
    \hline
    \multirow{2}{*}{\begin{tabular}{@{}c@{}}Lace-Weight \\ Acrylic \end{tabular}} & \begin{tabular}{@{}c@{}}Yarn per \\ stitch (mm)\end{tabular} & $ 6.46 \pm 0.62 $ & $ 6.46 \pm 0.62 $ & $ 7.27 \pm 0.70 $ & $ 7.00 \pm 0.67 $ \\
    \cline{2-6}
    & \begin{tabular}{@{}c@{}} Stitch Area \\ (mm\textsuperscript{2})\end{tabular} & $ 1.88 $ & $ 1.67 $ & $ 1.62 $ & $ 2.28 $ \\
    \hline
    \multirow{2}{*}{\begin{tabular}{@{}c@{}}Wool Blend \\ (glove)\end{tabular}}  & \begin{tabular}{@{}c@{}}Yarn per \\ stitch (mm)\end{tabular} & \begin{tabular}{@{}c@{}}$ 9.83^* \pm 0.32$ \\ $11.70 \pm 0.38 $\end{tabular} & $ 11.54 \pm 0.38 $ & $ 12.60 \pm 0.41 $ & $ 13.17 \pm 0.43 $ \\
    \cline{2-6}
    & \begin{tabular}{@{}c@{}} Stitch Area \\ (mm\textsuperscript{2})\end{tabular} & \begin{tabular}{@{}c@{}}$ 5.32^*$ \\ $7.74 $\end{tabular} & $ 6.05 $ & $ 6.18 $ & $ 8.50 $ \\
    \hline

\end{tabular}
}
}
\caption{\label{table:yps_stitchArea} \seg{The average yarn per stitch and stitch area for the lace weight samples and the therapeutic glove prototype samples. The lace weight samples were fabricated with a STOLL Industrial Knitting Machine and the glove samples were hand knit. The glove sample that is starred was made on 2.00 mm knitting needles (US size 0) and all remaining glove sample data was knit on 2.75 mm knitting needles (US size 2). The caliper used to measure the stitch areas had a measurement precision of 0.01 mm.}}
\end{table}

\subsection{Simulation method}

\subsubsection{General Methodology}

\krish{There have been a number of prior studies on yarn-level mechanics of knit stitches, including full 3D continuum elasticity models of yarn \cite{Vassiliadis2007,Liu2017}, as well as reduced-order models \cite{cirio_yarn-level_2017,Liu2018}.
Our simulation method was developed to examine stitch mechanics in a way that retains sufficient detail to explore the impact of stitch geometry (including clasp geometry in the entangled regions, as well as yarn sliding effects), while involving a coarse set of yarn properties (e.g., bending modulus, resistance to compression) to enable a materials-agnostic study.} \seg{Considering knit stitches as elastica -- a continuous curve with bending energy -- is a well-established method to consider knit fabric geometry \cite{postle_24analysis_1967, semnani_new_2003, kaldor8, RAMGULAM201148, Abel2012}. Elastica methods are the middle ground between full three-dimensional continuum elastic models (FEA of the yarn itself) \cite{kyosevFEM2012, abghary1026} and simplified bead-spring models \cite{htoo_3-dimension_2017, sha2021}, originally designed for molecular dynamics of polymers and a method that imposes a non-realistic contact geometry between clasped yarns.}

We approximated the yarn as an arclength-parametrized space curve $\bm{\gamma}(s)$ embedded in Euclidean $\mathbbm{R}^3$. 
Equilibrium configurations of the yarn balance stresses due to the (i) bending rigidity of the yarn, (ii) contact interactions of the yarn against itself, and (iii) external, or applied, forces. To this end, we modeled the yarn as inextensible elastica with an interaction energy such that the total energy is given by
\begin{equation}
    E_{\rm yarn} = \int_0^L{\rm d}s\,\left\{\frac{B}{2}\left|\partial_s\hat{\mathbf{t}}(s)\right|^2 + T + V_{\rm int}[\bm{\gamma}; s]\right\}
    \label{eq:simenergy}
\end{equation}
where the unit tangent vector is given by $\hat{\mathbf{t}}(s) = \partial_s\bm{\gamma}(s)$, $T$ is a Lagrange multiplier describing an overall tension that maintains the curve at a constant length $L$, and the interaction energy is given by
\begin{equation}
    V_{\rm int}[\bm{\gamma}; s] = \frac{1}{2}\int_0^L{\rm d}s'\,\mathcal{V}_{\rm int}\left(\left|\bm{\gamma}(s) - \bm{\gamma}(s')\right|\right) \, .
\end{equation}
The interaction energy density $\mathcal{V}_{\rm int}(R)$, with $R(s,s')=|\bm{\gamma}(s)-\bm{\gamma}(s')|$ is derived from the contact force model with $f_{\rm int}=-\partial \mathcal{V}_{\rm int}/\partial R$. 
Note that we must be careful when integrating the total interaction energy to only count interactions with a minimum separation $\Delta s$ along the arclength of the yarn. 
This prevents nearby points from adding divergent contributions to the interaction energy. 
Refer to \ref{table:sim_params} for a list of parameters used in the simulations.

In order to handle the complicated geometry of a knit stitch, we decompose the curve $\bm{\gamma}(s)$ into a sequence of curve segments $\left\{\bm{\gamma}_{\sigma}(s)\right\}$ with identified endpoints $\bm{\gamma}_{\sigma}(s_{{\rm end},\sigma}) = \bm{\gamma}_{\sigma+1}(s_{{\rm start},\sigma+1})$.

To numerically minimize the total energy, we represented the curve $\bm{\gamma}_{\sigma}(s)$ as a B\'ezier curve, expanding in the Bernstein polynomial basis, namely
\begin{equation}
    \bm{\gamma}_{\sigma}\left(s(t)\right) = \sum_{n=0}^N\mathbf{k}_{\sigma,n}\beta^N_n(t)\, ,
\end{equation}
where $\left\{\mathbf{k}_{\sigma,n}\right\}$ are the \textit{control points} of the curve and
\begin{equation}
    \beta_n^N(t) \equiv \frac{N!}{n!(N-n)!}t^n(1-t)^{N-n}
\end{equation}
are Bernstein polynomials.
The parameter $t \in [0,1]$ is a re-parametrization of the arclength parameter $s\in[s_{{\rm start},\sigma}, s_{{\rm end},\sigma}]$ for each segment $\sigma$ of the resulting B\'ezier spline curve. 
However, merely requiring the global curve to be continuous allows for kinks to be introduced into the joints between curve segments. 
In order to generate realistic results, we additionally require that the unit tangent vector $\hat{\mathbf{t}}(s)$ and its derivative $\partial_s\hat{\mathbf{t}}(s) = \kappa(s)\hat{\mathbf{n}}(s)$ are continuous in space, where $\kappa(s)$ is the curvature and $\hat{\mathbf{n}}(s)$ is the unit normal vector at each point along the curve. 
These joining conditions between curve segments ensure that the global curve lies within the $\mathcal{C}^3$ continuity class, where $\partial^3_s\bm{\gamma}(s)$ is continuous everywhere along the spline. 
We chose to represent each curve segment by degree-5 ($N=5$) B\'ezier curves, each specified by \seg{six} control points. 
The choice of degree-5 B\'ezier curves simultaneously gives sufficient flexibility for our simulations whilst maintaining a relatively small number of degrees of freedom and ensuring that the simulated curves remain in the $\mathcal{C}^3$ continuity class \cite{eck_b-spline-bezier_1992}.

\begin{figure}[h!]
\centering
\includegraphics[width=\textwidth]{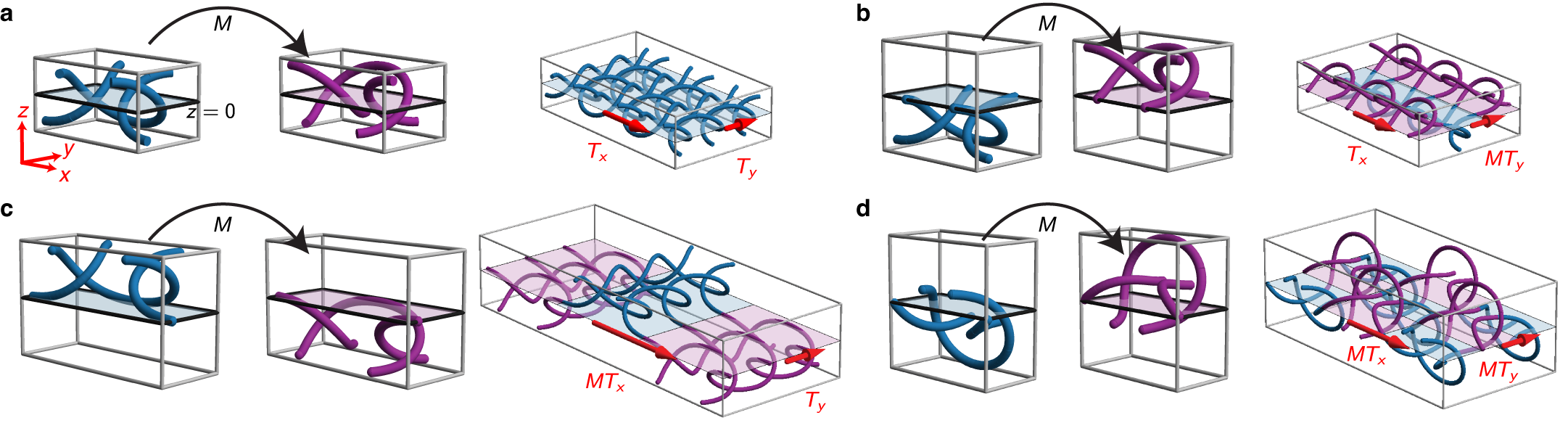}
\caption{\label{fig:comp_schematic} Depictions of the individual stitch cell construction of each of the four fabrics. \seg{These diagrams were created using sample outputs of the stitch-level simulations. The yarn radius shown is reduced significantly for clarity.} The left-most panels show the geometry of a single relaxed knit stitch contained in a box that outlines the spatial extent of the cell. The center panels show a single relaxed purl stitch, obtained from the knit stitch via the mirror operation $M$ through the $z=0$ plane, depicted by the colored plane cutting through each of the boxes. The right-most panels show the construction of (\textbf{a}) stockinette, (\textbf{b}) garter, (\textbf{c}) rib, and (\textbf{d}) seed fabrics. Each fabric is generated by the repeated action of a combination of mirror operations $M$, as well as translation operations $T_x$ and $T_y$, on the knit stitch shown in the left-most panels.}
\end{figure}

Even though this representation allows control over curve smoothness, variations in control parameters give non-local control over curve shape, giving rise to a large number of local energy minima. 
This is particularly problematic as a curve segment approaches a straight configuration, due to a degeneracy of control point arrangements for a straight curve. 
This leads to ``vertex bunching,'' a common problem in geometry optimization \cite{DeBenedictis2016}. 
To alleviate this problem, we introduced a cost functional that penalizes such configurations, characterized by small amplitude ``wiggles'' in the curve shape. 
We incorporated a regularizing energy of the form
\begin{equation}
    E_{\rm reg} = \frac{C_{\rm reg}}{2}\int_0^L{\rm d}s\,|\partial_{ss}\hat{\mathbf{t}}|^2
\end{equation}
where $C_{\rm reg}$ is a constant that controls the strength of the regularizing energy.

Rather than simulating a finite swatch of knitted fabric with boundaries, we took advantage of the symmetries of an infinite fabric without boundaries. 
This enabled a reduction in the scale of the simulation to a single stitch. 
Since the fabric is a rectangular grid, this individual stitch cell is a rectangular region of dimensions $\ell_x$ and $\ell_y$ (\ref{fig:comp_schematic}), with lattice positions indexed by a pair of integers $(m,n)$, representing the position of a cell relative to a reference cell at $m,n=0$. 
The central curve then has a periodic structure given by
\begin{equation}
    \bm{\gamma}_{(m,n)}(s) = M^{f(m,n)}T_y^nT_x^m\bm{\gamma}_{(0,0)}(s)\, ,
\end{equation}
where $T_x: \bm{\gamma}\mapsto \bm{\gamma}+\ell_x \hat{\mathbf{x}}$ and $T_y: \bm{\gamma}\mapsto \bm{\gamma}+\ell_y \hat{\mathbf{y}}$ are translation operations between stitch cells and $M \equiv (\mathbbm{1} - 2\hat{\mathbf{z}}\otimes\hat{\mathbf{z}})$ represents a mirror operation that reflects the stitch path through the midplane of the fabric, converting knits to purls, as depicted in \ref{fig:comp_schematic}. 
The function $f(m,n)$ sets the number of mirror operations $M$ that are applied at each cell and thus provides information regarding the pattern. 
It is given by $f(m,n)=0$ for stockinette fabric, $f(m,n)=n$ for garter fabric, $f(m,n)=m$ for rib fabric, and $f(m,n)=m+n$ for seed fabric.
Within a single cell, the path $\bm{\gamma}_{(0,0)}(s)$ obeys a form of periodic boundary conditions, where the mirror operation may be applied to the unit vector $\hat{\mathbf{t}}(s)$ at each boundary, depending on which stitch pattern being studied. 
We matched simulations to experiments by setting the length $L$ of the path $\bm{\gamma}_{(0,0)}(s)$ within a single simulated stitch to the measured yarn length per stitch for each manufactured sample.

We simulated the effect of fabric stretching in the $x$-direction ($y$-direction) by numerically minimizing the total yarn energy $E_{\rm yarn}[\gamma_{(0,0)}]$ at fixed stitch cell dimension $\ell_x$ ($\ell_y$), while allowing the transverse dimension $\ell_y$ ($\ell_x$) to vary.
This minimization was performed using the Sequential Least Squares Programming (SLSQP) method in the \texttt{scipy.optimize} Python package (\url{https://docs.scipy.org/doc/scipy/reference/optimize.html}), which is a gradient-free optimization algorithm that allows a number of equality and inequality constraints to be specified; in particular, we fixed the yarn length $L$ to be constant.
To generate a 1D energy landscape $E(\ell_\mu)$ (with $\mu = x,y$), we first started with a guess for an initial, un-stretched configuration, at an initial stitch dimension $\ell_{\mu}$, and numerically minimized that configuration.
We then incremented or decremented the stitch dimension $\ell_\mu \to \ell_\mu \pm \Delta \ell$ and used the result of the minimization as new initial conditions for minimizing the energy over this new cell dimension; we generated full 1D landscapes using this 0\textsuperscript{th} order parametric continuation, making sure to sweep in both $\pm \Delta \ell$ directions to bracket an energy minimum.
Since this minimization approach is prone to finding local, metastable energy minima, we performed this sweep on four different initial stitch configurations, accepting the lowest energy value as the accepted simulation result, in order to search for better approximations to the ``true ground state'' of the stitch.
Simulated annealing methods may get closer to this global minimum. 
However, we found that the B\'ezier curve representation suffers from a large number of near-degenerate configurations, complicating the application of simulated annealing methods.

With a given energy landscape $E(\ell_\mu)$, we applied a discrete, midpoint derivative and found the force profile $f_x(\ell_x) = {\rm d}E/{\rm d}\ell_x$ (or $f_y(\ell_y) = {\rm d}E/{\rm d}\ell_y$).
The completely relaxed, force-free configuration of the stitch corresponds to the point where $f_x = f_y = 0$, which can equivalently be found in either the $E(\ell_x)$ or the $E(\ell_y)$ landscapes since allowing the transverse dimension to vary freely in minimization is equivalent to specifying a zero-force condition on that dimension. The force-free configuration was found with a using 3\textsuperscript{rd} order polynomial interpolation on data where high-energy compression simulations were eliminated.
Denoting the stitch cell dimensions of that completely relaxed configuration as $\ell_{x,0}$ and $\ell_{y,0}$, we converted force data to nominal stress via $\sigma_{xx} = f_x/\ell_{y,0}$ and $\sigma_{yy} = f_y/\ell_{x,0}$, expressed as a function of the linear strain components $\varepsilon_{xx} = (\ell_x - \ell_{x,0})/\ell_{x,0}$ and $\varepsilon_{yy} = (\ell_y - \ell_{y,0})/\ell_{y,0}$.

\begin{table}[h!]
\centering
{
\setlength{\tabcolsep}{2mm}
\begin{tabular}{|c|c|c|c|c|c|c|c|}
    \hline
     & $B$ (mN mm\textsuperscript{2}) & $k$ (mN mm\textsuperscript{-2}) & $p$ & $r$ (mm) & $r_{\rm core}$ (mm) & $L$ (mm) & $r_{\rm core}/r$\\
    \hline\hline 
    \begin{tabular}{@{}c@{}}Stockinette \\ (acrylic)\end{tabular} & 46 & 0.6 & 2.4 & 0.74 & 0.335 & 11.3 & 0.453 \\
    \hline
    \begin{tabular}{@{}c@{}}Garter \\ (acrylic)\end{tabular} & 46 & 0.6 & 2.4 & 0.54 & 0.245 & 10.6 & 0.454 \\
    \hline
    \begin{tabular}{@{}c@{}}Rib \\ (acrylic)\end{tabular} & 46 & 0.6 & 2.4 & 1.18 & 0.415 & 16.1 & 0.352  \\
    \hline
    \begin{tabular}{@{}c@{}}Seed \\ (acrylic)\end{tabular} & 46 & 0.6 & 2.4 & 1.20 & 0.290 & 15.7 & 0.242 \\
    \hline
    \begin{tabular}{@{}c@{}}Stockinette \\ (cotton)\end{tabular} & 70 & 11.5 & 2.9 & 0.66 & 0.310 & 12.28 & 0.470 \\
    \hline
    \begin{tabular}{@{}c@{}}Garter \\ (cotton)\end{tabular} & 70 & 11.5 & 2.9 & 0.80 & 0.420 & 16.05 & 0.525 \\
    \hline
    \begin{tabular}{@{}c@{}}Rib \\ (cotton)\end{tabular} & 70 & 11.5 & 2.9 & 0.69 & 0.405 & 17.32 & 0.587 \\
    \hline
    \begin{tabular}{@{}c@{}}Seed \\ (cotton)\end{tabular} & 70 & 11.5 & 2.9 & 0.75 & 0.430 & 17.94 & 0.573 \\
    \hline
\end{tabular}
}
\caption{\label{table:sim_params} List of yarn material parameters used in simulations. We adjusted the hard-core radius $r_{\rm core}$ to obtain better agreement with the stress-vs-strain curves obtained from experiments. \krish{For the acrylic yarn, the core radius averages $37.5\%$ of the yarn radius with a standard deviation of $\pm 10.1\%$. For the cotton yarn, the core radius averages $53.9\%$ of the yarn radius with a standard deviation of $\pm 5.3\%$. This suggests the core radius has a dependence on the yarn type and is slightly influenced by the fabric type. How the fabric manufacturing process for different fabric types affects the core radius is currently unknown. Generally, increasing the core radius in the simulation leads to a stiffer fabric for all four fabric types, but the exact dependency of the constitutive model on the core radius is a subject of further study. Existing simulations looking to replicate experimental stretching responses often have many more fitting parameters with complex optimization schemes \cite{sperl_estimation_2022}.}}
\end{table}

\subsubsection{Restricting sliding with arclength constraints}

Our simulation method of minimizing an elastica energy functional over topologically-constrained configurations of yarn does not incorporate effects of friction.
Given the wispy, corrugated texture of the yarn, we expect that friction may play an important role in reducing the ability of the yarn to slide against itself in certain configurations.
We hypothesize that this effect may be particularly relevant for seed stitch, as their cross-over regions do not clasp as completely as other stitches, which can be seen in \ref{fig:comp_schematic}.
In particular, seed possesses relatively straight segments oriented along $\hat{\mathbf{y}}$, even in the un-stretched configuration, allowing for a soft sliding motion that is distinct from the soft near-rigid rotation of odd connecting yarn segments \mbox{(described in \ref{sec:RS_model})}.

To demonstrate the effect of contact sliding, we consider the extreme limit of quenched sliding.
This is implemented in the energy minimization through as pair of constraints that break the reptation symmetry of the yarn.
Using our decomposition of yarn into cross-overs and connections, we can approximate the arclength coordinate of the $i^{\rm th}$ contact point, $s_i$, as ``half-way'' between the ends of the yarn in the corresponding cross-over region, which are at points $s_{i,0}$ and $s_{i,1}$, taking $s_i = (s_{i,0} + s_{i,1})/2$.
The slide-quenching constraints amount to ensuring that the total arclength of yarn joining two neighboring cross-over regions remains constant under deformation, i.e. $s_{i+1} - s_i = s^0_{i+1} - s^0_{i}$, where $s_i^0$ are the corresponding arclength positions in the un-stretched state.
Within each stitch there are \seg{four} such contact points, so in principle there need to be \seg{four} such constraints. 
However, the mirror symmetry of the stitch about the $yz$-plane passing through its middle relates two of the arclength coordinates, and the total arclength constraint implies that $\sum_i s_i = L$, leaving only two constraints that need to be enforced, namely
\begin{equation}
\begin{split}
    s_2 - s_1 &= s^0_2 - s^0_1 \,\,\, {\rm and} \\
    s_3 - s_2 &= s^0_3 - s^0_2 \,\,\, .
\end{split}
\end{equation}

Note that the addition of these constraints require that stretched stitches inherit information about the un-stretched equilibrium, namely $\{s_i^0\}$.
This dependence on a reference configuration distinguishes the elasticity of slide-quenched fabrics from those that allow for sliding in a way reminiscent to the difference between elastomeric materials (e.g.~polymer gels and rubbers) that attain rigidity via permanent cross-links, versus so-called ``topological'' constraints.
We leave further explorations of quenched versus annealed sliding for future studies.

\subsubsection{Simulation energy analysis}

\begin{figure}
    \centering
    \includegraphics[width=\textwidth]{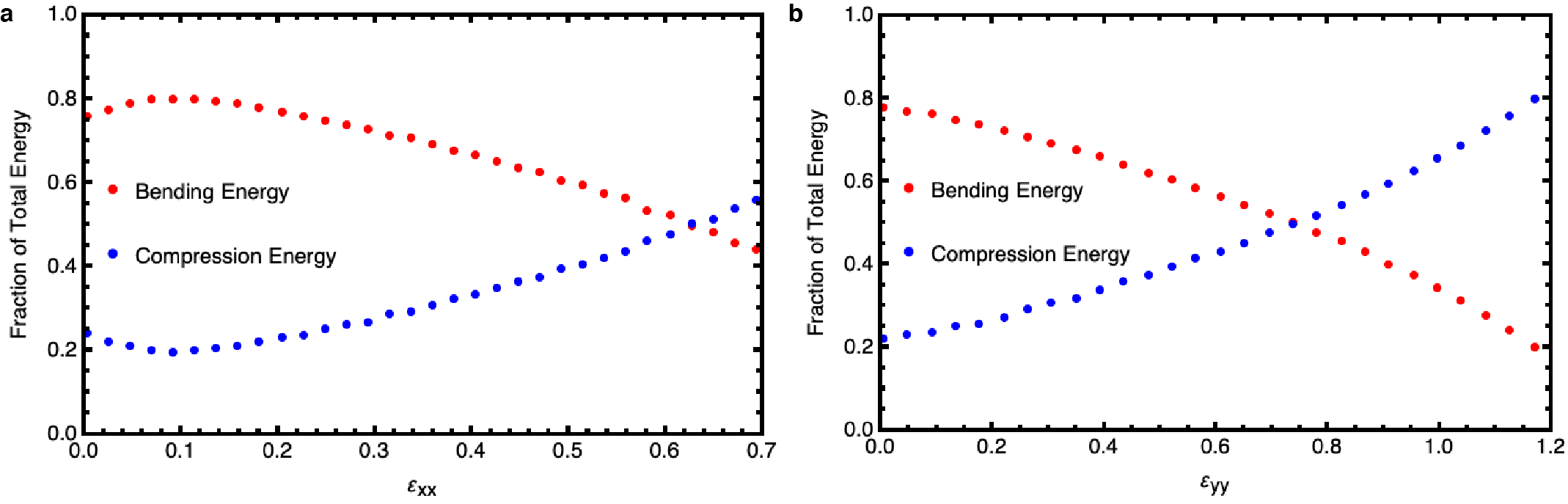}
    \caption{Energy breakdown for cotton garter sample stretched in the x- (\textbf{a}) and y-directions (\textbf{b}). The compression (blue) and bending (red) energies are given as percentages of the total energy at each value of strain.}
    \label{fig:garterenergy}
\end{figure}

\krish{Some of the value in the yarn-level simulations is the ability to determine the components of the energy of the yarn as given in \ref{eq:simenergy}. For each simulation, we can extract the components of energy due to yarn bending and yarn compression. Generally, as seen in \ref{fig:garterenergy} with the cotton garter sample, the bending energy dominates at low strain. As the strain increases, the proportion of compression energy increases, finally overcoming the bending contribution at high enough strain. This transition is typically smooth, though an exception to this is discussed in \seg{\sout{the next section,}} \ref{sec:instability}, for acrylic seed. Values for the total energy per stitch and energy ratios for all fabric types and both yarn types can be found in \ref{table:sim_energies}. This simulation data supports the Reduced Symmetry model, where bending is the dominant energy contribution at low strain.}

\begin{table}[h!]
\centering
{
\setlength{\tabcolsep}{2mm}
\begin{tabular}{|c|c|c|c|}
    \hline
     & $E_{\rm total}$ (J) & $E_{\rm compression}/E_{\rm total}$  & $E_{\rm bending}/E_{\rm total}$ \\
    \hline\hline 
    \begin{tabular}{@{}c@{}}Stockinette \\ (acrylic)\end{tabular} & 0.249 & 0.130 & 0.870  \\
    \hline
    \begin{tabular}{@{}c@{}}Garter \\ (acrylic)\end{tabular} & 0.175 & 0.026 & 0.974 \\
    \hline
    \begin{tabular}{@{}c@{}}Rib \\ (acrylic)\end{tabular} & 0.189 & 0.391 & 0.609  \\
    \hline
    \begin{tabular}{@{}c@{}}Seed \\ (acrylic)\end{tabular} & 0.288 & 0.459 & 0.541 \\
    \hline
    \begin{tabular}{@{}c@{}}Stockinette \\ (cotton)\end{tabular} & 0.401 & 0.233 & 0.767 \\
    \hline
    \begin{tabular}{@{}c@{}}Garter \\ (cotton)\end{tabular} & 0.254 & 0.216 & 0.784 \\
    \hline
    \begin{tabular}{@{}c@{}}Rib \\ (cotton)\end{tabular} & 0.139 & 0.044 & 0.956 \\
    \hline
    \begin{tabular}{@{}c@{}}Seed \\ (cotton)\end{tabular} & 0.262 & 0.172 & 0.828 \\
    \hline
\end{tabular}
}
\caption{\label{table:sim_energies} List of total energy per stitch, the ratio of compression energy to total energy, and the ratio of bending energy to total energy as given by zero-force simulations. On average over all fabric types in both yarn types, the bending energy is $79\% \pm 15\%$ of the total energy. }
\end{table}

\subsubsection{Observed ``jamming'' response in low-stress regime}

\krish{At very low stresses, we sometimes observe an initial high-rigidity response before the fabric softens into a linear stress-strain response. This behavior is seen in both experimental (see acrylic garter pulled in the x-direction and rib in the y-direction (\ref{fig:rawData}a), cotton stockinette and garter pulled in the x-direction (\ref{fig:rawData}b), and the therapeutic glove test samples for stockinette in the x-direction and stockinette and rib in the y-direction (\ref{fig:rawData}c)) and simulation (see stockinette and seed in the x-direction and seed in the y-direction in \ref{fig:cotton}) data. Other groups studying knits made with incompressible yarn have seen similar low-stress, high-rigidity regions \cite{duhovic_simulating_2006}. Postle \cite{Postle2002} described this behavior as a jammed regime where forces normal to the stretching direction prevent the yarn from rearranging and the fabric from extending. Since jamming is a contact-dependent phenomenon, jamming behavior present in simulation results that is not found in experiments may be a result of our contact model.}

\subsubsection{Instability in acrylic seed}
\label{sec:instability}

\begin{figure}
    \centering
    \includegraphics[width=\textwidth]{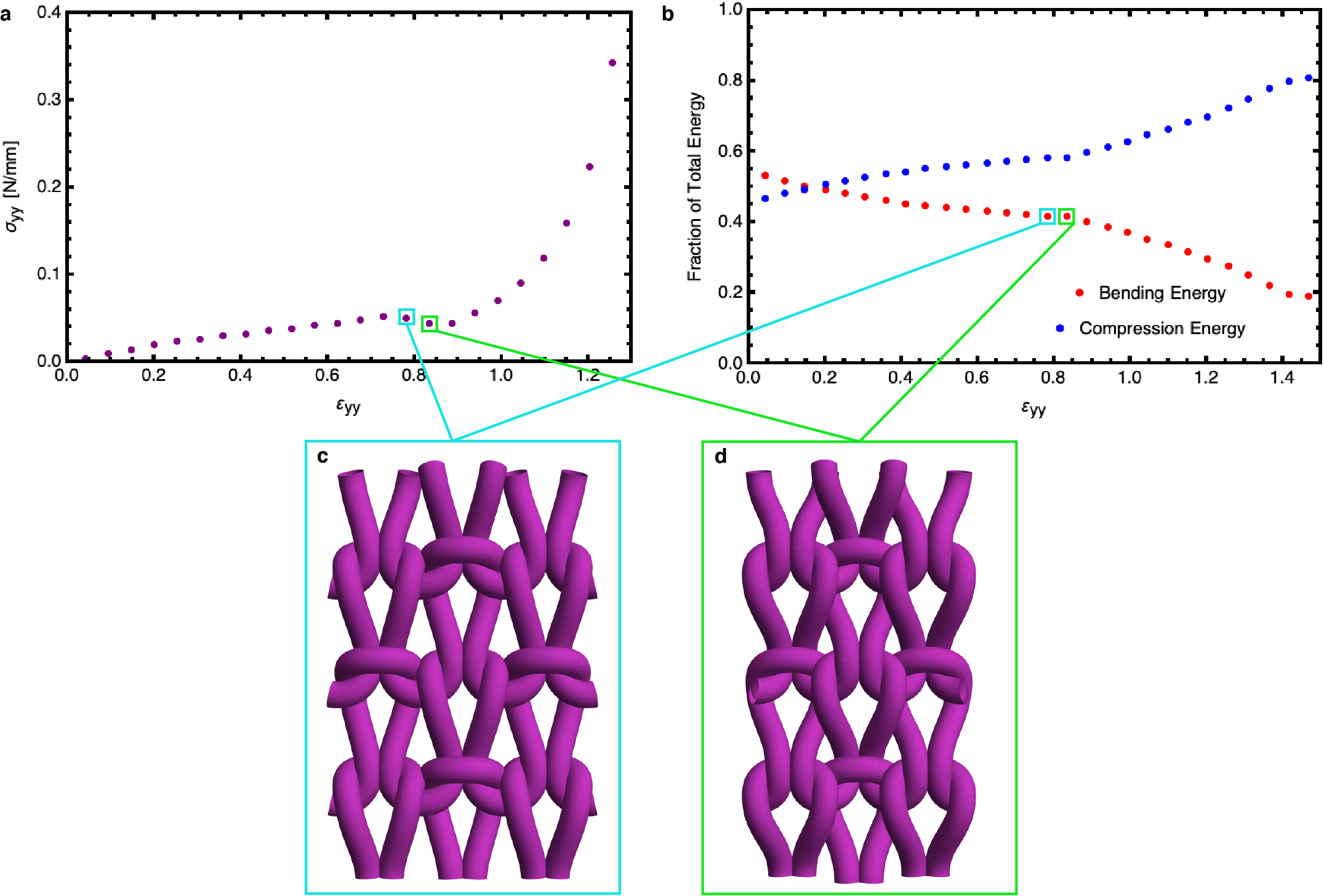}
    \caption{Energy breakdown (\textbf{b}) and stitch configurations (\textbf{c}, \textbf{d}) for the buckling instability in seed acrylic simulations that correlate to the abnormality in the stress-strain plot (\textbf{a}). The stitch configurations show a tiling of 3 by 3 stitches in the $x-y$ plane before (\textbf{c}) and after (\textbf{d}) the buckling instability occurs. The buckling is characterized by sudden overlaps of the entangled regions of neighboring stitches. These strain locations in the energy breakdown in \textbf{b} show that the onset of buckling is correlated with a decrease in the relative compression energy while the relative bending energy increases, counter to the general trends seen for each component of the energy. \seg{These renderings (\textbf{c,d}) were made using the outputs of the seed simulations.}}
    \label{fig:seedbuckle}
\end{figure}

\krish{Shown in Fig.~2, the simulation stress-strain curve for acrylic seed differs from the experimental result. Further investigation into the simulation shows a buckling instability at a strain of $\approx 0.8$, where the stitches move out of the $z$-plane to reduce the compression energy as shown in \ref{fig:seedbuckle}. Due to the checkerboard pattern of knit and purl stitches, seed fabric is uniquely able to express this out-of-plane buckling instability in comparison to other fabric types. For hard colloidal spheres, out-of-plane buckling (from two-dimensions to three-dimensions) results in a square lattice \cite{pieranski_thin_1983}. Neighboring particles want to go opposite directions out of the plane, such that a neighboring pair have one particle above the plane and one below. The stitch configuration of seed enables this transition, whereas the stitch configurations of the three other fabric types prevent it. Simulations for other stitch patterns with more limited frustration may also allow this buckling. Seed also has the highest proportion of compression energy of all the fabric types, as seen in \ref{table:sim_energies}, which would make it more susceptible to contact-induced buckling. As seen in \ref{table:acrylic_youngpoisson}, the Poisson ratio $\nu_{xy}$ for the simulation is approximately four times larger than experiment. This larger Poisson ratio is a marker of the simulation buckling, but our current analysis is insufficient to determine causality. The non-monotonic behavior of the contact energy in \ref{fig:seedbuckle}b indicates that there is an instability in the numerics caused by the contact model.} 


\seg{
\subsubsection{Accounting for Manufacturing Tension}

In the simulation, manufacturing tension is relevant during initialization of the fabric before it is stretched or deformed in any way. Others have simulated. the actual knitting process \cite{Duhovic2006}, implemented a shrinking factor that reduces the arc-length of segments of yarn until the fabric settles into a rest-state \cite{kaldor8}, or taken a picture of a physical sample and used that geometry as an input of the simulation that is then relaxed to near force-balance \cite{sperl_estimation_2022}. Many don't consider tension at all \cite{ru_modeling_2023}. Of these initialization strategies, the method we use is closest to that of Sperl et al. \cite{sperl_estimation_2022}; we start with a input geometry inspired by the actual geometry of the stitches within the fabric. We then impose constraints and yarn properties. The length constraint, which fixes the length of yarn per stitch, is how we control how tightly the stitches are manufactured. Once these input properties are imposed, we allow the simulation to find the minimum energy configuration that fulfills these constraints. In \ref{fig:tension}, we show how we can control the tightness of the stitches by changing the length of yarn per stitch.

We also consider how tension may affect the yarn structure by allowing the core radius of the yarn to vary. In this way, we account for how the yarn may change its compressibility under tension without a physical model for that phenomenon, which is currently poorly understood. We also take measurements of the yarn radius \textit{in situ} to represent changes in yarn radius under tension. Worsted weight yarn, such as the acrylic and cotton yarns used in our samples, often visibly change radius under tension.

\begin{figure}
    \centering
    \includegraphics[width=\textwidth]{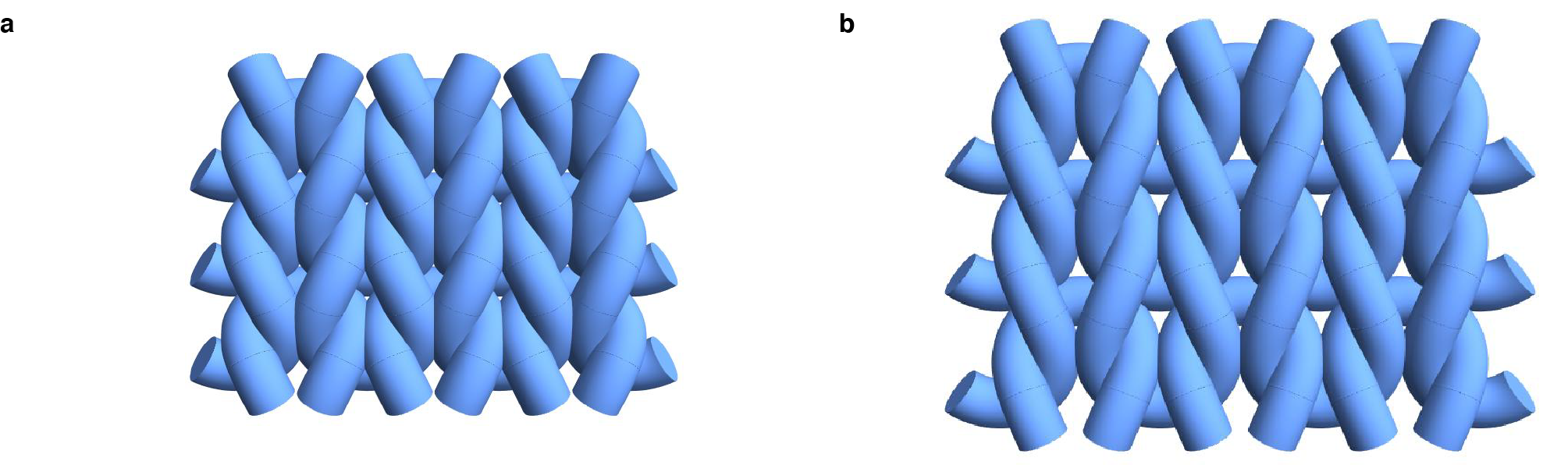}
    \caption{\seg{Stitch configurations of simulation outputs of stockinette fabric with different length of yarn per stitch constraints, \textbf{(a)} 10.5 mm  and \textbf{(b)} 12.5 mm, to represent different levels of manufacturing tension. These stitch configurations are made from multiple tilings of the stitch unit cell to create a fabric that is \seg{three} stitches wide by \seg{three} stitches high.}}
    \label{fig:tension}
\end{figure}

\subsubsection{Friction}

Friction is often included in dynamic simulations of knitted fabrics \cite{kaldor8, cirio_yarn-level_2017, duhovic_simulating_2006}. We do not include friction in our stitch-level simulations, primarily because our simulations are static. Friction can not be incorporated into an energy minimization scheme. Including a dissipative term such as friction into a static simulation is not supported by both the general simulation method and the specific way we minimize the energy.

For each set of stitch cell dimensions, we iteratively change the shape of the stitch to find the minimum energy configuration \cite{sokolnikoff1956mathematical}, as previously described. We are not continuously stretching the stitch cell. Each set of stitch cell dimensions is a single simulation, unconnected to other simulations of different stitch cell dimensions. Static simulations of this kind well suit our purposes to use simulations to investigate the role of topology on fabric mechanics. By finding the mechanical equilibrium point of the stitch cell for each set of given dimensions, we well represent the mechanics of our experiments. Our stress-strain simulation results are able to replicate the shape of both the linear and non-linear elastic responses of knit fabrics, which has yet to be achieved for knits made of compressible yarns.

As mentioned in Supplementary Note 4.1, we use the Sequential Least Squares Programming (SLSQP) method in the \texttt{scipy.optimize} Python package (\url{https://docs.scipy.org/doc/scipy/reference/optimize.html}) to conduct our energy minimization. This method of optimization does not use gradients and cannot incorporate a dissapative energy term. We chose this method due to its suitability for our specific static simulations; this minimization method is able to take large steps in the energy landscape to converge faster and can often recover from divergent energy configurations. To include a dissapative energy term like friction, we would have to move to a dynamic simulation and use a different optimization scheme, such as gradient descent.

Prior research on rib fabric made of incompressible yarn shows that energy lost to friction is very small, at most the totalling the energy of the third-largest contribution for the entirety of the knit's elastic response \cite{duhovic_simulating_2006}. Since we are using a compressible yarn with an appreciable energy contribution from yarn compression (see \ref{fig:garterenergy}) and stretching fabric in the quasi-static regime, we estimate that friction has a similarly small, if not smaller, contribution to our fabric.

We do not see frictional effects on the elasticity of the knit fabrics when we iteratively repeat extension experiments, as seen in \ref{fig:rawData}. This lack of measurable frictional effect on the experimental samples indicates that friction must be a very small contribution. This is supported by the fact that we can wear clothes multiple times without them losing their elasticity. Socks in particular retain their elasticity over multiple wears even though they are constantly being stretched and deformed with every step. If friction had a large role in knit fabric mechanics, socks would become single-use items.

}

\subsection{Transverse stress-strain behavior}

In the uniaxial stretching experiments, the fabrics attain an hourglass-like waist as the force acting along one direction causes a response in the transverse direction. 
This response is a result of the fabric’s Poisson effect, i.e.~stretching the fabric in one direction thins it in transverse directions. 
For example, if the fabric is stretched in the $x$-direction so that the strain component $\varepsilon_{xx}$ is positive, then it thins in the $y$-direction. 
Due to the constraints on transverse deformation imposed by clamps on two of the edges, the strain field component $\varepsilon_{yy}<0$ varies along the fabric, and the magnitude of the strain reaches a maximum where the waist narrows. 
The pin tracking approach we employed measures the deformation in this region.

Due to the anisotropy of each fabric, the transverse deformation is characterized by two Poisson ratios, $\nu_{yx}$ and $\nu_{xy}$.
In principle, these ratios describe the linear response of the fabric under two different deformation protocols: if the fabric is stretched in the $x$-direction (i.e.~$\varepsilon_{xx}>0$) then the transverse response is $\varepsilon_{yy} = -\nu_{yx}\varepsilon_{xx}$; if the fabric is stretched in the $y$-direction then the transverse response is $\varepsilon_{xx} = -\nu_{xy}\varepsilon_{yy}$. We measured these ratios for experimental and simulation results, and they are reported in \ref{table:acrylic_youngpoisson}, \ref{table:cotton_youngpoisson} and \ref{table:baby_youngpoisson}.

\subsection{Nonlinear constitutive model}

\begin{figure}
    \centering
    \includegraphics{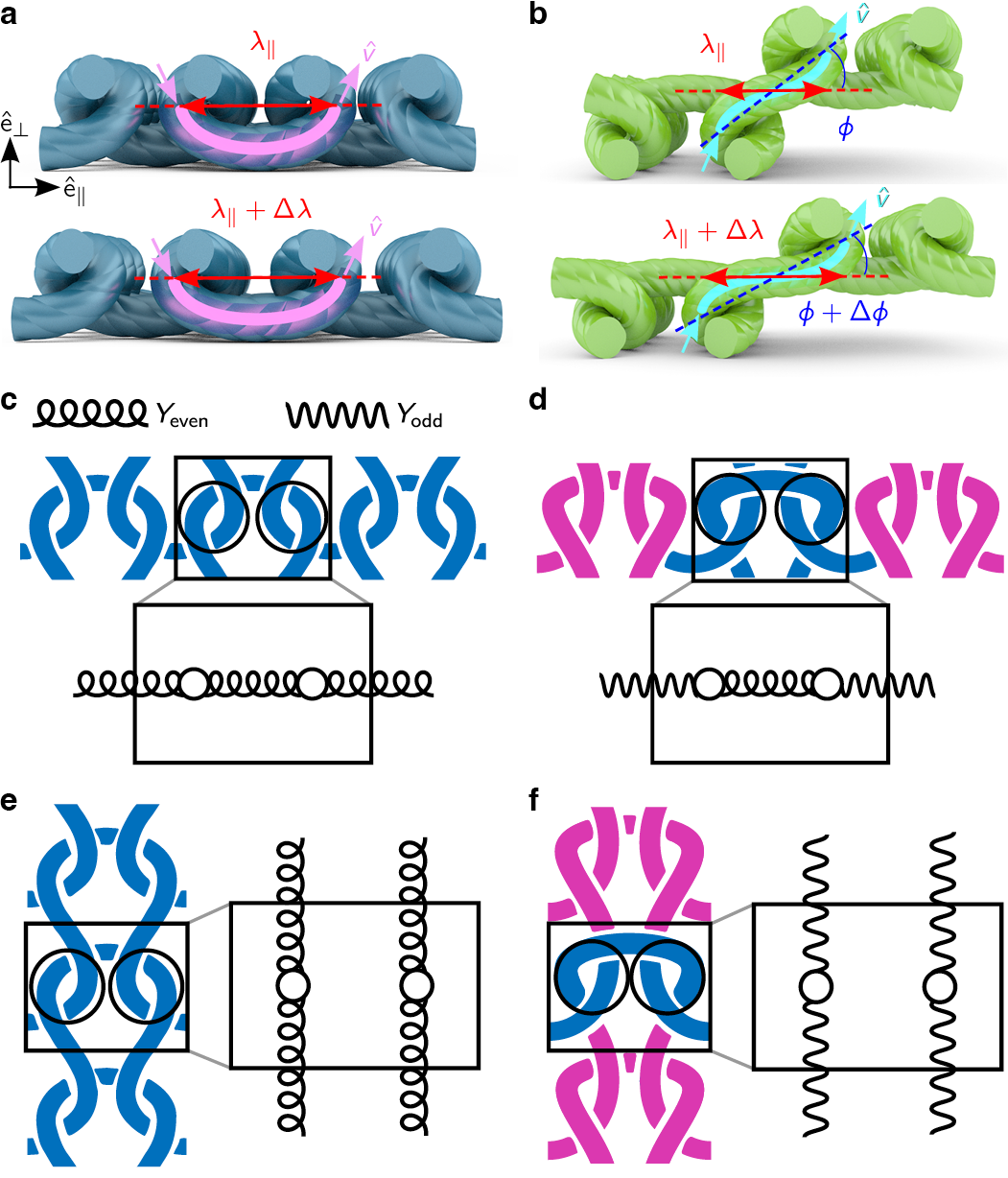}
    \caption{(\textbf{a},\textbf{b}) Diagrams of the even (\textbf{a}) and odd (\textbf{b}) regions of stockinette and rib, respectively, with labels for the re-parameterized geometric variables used in \ref{sec:RS_model}. \seg{These renderings we done using sample outputs of the stitch-level simulations.} For (\textbf{a}, \textbf{b}), horizontal is the $x$-direction of the fabric and into-the-plane is the $y$-direction. (\textbf{c}, \textbf{d}, \textbf{e}, \textbf{f}) Effective spring network elements for like-stitch (\textbf{K}-\textbf{K} or \textbf{P}-\textbf{P}) neighbors in the $x$-direction (\textbf{c}), unlike-stitch (\textbf{K}-\textbf{P}) neighbors in the $x$-direction (\textbf{d}), like-stitch neighbors in the $y$-direction (\textbf{e}), and unlike-stitch neighbors in the $y$-direction (\textbf{f}). For (\textbf{c}, \textbf{d}, \textbf{e}, \textbf{f}), horizontal is the $x$-direction of the fabric and vertical is the $y$-direction. Circled areas indicate entangled regions.}
    \label{fig:rs-figure}
\end{figure}

Here, we provide a scaling rationale for the constitutive relationship.
The results presented here are consistent with the RS model of \ref{sec:RS_model}, including the form of the strain-stiffening term.
First, consider the various lengthscales that describe a knit stitch.
These include the yarn radius $r$, the length of yarn per stitch $L$, and a ``bending lengthscale'' $\ell_{\rm b} \sim \sqrt{B/T}$, where $B$ is the yarn's bending modulus and $T$ is the tension of the yarn, which is obtained from dimensional analysis.
In fact, $\ell_{\rm b}$ has a simple physical interpretation: if one considers an arc of radius $R$, the work done in stretching the arc's radius to $R + \delta R$ under constant tension scales as $\sim T\delta R$, which is counteracted by a change in the bending energy, which scales as $\sim -B\delta R/R^2$, and the two generalized forces are in equilibrium if $R \sim \ell_{\rm b} \sim \sqrt{B/T}$.
Thus, $\ell_{\rm b}$ can be regarded as the radius of curvature that dominates the bending energy of a curve in mechanical equilibrium under tension.
As $T$ increases, $\ell_{\rm b}$ decreases, so that the bending energy of a stitch is increasingly concentrated to small regions of high curvature, which must be located at the entangled regions.
Note that in the arc approximation, the $\ell_{\rm b}$ is both the radius of curvature and the arclength of the curved regions, so that $E_{\rm bend} \sim 1/\ell_{\rm b}$.
Since the radius $r$ of the yarn represents a lower limit of the radius of curvature and the stitch length $L$ determines the periodicity of yarn shape, $\ell_{\rm b}$ is bounded by these two lengths: $r \lesssim \ell_{\rm b} \lesssim L$.

The limits of this region correspond to two physical cases. 
Case one, $\ell_{\rm b}\approx L$, occurs when there is little to no applied external force. 
Under small strains $\varepsilon$, the bending length decreases linearly, such that changes in $\ell_{\rm b}$ go as $\Delta \ell_{\rm b} \sim -\varepsilon \ell_{{\rm b}_0}$, where $\ell_{{\rm b}_0}$ is the bending length at zero strain. 
This results in the linear stress-strain relationship $\sigma_{\rm low} \propto \varepsilon$. 
Case two, $\ell_{\rm b} \approx r$, occurs when the fabric is under high external load. 
Here, the yarn segments within each crossover region clasp increasingly tightly around each other. 
In this regime, the total length of yarn $L$ can be approximated as $L \approx c(\lambda_{\rm max} + r)$, where $c$ is a numerical prefactor of $\mathcal{O}(1)$ and $\lambda_{\rm max} \approx (1 + \varepsilon)\lambda_0$, the maximal separation between crossover regions, varies linearly with the average separation between crossover regions in the unstrained case, $\lambda_0$. 
Therefore, the bending length can be approximated as $\ell_{\rm b} \approx r \approx (L/c) - \lambda_{\rm max} \approx A(1-\alpha\varepsilon)$, where $A$ and $\alpha$ are constants determined by the material properties of the yarn. 
The bending energy scales as $E_{\rm bend}\sim \ell_{\rm b}^{-1} \sim (1-\alpha \varepsilon)^{-1}$.
This implies that the high-stress regime scales as $\sigma_{\rm high} \sim \partial E_{\rm bend}/\partial \varepsilon \sim (1 - \alpha \varepsilon)^{-2}$.
This is consistent with our elastica analysis in the preceding section, where the high-tension limit $q\gg 1$ recovers the same bending energy scaling form.

While the low-stress regime is determined by topology, the high-stress regime is dominated by the material properties of the yarn. 
Combining these limiting behaviors leads us to a stress-strain relationship $\sigma(\varepsilon) \approx \sigma_{\rm low}(\varepsilon) + \sigma_{\rm high}(\varepsilon)$, which fits our experimental and simulation data quite well, as shown in {Fig.~2}, \ref{fig:stressstrain}, and \ref{fig:cotton}.

\subsection{Uniaxial constitutive model and fitting}

Our constitutive model for a sample of knitted fabric under uniaxial stress is given by
\begin{equation}\label{eq:constitutive_model}
    \left\{\begin{array}{c}
    \sigma_{xx}(\varepsilon_{xx}, \varepsilon_{yy}) = C^0_{xxxx}\varepsilon_{xx} + C^0_{xxyy}\varepsilon_{yy} + \beta_{xx}\left(\frac{1}{(1 - \alpha_{xx}\varepsilon_{xx})^2} - 1 - 2 \alpha_{xx}\varepsilon_{xx}\right) \\[10pt]
    \sigma_{yy}(\varepsilon_{xx}, \varepsilon_{yy}) = C^0_{yyyy}\varepsilon_{yy} + C^0_{yyxx}\varepsilon_{xx} + \beta_{yy}\left(\frac{1}{(1 - \alpha_{yy}\varepsilon_{yy})^2} - 1 - 2\alpha_{yy}\varepsilon_{yy}\right)
    \end{array}\right. \, ,
\end{equation}
where $C_{xxxx}^0$, $C_{yyyy}^0$, $C_{xxyy}^0$, and $C_{yyxx}^0$ are components of the rigidity tensor $C_{ijkl} \equiv \partial \sigma_{ij}/\partial \varepsilon_{kl}$, evaluated in the low-strain, linear elastic limit wherein $\sigma_{ij} \approx C_{ijkl}^0 \varepsilon_{kl}$. 
The constants $\alpha_{xx}$ and $\alpha_{yy}$ characterize the finite extensibility of knitted fabric when stretched in orthogonal directions, with $\beta_{xx}$ and $\beta_{yy}$ setting the stress scale of the strain-stiffening regime.
Our experimental results find significant asymmetry between $C_{xxyy}^0$ and $C_{yyxx}^0$ components, see \ref{table:acrylic_constitutive}, \ref{table:cotton_constitutive}, and \ref{table:baby_constitutive}. This is backed up by simulations showing significant asymmetry between $x$-strain and $y$-strain. Therefore, we do not enforce the standard symmetry $C_{xxyy}^0 = C_{yyxx}^0$ in our model. 
We hypothesize that this is due to changes in non-local contact interactions that occur when the fabric is strained in different directions.

\begin{figure}
    \centering
    \includegraphics[width=8.9cm]{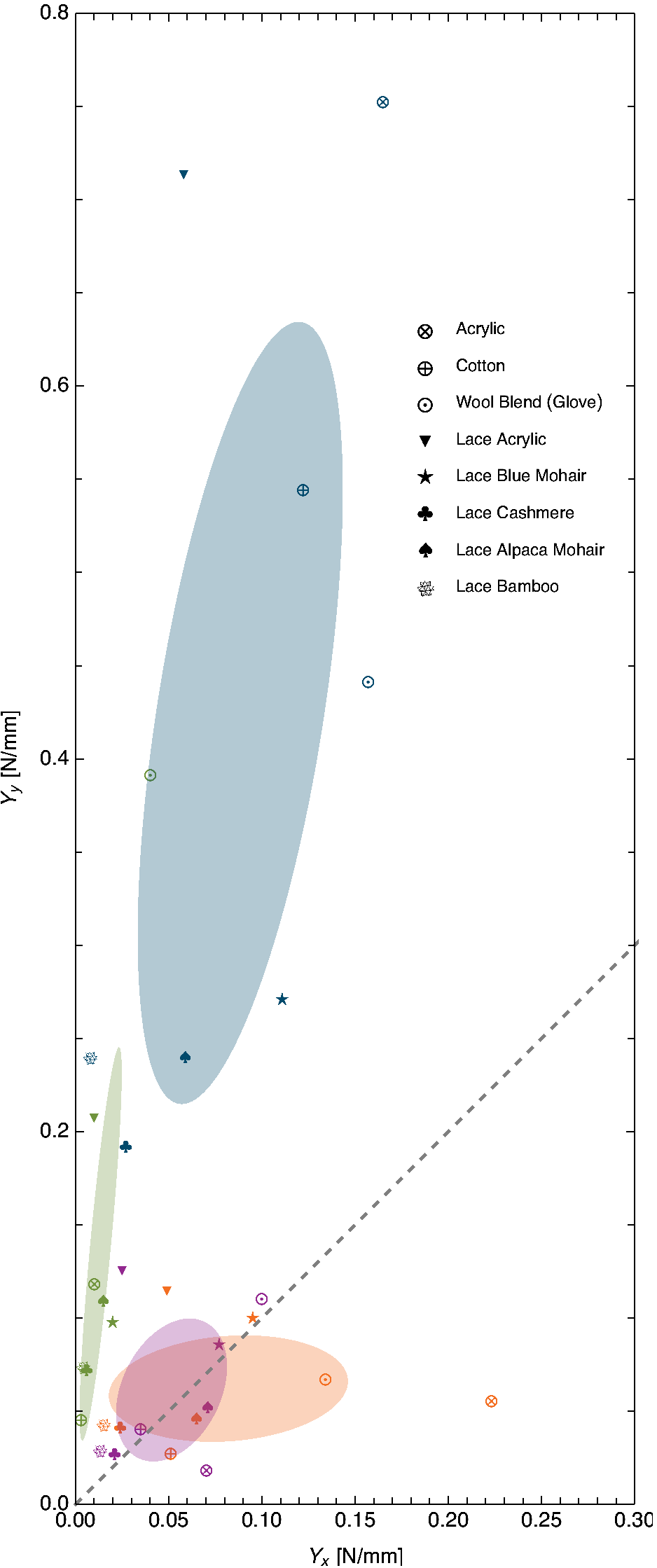}
    \caption{\seg{Rigidity-rigidity plot for all fabric samples, where $Y_i$ is the Young's modulus in the $i$\textsuperscript{th} direction. The colored ellipses represent one standard deviation for each of the four fabric types and are oriented along the principal axes: stockinette in blue, garter in orange, rib in green, and seed in purple. The gray dashed line represents a isotropic mechanical response.}}
    \label{fig:rig-rig}
\end{figure}

We obtained values of these eight parameters for each fabric by fitting the constitutive relations to data via a least-squares scheme. 
For each fabric, we obtained two independent data series from the uniaxial stretching experiments: (i) measured triplets $S_x=\left\{\left(\varepsilon_{xx}^{\rm data},\varepsilon_{yy}^{\rm data},\sigma_{xx}^{\rm data}\right)\right\}$ of strain and stress in the x-direction and (ii) measured triplets $S_y=\left\{\left(\varepsilon_{yy}^{\rm data},\varepsilon_{xx}^{\rm data},\sigma_{yy}^{\rm data}\right)\right\}$ of strain and stress in the y-direction. 
We then minimize the functional
\begin{equation}
    I = \mkern-36mu \sum_{(\varepsilon_{xx}^{\rm data},\varepsilon_{yy}^{\rm data},\sigma_{xx}^{\rm data}) \in S_x} \mkern-36mu \left(\sigma_{xx}(\varepsilon_{xx}^{\rm data}, \varepsilon_{yy}^{\rm data}) - \sigma_{xx}^{\rm data}\right)^2
    + \mkern-36mu\sum_{(\varepsilon_{yy}^{\rm data},\varepsilon_{xx}^{\rm data},\sigma_{yy}^{\rm data}) \in S_y}\mkern-36mu \left(\sigma_{yy}(\varepsilon_{xx}^{\rm data}, \varepsilon_{yy}^{\rm data}) - \sigma_{yy}^{\rm data}\right)^2
\end{equation}
with respect to the seven unknown parameters in the constitutive relation. 
However, minimizing this functional alone is insufficient because it ignores constraints imposed by the boundary conditions of the fabric. 
The stress-free boundary conditions couple longitudinal and transverse strains, giving rise to the Poisson effect.
In order to introduce this coupling when fitting the data, we determine the pair of Poisson ratios, $\nu_{yx}$ and $\nu_{xy}$, via linear fits to data sets $S_x$ and $S_y$, respectively. Then we minimize the functional $I$ under the constraints that the linear rigidity tensor components are consistent with these Poisson ratios via the relationships $\nu_{yx} = C^0_{yyxx}/C^0_{yyyy}$ and $\nu_{xy} = C^0_{xxyy}/C^0_{xxxx}$.
Values obtained for the fitting parameters are shown in \ref{table:acrylic_constitutive}, \ref{table:cotton_constitutive}, and \ref{table:baby_constitutive}. 
The Young's moduli are then given by
\begin{align}
\begin{split}
Y_x &= C^0_{xxxx} - \frac{C^0_{xxyy}C^0_{yyxx}}{C^0_{yyyy}} = (1 - \nu_{xy}\nu_{yx})C^0_{xxxx} \\
Y_y &= C^0_{yyyy} - \frac{C^0_{yyxx}C^0_{xxyy}}{C^0_{xxxx}} = (1 - \nu_{yx}\nu_{xy})C^0_{yyyy}
\end{split}
\end{align}
where we see that $Y_x/Y_y = C^0_{xxxx}/C^0_{yyyy}$, yet $Y_x \neq C^0_{xxxx}$ and $Y_y \neq C^0_{yyyy}$.
Values of the Young's moduli, along with error estimates based on the variance obtained from least squares fitting, are shown in \ref{table:acrylic_youngpoisson} and \ref{table:cotton_youngpoisson}. \seg{These figures are plotted in \ref{fig:rig-rig}.}

\begin{table}[h!]
\centering
{
\setlength{\tabcolsep}{2mm}
\begin{tabular}{|c|c|c|c|c|c|c|c|c|}
    \hline
     & \begin{tabular}{@{}c@{}}$C^0_{xxxx}$ \\ (N/mm)\end{tabular} & \begin{tabular}{@{}c@{}}$C^0_{yyyy}$ \\ (N/mm)\end{tabular} & \begin{tabular}{@{}c@{}}$C^0_{xxyy}$ \\ (N/mm)\end{tabular} & \begin{tabular}{@{}c@{}}$C^0_{yyxx}$ \\ (N/mm)\end{tabular} & $\alpha_{xx}$ & $\alpha_{yy}$ & \begin{tabular}{@{}c@{}}$\beta_{xx}$ \\ (N/mm)\end{tabular}  & \begin{tabular}{@{}c@{}}$\beta_{yy}$ \\ (N/mm)\end{tabular}\\
    \hline\hline 
    \begin{tabular}{@{}c@{}}Stockinette \\ (experiment)\end{tabular} & \fin{0.204} & \fin{0.930} & \fin{0.088} & \fin{0.413} & \fin{1.111} & \fin{2.537} & \fin{0.046} & \fin{0.010}\\
    \hline
    \begin{tabular}{@{}c@{}}Stockinette \\ (simulation)\end{tabular} & \fin{0.200} & \fin{0.753} & \fin{0.040} & \fin{0.341} & \fin{1.391} & \fin{2.971} & \fin{0.005} & \fin{0.013} \\
    \hline
    \begin{tabular}{@{}c@{}}Garter \\ (experiment)\end{tabular} & \fin{0.241} & \fin{0.060} & \fin{0.036} & \fin{0.029} & \fin{1.170} & \fin{0.802} & \fin{0.022} & \fin{0.022} \\
    \hline
    \begin{tabular}{@{}c@{}}Garter \\ (simulation)\end{tabular} & \fin{0.252} & \fin{0.038} & \fin{0.103} & \fin{0.019} & \fin{1.073} & \fin{1.135} & \fin{0.002} & \fin{0.006} \\
    \hline
    \begin{tabular}{@{}c@{}}Rib \\ (experiment)\end{tabular} & \fin{0.011} & \fin{0.126} & \fin{0.003} & \fin{0.026} & \fin{0.385} & \fin{1.251} & \fin{0.011} & \fin{0.034} \\
    \hline
    \begin{tabular}{@{}c@{}}Rib \\ (simulation)\end{tabular} & \fin{0.024} & \fin{0.142} & \fin{0.011}  & \fin{0.028} & \fin{0.446} & \fin{1.411} & \fin{0.004} & \fin{0.014} \\
    \hline
    \begin{tabular}{@{}c@{}}Seed \\ (experiment)\end{tabular} & \fin{0.074} & \fin{0.020} & \fin{0.010} & \fin{0.009} & \fin{1.148} & \fin{0.568} & \fin{0.027} & \fin{0.017} \\
    \hline
    \begin{tabular}{@{}c@{}}Seed \\ (simulation)\end{tabular} & \fin{0.128} & \fin{0.057} & \fin{0.066} & \fin{0.021} & \fin{0.940} & \fin{0.693} & \fin{0.010} & \fin{0.005} \\
    \hline
\end{tabular}
}
\caption{\label{table:acrylic_constitutive} List of parameters obtained by fitting the constitutive model to experimental and simulation data representing fabric made from the acrylic yarn. }
\end{table}

\begin{table}[h!]
\centering
{
\setlength{\tabcolsep}{2mm}
\begin{tabular}{|c|c|c|c|c|}
    \hline
     & \begin{tabular}{@{}c@{}}$Y_x$ \\ (N/mm)\end{tabular} & \begin{tabular}{@{}c@{}}$Y_y$ \\ (N/mm)\end{tabular} & $\nu_{yx}$ & $\nu_{xy}$ \\
    \hline\hline 
    \begin{tabular}{@{}c@{}}Stockinette \\ (experiment)\end{tabular} & $0.165 \pm 0.011$ & $0.753 \pm 0.034$ & $0.444 \pm 0.005$ & $0.430 \pm 0.015$ \\
    \hline
    \begin{tabular}{@{}c@{}}Stockinette \\ (simulation)\end{tabular} & 0.182 & 0.684 & 0.453 & 0.202 \\
    \hline
    \begin{tabular}{@{}c@{}}Garter \\ (experiment)\end{tabular} & $0.223 \pm 0.021$ & $0.056 \pm 0.015$ & $0.481 \pm 0.004$ & $0.150 \pm 0.003$ \\
    \hline
    \begin{tabular}{@{}c@{}}Garter \\ (simulation)\end{tabular} & 0.200 &  0.030 & 0.504 & 0.407 \\
    \hline
    \begin{tabular}{@{}c@{}}Rib \\ (experiment)\end{tabular} & $0.010 \pm 0.006$ & $0.119 \pm 0.012$ & $0.208 \pm 0.001$ & $0.294 \pm 0.006$ \\
    \hline
    \begin{tabular}{@{}c@{}}Rib \\ (simulation)\end{tabular} & 0.022 & 0.129 & 0.200 & 0.461 \\
    \hline
    \begin{tabular}{@{}c@{}}Seed \\ (experiment)\end{tabular} & $0.070 \pm 0.007$ & $0.019 \pm 0.001$ & $0.471 \pm 0.004$ & $0.138 \pm 0.002$ \\
    \hline
    \begin{tabular}{@{}c@{}}Seed \\ (simulation)\end{tabular} & 0.103 & 0.046 & 0.373 & 0.515 \\
    \hline
\end{tabular}
}
\caption{\label{table:acrylic_youngpoisson} \seg{List of parameters obtained by fitting the Young's moduli and Poisson ratios to experimental and simulation data representing fabric made from the acrylic yarn.}}
\end{table}

\begin{table}[h!]
\centering
{
\setlength{\tabcolsep}{2mm}
\begin{tabular}{|c|c|c|c|c|c|c|c|c|}
    \hline
     & \begin{tabular}{@{}c@{}}$C^0_{xxxx}$ \\ (N/mm)\end{tabular} & \begin{tabular}{@{}c@{}}$C^0_{yyyy}$ \\ (N/mm)\end{tabular} & \begin{tabular}{@{}c@{}}$C^0_{xxyy}$ \\ (N/mm)\end{tabular} & \begin{tabular}{@{}c@{}}$C^0_{xxyy}$ \\ (N/mm)\end{tabular} & $\alpha_{xx}$ & $\alpha_{yy}$ & \begin{tabular}{@{}c@{}}$\beta_{xx}$ \\ (N/mm)\end{tabular}  & \begin{tabular}{@{}c@{}}$\beta_{yy}$ \\ (N/mm)\end{tabular}\\
    \hline\hline 
    \begin{tabular}{@{}c@{}}Stockinette \\ (experiment)\end{tabular} & \fin{0.147} & \fin{0.659} & \fin{0.061} & \fin{0.277} & \fin{1.388} & \fin{2.440} & \fin{0.007} & \fin{0.064} \\
    \hline
    \begin{tabular}{@{}c@{}}Stockinette \\ (simulation)\end{tabular} & \fin{0.354} & \fin{0.637} & \fin{0.127} & \fin{0.280} & \fin{1.250} & \fin{2.102} & \fin{0.006} & \fin{0.012} \\
    \hline
    \begin{tabular}{@{}c@{}}Garter \\ (experiment)\end{tabular} & \fin{0.057} & \fin{0.031} & \fin{0.010} & \fin{0.017} & \fin{1.225} & \fin{0.700} & \fin{0.020} & \fin{0.021} \\
    \hline
    \begin{tabular}{@{}c@{}}Garter \\ (simulation)\end{tabular} & \fin{0.208} & \fin{0.052} & \fin{0.044} & \fin{0.024} & \fin{1.093} & \fin{0.719} & \fin{0.024} & \fin{0.016} \\
    \hline
    \begin{tabular}{@{}c@{}}Rib \\ (experiment)\end{tabular} & \fin{0.003} & \fin{0.049} & \fin{0.001} & \fin{0.010} & \fin{0.388} & \fin{1.439} & \fin{0.003} & \fin{0.017} \\
    \hline
    \begin{tabular}{@{}c@{}}Rib \\ (simulation)\end{tabular} & \fin{0.009} & \fin{0.028} & \fin{0.004} & \fin{0.006} & \fin{0.393} & \fin{1.243} & \fin{0.002} & \fin{0.007} \\
    \hline
    \begin{tabular}{@{}c@{}}Seed \\ (experiment)\end{tabular} & \fin{0.038} & \fin{0.044} & \fin{0.006} & \fin{0.019} & \fin{1.351} & \fin{0.951} & \fin{0.010} & \fin{0.022} \\
    \hline
    \begin{tabular}{@{}c@{}}Seed \\ (simulation)\end{tabular} & \fin{0.114} & \fin{0.192} & \fin{0.019} & \fin{0.069} & \fin{1.102} & \fin{1.212} & \fin{0.017} & \fin{0.008} \\
    \hline
\end{tabular}
}
\caption{\label{table:cotton_constitutive} List of parameters obtained by fitting the constitutive model to experimental and simulation data representing fabric made from the cotton yarn.}
\end{table}

\begin{table}[h!]
\centering
{
\setlength{\tabcolsep}{2mm}
\begin{tabular}{|c|c|c|c|c|}
    \hline
     & \begin{tabular}{@{}c@{}}$Y_x$ \\ (N/mm)\end{tabular} & \begin{tabular}{@{}c@{}}$Y_y$ \\ (N/mm)\end{tabular} & $\nu_{yx}$ & $\nu_{xy}$ \\
    \hline\hline 
    \begin{tabular}{@{}c@{}}Stockinette \\ (experiment)\end{tabular} & $0.122 \pm 0.018$ & $0.545 \pm 0.038$ & $0.420 \pm 0.003$ & $0.412 \pm 0.020$ \\
    \hline
    \begin{tabular}{@{}c@{}}Stockinette \\ (simulation)\end{tabular} & 0.298 & 0.536 & 0.441 & 0.359 \\
    \hline
    \begin{tabular}{@{}c@{}}Garter \\ (experiment)\end{tabular} & $0.051 \pm 0.006$ & $0.028 \pm 0.002$ & $0.561 \pm 0.006$  & $0.177 \pm 0.002$ \\
    \hline
    \begin{tabular}{@{}c@{}}Garter \\ (simulation)\end{tabular} & 0.188 & 0.047 & 0.459 & 0.210 \\
    \hline
    \begin{tabular}{@{}c@{}}Rib \\ (experiment)\end{tabular} & $0.003 \pm 0.005$ & $0.046 \pm 0.006$ & $0.205 \pm 0.003$ & $0.289 \pm 0.006$ \\
    \hline
    \begin{tabular}{@{}c@{}}Rib \\ (simulation)\end{tabular} & 0.008 & 0.026 & 0.195 & 0.439 \\
    \hline
    \begin{tabular}{@{}c@{}}Seed \\ (experiment)\end{tabular} & $0.035 \pm 0.005$ & $0.041 \pm 0.003$ & $0.435 \pm 0.006$ & $0.165 \pm 0.008$ \\
    \hline
    \begin{tabular}{@{}c@{}}Seed \\ (simulation)\end{tabular} & 0.107 & 0.180 & 0.359 & 0.168 \\
    \hline
\end{tabular}
}
\caption{\label{table:cotton_youngpoisson} \seg{List of parameters obtained by fitting Young's moduli and Poisson ratios to experimental and simulation data representing fabric made from the cotton yarn.}}
\end{table}

\begin{table}[h!]
\centering
{
\setlength{\tabcolsep}{2mm}
\begin{tabular}{|c|c|c|c|c|c|c|c|c|}
    \hline
     & \begin{tabular}{@{}c@{}}$C^0_{xxxx}$ \\ (N/mm)\end{tabular} & \begin{tabular}{@{}c@{}}$C^0_{yyyy}$ \\ (N/mm)\end{tabular} & \begin{tabular}{@{}c@{}}$C^0_{xxyy}$ \\ (N/mm)\end{tabular} & \begin{tabular}{@{}c@{}}$C^0_{yyxx}$ \\ (N/mm)\end{tabular} & $\alpha_{xx}$ & $\alpha_{yy}$ & \begin{tabular}{@{}c@{}}$\beta_{xx}$ \\ (N/mm)\end{tabular}  & \begin{tabular}{@{}c@{}}$\beta_{yy}$ \\ (N/mm)\end{tabular}\\
    \hline\hline 
    \multicolumn{9}{|c|}{\seg{Lace-Weight Acrylic}} \\
    \hline
    \seg{Stockinette} & \seg{0.119} & \seg{1.454} & \seg{0.130} & \seg{0.678} & \seg{0.655} & \seg{1.870} & \seg{0.078} & \seg{0.275} \\
    \hline
    \seg{Garter} & \seg{0.066} & \seg{0.154} & \seg{0.034} & \seg{0.075} & \seg{0.657} & \seg{1.380} & \seg{0.050} & \seg{0.056}\\
    \hline
    \seg{Rib} & \seg{0.012} & \seg{0.261} & \seg{0.009} & \seg{0.074} & \seg{0.342} & \seg{1.573} & \seg{0.018} & \seg{0.083} \\
    \hline
    \seg{Seed} & \seg{0.037} & \seg{0.189} & \seg{0.023} & \seg{0.103} & \seg{0.589} & \seg{1.451} & \seg{0.031} & \seg{0.037} \\

    \hline\hline 
    \multicolumn{9}{|c|}{\seg{Lace-Weight Blue Mohair}} \\
    \hline
    \seg{Stockinette} & \seg{0.168} & \seg{0.410} & \seg{0.134} & \seg{0.174} & \seg{0.670} & \seg{1.644} & \seg{0.050} & \seg{0.125} \\
    \hline
    \seg{Garter} & \seg{0.146} & \seg{0.069} & \seg{0.075} & \seg{0.048} & \seg{0.847} & \seg{1.020} & \seg{0.028} & \seg{0.034} \\
    \hline
    \seg{Rib} & \seg{0.026} & \seg{0.126} & \seg{0.020} & \seg{0.037} & \seg{0.327} & \seg{1.323} & \seg{0.014} & \seg{0.039} \\
    \hline
    \seg{Seed} & \seg{0.116} & \seg{0.130} & \seg{0.060} & \seg{0.086} & \seg{0.803} & \seg{0.986} & \seg{0.020} & \seg{0.028} \\

    \hline\hline 
    \multicolumn{9}{|c|}{\seg{Lace-Weight Cashmere}} \\
    \hline
    \seg{Stockinette} & \seg{0.044} & \seg{0.309} & \seg{0.040} & \seg{0.129} & \seg{0.622} & \seg{1.575} & \seg{0.035} & \seg{0.098} \\
    \hline
    \seg{Garter} & \seg{0.034} & \seg{0.060} & \seg{0.019} & \seg{0.033} & \seg{0.619} & \seg{1.054} & \seg{0.028} & \seg{0.028} \\
    \hline
    \seg{Rib} & \seg{0.007} & \seg{0.090} & \seg{0.005} & \seg{0.026} & \seg{0.304} & \seg{1.202} & \seg{0.012} & \seg{0.037} \\
    \hline
    \seg{Seed} & \seg{0.030} & \seg{0.038} & \seg{0.013} & \seg{0.025} & \seg{0.676} & \seg{0.780} & \seg{0.022} & \seg{0.025} \\

    \hline\hline 
    \multicolumn{9}{|c|}{\seg{Lace-Weight Alpaca Mohair}} \\
    \hline
    \seg{Stockinette} & \seg{0.099} & \seg{0.399} & \seg{0.870} & \seg{0.181} & \seg{0.633} & \seg{1.721} & \seg{0.044} & \seg{0.116} \\
    \hline
    \seg{Garter} & \seg{0.092} & \seg{0.064} & \seg{0.044} & \seg{0.038} & \seg{0.769} & \seg{1.032} & \seg{0.034} & \seg{0.0311} \\
    \hline
    \seg{Rib} & \seg{0.021} & \seg{0.152} & \seg{0.018} & \seg{0.051} & \seg{0.389} & \seg{1.476} & \seg{0.012} & \seg{0.051} \\
    \hline
    \seg{Seed} & \seg{0.104} & \seg{0.076} & \seg{0.048} & \seg{0.052} & \seg{0.817} & \seg{0.869} & \seg{0.018} & \seg{0.026} \\

    \hline\hline 
    \multicolumn{9}{|c|}{\seg{Lace-Weight Bamboo}} \\
    \hline
    \seg{Stockinette} & \seg{0.015} & \seg{0.432} & \seg{0.016} & \seg{0.186} & \seg{0.605} & \seg{2.095} & \seg{0.012} & \seg{0.103} \\
    \hline
    \seg{Garter} & \seg{0.023} & \seg{0.068} & \seg{0.015} & \seg{0.037} & \seg{0.638} & \seg{1.542} & \seg{0.018} & \seg{0.022} \\
    \hline
    \seg{Rib} & \seg{0.005} & \seg{0.100} & \seg{0.004} & \seg{0.030} & \seg{0.323} & \seg{1.787} & \seg{0.008} & \seg{0.028} \\
    \hline
    \seg{Seed} & \seg{0.019} & \seg{0.043} & \seg{0.010} & \seg{0.025} & \seg{0.669} & \seg{1.019} & \seg{0.014} & \seg{0.021} \\
    \hline
\end{tabular}
}
\caption{\label{table:lace_constitutive} \seg{List of parameters obtained by fitting the constitutive model to experimental data from samples made from lace weight yarn.}}
\end{table}


\begin{table}[h!]
\centering
{
\setlength{\tabcolsep}{2mm}
\begin{tabular}{|c|c|c|c|c|}
    \hline
     & \begin{tabular}{@{}c@{}}$Y_x$ \\ (N/mm)\end{tabular} & \begin{tabular}{@{}c@{}}$Y_y$ \\ (N/mm)\end{tabular} & $\nu_{yx}$ & $\nu_{xy}$ \\
    \hline\hline 
    \multicolumn{5}{|c|}{\seg{Lace-Weight Acrylic}} \\
    \hline
    \seg{Stockinette} & \seg{0.058} & \seg{0.714} & \seg{0.466} & \seg{1.092} \\
    \hline
    \seg{Garter} & \seg{0.049} & \seg{0.115} & \seg{0.488} & \seg{0.518} \\
    \hline
    \seg{Rib} & \seg{0.010} & \seg{0.208} & \seg{0.284} & \seg{0.723} \\
    \hline
    \seg{Seed} & \seg{0.025} & \seg{0.126} & \seg{0.545} & \seg{0.610} \\
    \hline\hline 
    \multicolumn{5}{|c|}{\seg{Lace-Weight Blue Mohair}} \\
    \hline
    \seg{Stockinette} & \seg{0.111} & \seg{0.271} & \seg{0.425} & \seg{0.800} \\
    \hline
    \seg{Garter} & \seg{0.095} & \seg{0.100} & \seg{0.679} & \seg{0.515} \\
    \hline
    \seg{Rib} & \seg{0.020} & \seg{0.098} & \seg{0.291} & \seg{0.781} \\
    \hline
    \seg{Seed} & \seg{0.077} & \seg{0.086} & \seg{0.661} & \seg{0.514} \\
    \hline\hline 
    \multicolumn{5}{|c|}{\seg{Lace-Weight Cashmere}} \\
    \hline
    \seg{Stockinette} & \seg{0.027} & \seg{0.192} & \seg{0.417} & \seg{0.912} \\
    \hline
    \seg{Garter} & \seg{0.024} & \seg{0.041} & \seg{0.552} & \seg{0.552} \\
    \hline
    \seg{Rib} & \seg{0.006} & \seg{0.072} & \seg{0.286} & \seg{0.695} \\
    \hline
    \seg{Seed} & \seg{0.021} & \seg{0.027} & \seg{0.652} & \seg{0.445} \\
    \hline\hline 
    \multicolumn{5}{|c|}{\seg{Lace-Weight Alpaca Mohair}} \\
    \hline
    \seg{Stockinette} & \seg{0.059} & \seg{0.240} & \seg{0.454} & \seg{0.879} \\
    \hline
    \seg{Garter}Garter & \seg{0.065} & \seg{0.046} & \seg{0.596} & \seg{0.482} \\
    \hline
    \seg{Rib} & \seg{0.015} & \seg{0.109} & \seg{0.337} & \seg{0.841} \\
    \hline
    \seg{Seed} & \seg{0.071} & \seg{0.052} & \seg{0.683} & \seg{0.461} \\
    \hline\hline 
    \multicolumn{5}{|c|}{\seg{Lace-Weight Bamboo}} \\
    \hline
    \seg{Stockinette} & \seg{0.008} & \seg{0.240} & \seg{0.430} & \seg{1.038} \\
    \hline
    \seg{Garter} & \seg{0.015} & \seg{0.043} & \seg{0.549} & \seg{0.653} \\
    \hline
    \seg{Rib} & \seg{0.004} & \seg{0.074} & \seg{0.301} & \seg{0.834} \\
    \hline
    \seg{Seed} & \seg{0.013} & \seg{0.029} & \seg{0.598} & \seg{0.537} \\
    \hline
\end{tabular}
}
\caption{\label{table:lace_youngpoisson} \seg{List of parameters obtained by fitting the Young's moduli and Poisson ratios to experimental data from samples made from lace weight yarn.}}
\end{table}

\newpage
\subsection{Composite elasticity from reduced-symmetry model (RS model) \label{sec:RS_model}}

To connect between the micromechanics of the yarn and the fabric's macroscopic response, we developed a reduced-symmetry (RS) model of stitch mechanics.
This model starts with the full 3D elastica model and determines the change in bending energy due to deflecting the shape of individual yarn segments from their original shape, as they sit in an un-stretched sample of fabric.
We distill this shape-response to stretching into a dependence on the spatial symmetry of yarn joining neighboring entangled regions.
To this end, we approximate the yarn segment shapes as 2D curves, given by their projections onto either the $xz$-plane or the $yz$-plane (as shown in \ref{fig:rs-figure}). 
We can approximate the shape of these 3D curves as the image of 2D parametric curves $\mathbf{r}(u) = r_{\parallel}(u)\hat{\mathbf{e}}_{\parallel} + r_{\perp}\hat{\mathbf{e}}_{\perp}$ where $-1/2 \leq u \leq 1/2$, $\hat{\mathbf{e}}_{\parallel}$ is a unit vector in the $xy$-plane and $\hat{\mathbf{e}}_\perp = \hat{\mathbf{z}}$ lies along the fabric’s thickness. 
Translating the coordinate system such that the endpoints of a segment are at antipodal values $\mathbf{r}(\pm1/2) = \pm \mathbf{r}_0 = \frac{\lambda}{2}(\cos\phi\, \hat{\mathbf{e}}_{\parallel} + \sin\phi\,\hat{\mathbf{e}}_{\perp}$),
where $\lambda$ is the separation of the endpoints.
We recognize that there are distinct curves with even symmetry $r_{\rm even}(-u)= r_{\rm even}(u)$ (and $\phi = 0$), and distinct curves with odd symmetry $r_{\rm odd}(-u)= - r_{\rm odd}(u)$ (and general $\phi \neq 0$). 
This follows from the mapping of knit stitches to purl stitches in 3D space induced by the action of the mirror operation $M = (\mathbbm{1} - 2\hat{\mathbf{e}}_\perp \otimes \hat{\mathbf{e}}_\perp)$ on the centerline of the yarn. 

The geometry of the entangled regions at the endpoints of the yarn segment constrains the shape of the curve at its endpoints by the requirement that the curve must clasp around another curve in the entangled region. 
The end of the yarn is forced to deflect out of the plane, following a given tangent vector $\hat{\mathbf{v}}$ adding an additional set of boundary conditions to the parametric curve $\mathbf{r}(u)$. 
For even curves, this boundary condition is $\left[\partial_u\mathbf{r}_{\rm even}/|\partial_u\mathbf{r}_{\rm even}|\right]_{\pm 1/2} = \hat{v}_{\parallel}\hat{\mathbf{e}}_{\parallel} \pm \hat{v}_{\perp}\hat{\mathbf{e}}_{\perp}$ and for odd curves, it is $\left[\partial_u\mathbf{r}_{\rm odd}/|\partial_u\mathbf{r}_{\rm odd}|\right]_{\pm 1/2} = \hat{\mathbf{v}}$. 

To simplify calculations, we will express the curve $\mathbf{r}(u)$ as a dimensionless deflection $\zeta(u)$ transverse to the end-to-end orientation $\bm{\lambda}(\phi) = \lambda(\cos\phi\,\hat{\mathbf{e}}_{\parallel} + \sin\phi\,\hat{\mathbf{e}}_{\perp})$ via
\begin{equation}
    \mathbf{r}(u) = u\bm{\lambda}(\phi) + \lambda\zeta(u)\hat{\mathbf{r}}_{0,\perp}(\phi)
\end{equation}
where $\hat{\bm{\zeta}} = -\sin\phi\,\hat{\mathbf{e}}_{\parallel} + \cos\phi\,\hat{\mathbf{e}}_\perp$ is the direction transverse to the end-to-end orientation. 
In this representation, the deflection function $\zeta(u)$ obeys the boundary conditions $\zeta(\pm1/2) = 0$. 
We additionally take the small-deflection approximation so that the tangent vector at each point is given by $\hat{\mathbf{t}} \approx \hat{\bm{\lambda}} + \hat{\bm{\zeta}}\partial_u\zeta + \mathcal{O}(\zeta^2)$. 
Therefore, for even connecting yarn segments, the slope of the deflection function at the ends is given by $\partial_u \zeta_{\rm even}(\pm 1/2) \approx \pm\hat{\mathbf{v}}\cdot\hat{\bm{\zeta}} = \pm\hat{v}_{\zeta}$. 
For odd connecting yarn segments, the slope is given by $\partial_u \zeta_{\rm odd} (\pm 1/2) \approx \hat{\mathbf{v}}\cdot\hat{\bm{\zeta}} = \hat{v}_{\zeta}$.

In the small-deflection approximation, the elastica energy is given by
\begin{equation}
    E[\zeta, \lambda, T] \approx \frac{\lambda}{2}\int_{-1/2}^{1/2}{\rm d}u\,\left[\frac{B}{\lambda^2}(\partial^2_u \zeta)^2 + T(\partial_u \zeta)^2 \right] + T(\lambda - L)
\end{equation}
where $T$ is a Lagrange multiplier, a tension that constrains the length of the curve to $L$. 
It is useful to rewrite the energy as
\begin{equation}
    E[\zeta, \lambda, q] \approx \frac{B}{2\lambda} \int_{-1/2}^{1/2} {\rm d}u\,\left[(\partial^2_u \zeta)^2 + 4q^2(\partial_u \zeta)^2 - 8 q^2\left(\frac{L}{\lambda} - 1\right)\right]
\end{equation}
where $q^2 \equiv T\lambda^2/(4B)$ is a dimensionless form of the Lagrange multiplier. 
The equilibrium deflection $\zeta(u)$ extremizes this energy function so that $\delta E/\delta \zeta = 0$ and therefore solves the differential equation $\partial_u^4\zeta - 4q^2\partial_u^2\zeta = 0$. 
Even solutions have the form $\zeta_{\rm even} = a + b\cosh 2qu$ and odd solutions have the form $\zeta_{\rm odd} = cu + d\sinh 2qu$, where the constants $a$, $b$, $c$, and $d$ are determined by the boundary conditions on $\zeta_{\rm even}$ and $\zeta_{\rm odd}$. 
Inserting these solutions back into the energy functional, the total energy for even connecting yarn segments $E_{\rm even}$ is given by
\begin{equation}\label{eq:even_energy}
    E_{\rm even}(\hat{v}_\zeta, \lambda, q) \approx \frac{2B}{\lambda} \left[ \hat{v}_{\zeta}^2q\coth q - 2q^2 \left(\frac{L}{\lambda} - 1\right)  \right]
\end{equation}
and the total energy for odd connecting yarn segments $E_{\rm odd}$ is given by
\begin{equation}\label{eq:odd_energy}
    E_{\rm odd}(\hat{v}_\zeta, \lambda, q) \approx \frac{2B}{\lambda} \left[\hat{v}_{\zeta}^2\frac{q^2 \sinh q}{q \cosh q - \sinh q} - 2q^2 \left(\frac{L}{\lambda} - 1\right)\right] \, .
\end{equation}
In order for the length constraint to be enforced, the dimensionless Lagrange multiplier $q$ is chosen to solve the equation $\partial E/\partial q = 0$. 
However, to solve for $q$, we require the solution to transcendental equations for both even and odd connecting yarn segments. 
To avoid this, we find it is sufficient to Taylor expand each energy function to quartic order in $q$, yielding
\begin{equation}
    E_{\rm even}(\hat{v}_\zeta, \lambda, q) \approx \frac{2B}{\lambda} \left[\hat{v}_{\zeta}^2\left(1 + \frac{q^2}{3} - \frac{q^4}{45} + \mathcal{O}(q^6)\right) - 2 q^2 \left(\frac{L}{\lambda} - 1\right)\right]
\end{equation}
and
\begin{equation}
    E_{\rm odd}(\hat{v}_\zeta, \lambda, q) \approx \frac{2B}{\lambda} \left[\hat{v}_{\zeta}^2\left(3 + \frac{q^2}{5} - \frac{q^4}{175} + \mathcal{O}(q^6)\right) - 2 q^2 \left(\frac{L}{\lambda} - 1\right)\right]
\end{equation}
which yield approximate polynomial equations for the constraining tension $q$. 
Using the solution for $q$, we find effective elastica energies 
\begin{equation}
    E_{\rm  even}(\hat{v}_\zeta, \lambda) \approx \frac{2B}{\lambda} \left[ \frac{9}{4}\hat{v}^2_{\zeta} + \frac{45}{\hat{v}_\zeta^2}\left(\frac{L}{\lambda} - 1\right)\left(\frac{L}{\lambda} - 1 - \frac{\hat{v}_{\zeta}^2}{3}\right) \right]
\end{equation}
and
\begin{equation}
    E_{\rm odd}(\hat{v}_\zeta, \lambda) \approx \frac{2B}{\lambda} \left[ \frac{19}{4}\hat{v}^2_{\zeta} + \frac{175}{\hat{v}_\zeta^2}\left(\frac{L}{\lambda} - 1\right)\left(\frac{L}{\lambda} - 1 - \frac{\hat{v}_{\zeta}^2}{5}\right) \right]
\end{equation}
which include the lowest-order correction to the bending energy arising from the enforced length constraint. 
The first term of the energy arises from an overall penalty from curvature, so that at fixed endpoint orientation $\hat{\mathbf{v}}$, the internal stress of the curve pushes its endpoint separation $\lambda$ to higher values. 
This stress is countered by the second term, representing the cost of concentrating curvature to the endpoints of the curve when the endpoint separation $\lambda$ approaches the total length $L$ of the curve. 
Therefore, there is an endpoint separation $\lambda^*$ that minimizes the elastica energy so $\partial E/\partial \lambda|_{\lambda^*} = 0$ at fixed endpoint orientation $\hat{\mathbf{v}}$.
For even connecting yarn segments, $\lambda^*_{\rm even}/L \approx 1 - \hat{v}_\zeta^2/6 + \mathcal{O}(\hat{v}_\zeta^4)$, and for odd connecting yarn segments, $\lambda^*_{\rm odd}/L \approx 1 - \hat{v}^2_{\zeta}/10 + \mathcal{O}(\hat{v}^4_\zeta)$. 
We will next assume that under low applied stress, these segments have separation length $\lambda$ that are almost the energy-minimizing length $\lambda^*$. 
Expanding the elastica energy to second order in $(\lambda - \lambda^*)/\lambda^*$, the even connecting yarn segment energy is approximately
\begin{equation}
    E_{\rm even}(\hat{v}_\zeta, \lambda) \approx \frac{2B}{\lambda}\left[ \hat{v}_{\zeta}^2 + \frac{45}{\hat{v}^2_\zeta}\left(\frac{\lambda - \lambda^*_{\rm even}}{\lambda^*_{\rm even}}\right)^2 \right]
\end{equation}
and the odd connecting yarn segment energy is approximately
\begin{equation}
    E_{\rm odd}(\hat{v}_\zeta, \lambda) \approx \frac{6B}{\lambda}\left[ \hat{v}_{\zeta}^2 + \frac{175}{3\hat{v}^2_\zeta}\left(\frac{\lambda - \lambda^*_{\rm odd}}{\lambda^*_{\rm odd}}\right)^2 \right]
\end{equation}
where we have kept only leading-order terms in $\hat{v}_\zeta$. 
Note that the cost of deforming each segment diverges when the endpoint orientations approach the orientation of the endpoint separation vector, i.e.~$\hat{v}_\zeta \to 0$.
In this limit, the energy-minimizing length approaches the total length of the curve, $\lambda^* \to L$, and due to the length constraint, the cost of stretching the curve beyond its total length should diverge. 
Finally, the bending energy for odd connecting yarn segments is generally larger than the bending energy for even connecting yarn segments, assuming each curve has identical values of length $L$, endpoint separation $\lambda$, and endpoint orientation $\hat{v}_\zeta$. 
This is reasonable since odd connecting yarn segments have two arches, each with a fraction of the radius of curvature of the single arch of an even connecting yarn segment.

Next, we determine the rigidity $Y \equiv (\partial^2 E/\partial \lambda^2_{\parallel})_{\hat{\mathbf{v}}}$ for extensile deformations of each curve along the fabric plane, where $\lambda_\parallel = \bm{\lambda}\cdot\hat{\mathbf{e}}_\parallel = \lambda \cos\phi$ is the $x$-axis projection of the endpoint separation. 
As shown in \ref{fig:rs-figure}, the endpoint orientation $\hat{\mathbf{v}}$ remains effectively fixed under such deformations. 
For even connecting yarn segments, the planar projection of the endpoint separation $\lambda_\parallel$ is identical to the full endpoint separation $\lambda$, since $\phi = 0$, so $Y_{\rm even} \equiv (\partial^2 E_{\rm even}/\partial \lambda^2)_{\hat{\mathbf{v}}}$. 
Therefore, the extensional rigidity for even connecting yarn segments is approximately
\begin{equation}
    Y_{\rm even} \approx \frac{180 B}{\hat{v}^2_{\zeta} L \lambda^2} = \frac{180 B}{\hat{v}^2_{\zeta}\lambda^3(1 - \delta_{\rm even})}
\end{equation}
where the divergence as $\hat{v}_{\zeta} \to 0$ is due to infinite energy cost for stretching the curve beyond its constrained length.
Here, $\delta_{\rm even} = 1 - (L/\lambda)$ is a geometric factor.
In general, odd connecting yarn segments align along an angle $\phi \neq 0$ and deformations of the $x$-axis projection of the endpoint separation $\lambda_{\parallel}$ can be achieved by changes in both the endpoint separation $\lambda$ and the angle $\phi$. 
It is evident that deformations involving changes in endpoint separation $\lambda$ are even more rigid than those for even connecting yarn segments. 
Therefore, odd connecting yarn segments undergo extensile deformations by rotating into the fabric plane (as shown in \ref{fig:rs-figure}b), i.e.~they change $\lambda_\parallel$ by changing the angle $\phi$ at fixed $\lambda$ so $Y_{\rm odd} \equiv (\partial^2 E_{\rm odd}/\partial \lambda^2_\parallel)_{\hat{\mathbf{v}}, \lambda}$ . Therefore, the extensional rigidity for odd connecting yarn segments is approximately
\begin{equation}
    Y_{\rm odd} \approx \frac{12 B}{L\lambda^2_{\perp}}(1 + \hat{v}_{\zeta}\cot\phi) = \frac{12 B}{\lambda^3(1 - \delta_{\rm odd})}(1 + \hat{v}_{\zeta}\cot\phi)
\end{equation}
where $\lambda_\perp = \bm{\lambda}\cdot\hat{\mathbf{e}}_\perp = \lambda\sin\phi$ is the out-of-plane projection of the endpoint separation $\lambda$, and $\delta_{\rm odd} = 1 - (L\sin\phi/\lambda)$.
Note that this energy diverges as $\phi \to 0$, for which the rotational freedom of the odd connecting yarn segments saturates and the curve must increase $\lambda$ in order to undergo extensile deformations, much like even connecting yarn segments. 
However, for garter, rib, and seed stitches, values of $\phi$ for the un-deformed stitch are closer to $45^\circ$, so that $\cot\phi = 1$. In this case, we find that the ratio of even connecting yarn segments rigidity to odd connecting yarn segments rigidity can be significantly greater than one, with $Y_{\rm even}/Y_{\rm odd} \sim \mathcal{O}(10)$ being typical.

\subsubsection{Effective stitch rigidities}

The RS model provides estimates for the linear stiffnesses of different yarn connecting segments, based on the ``rule of mixing'' from the theory of fiber composites.
To estimate the effective linear stiffness of the entire stitch, we treat each connecting segment as a spring element, either in series or in parallel with other springs comprising the stitch, as shown in \ref{fig:rs-figure}.
For stretching in the $x$-direction, we consider the effective stiffness of the connecting yarn segments that are oriented in the $x$-direction; likewise for stretching in the $y$-direction. 
Like-stitch neighbors (\textbf{K}-\textbf{K} or \textbf{P}-\textbf{P}) have a pair of spring elements, each with stiffness $Y_{\rm even}$, that add in series in the $x$-direction, giving an effective stiffness of $Y_{\rm eff} = Y_{\rm even}/2$; in the $y$-direction, two sets of in-series pairs are in parallel, giving an effective stiffness of $Y_{\rm eff} = Y_{\rm even}$.
Unlike-stitch neighbors (\textbf{K}-\textbf{P}) have different spring elements in the $x$-direction and $y$-direction.
In the $x$-direction, there is one even connecting yarn segment in series with an odd connecting yarn segment, leading to a effective stiffness of $Y_{\rm eff} = (Y_{\rm even}^{-1} + Y_{\rm odd}^{-1})^{-1} \approx Y_{\rm odd}$, since $Y_{\rm odd}/Y_{\rm even} \ll 1$.
In the $y$-direction, the springs add similarly to the like-stitch case, giving an effective stiffness of $Y_{\rm eff} = Y_{\rm odd}$.

\begin{table}[h!]
\centering
{
\setlength{\tabcolsep}{2mm}
\begin{tabular}{|c|c|c|c|c|}
    \hline
     & \begin{tabular}{@{}c@{}}$\lambda$ (mm)\end{tabular} & \begin{tabular}{@{}c@{}}$\delta$ \end{tabular} & $\hat{v}_\zeta$ & $\phi$ \\
    \hline\hline 
    \begin{tabular}{@{}c@{}}Stockinette \\ ($x$; even) \end{tabular} & 2.187 & -0.884 & 1.000 & - \\
    \hline
    \begin{tabular}{@{}c@{}}Stockinette \\ ($y$; even) \end{tabular} & 5.289 & -0.490 & 0.686 & - \\
    \hline
    \begin{tabular}{@{}c@{}}Garter \\ ($x$; even) \end{tabular} & 2.120  & -0.583 & 1.000 & - \\
    \hline
    \begin{tabular}{@{}c@{}}Garter \\ ($y$; odd) \end{tabular} & 3.499 & 0.257 & 0.686 & 0.749 \\
    \hline
    \begin{tabular}{@{}c@{}}Rib \\ ($x$; odd) \end{tabular} & 3.093 & -0.156 & 1.000 & 1.181 \\
    \hline
    \begin{tabular}{@{}c@{}}Rib \\ ($y$; even) \end{tabular} & 4.846 & -0.334 & 0.999 & - \\
    \hline
    \begin{tabular}{@{}c@{}}Seed \\ ($x$; odd) \end{tabular} & 3.053 & 0.462 & 0.073 & 0.450 \\
    \hline
    \begin{tabular}{@{}c@{}}Seed \\ ($y$; odd) \end{tabular} & 3.776 & -0.020 & 0.720 & 0.821 \\
    \hline
\end{tabular}
}
\caption{\label{table:RS_acrylic} List of geometric parameters for use in RS model calculations, obtained from simulations of acrylic yarn.}
\end{table}

\begin{table}[h!]
\centering
{
\setlength{\tabcolsep}{2mm}
\begin{tabular}{|c|c|c|c|c|}
    \hline
     & \begin{tabular}{@{}c@{}}$\lambda$ (mm)\end{tabular} & \begin{tabular}{@{}c@{}}$\delta$ \end{tabular} & $\hat{v}_\zeta$ & $\phi$ \\
    \hline\hline 
    \begin{tabular}{@{}c@{}}Stockinette \\ ($x$; even) \end{tabular} & 2.442 & -0.421 & 0.988 & - \\
    \hline
    \begin{tabular}{@{}c@{}}Stockinette \\ ($y$; even) \end{tabular} & 3.559 & -0.513 & 0.783 & - \\
    \hline
    \begin{tabular}{@{}c@{}}Garter \\ ($x$; even) \end{tabular} & 3.488 & -0.493 & 1.000 & - \\
    \hline
    \begin{tabular}{@{}c@{}}Garter \\ ($y$; odd) \end{tabular} & 4.895 & 0.202 & 0.696 & 0.770 \\
    \hline
    \begin{tabular}{@{}c@{}}Rib \\ ($x$; odd) \end{tabular} & 2.834 & -0.565 & 1.000 & 1.355 \\
    \hline
    \begin{tabular}{@{}c@{}}Rib \\ ($y$; even) \end{tabular} & 5.744 & -0.338 & 1.000 & - \\
    \hline
    \begin{tabular}{@{}c@{}}Seed \\ ($x$; odd) \end{tabular} & 3.344 & 0.432 & 0.100 & 0.474 \\
    \hline
    \begin{tabular}{@{}c@{}}Seed \\ ($y$; odd) \end{tabular} & 4.181 & -0.102 & 0.816 & 0.864 \\
    \hline
\end{tabular}
}
\caption{\label{table:RS_cotton}  List of geometric parameters for use in RS model calculations, obtained from simulations of cotton yarn.}
\end{table}

\begin{table}[h!]
\centering
{
\setlength{\tabcolsep}{2mm}
\begin{tabular}{|c|c|c|c|c|}
    \hline
     & \begin{tabular}{@{}c@{}}$Y_x$ \\ (N/mm)\end{tabular} & \begin{tabular}{@{}c@{}}$Y_y$ \\ (N/mm)\end{tabular} & $\nu_{yx}$ & $\nu_{xy}$ \\
    \hline\hline 
    \begin{tabular}{@{}c@{}}Stockinette \\ (experiment)\end{tabular} & $0.165 \pm 0.011$ & $0.753 \pm 0.034$ & $0.444 \pm 0.005$ & $0.430 \pm 0.015$ \\
    \hline
    \begin{tabular}{@{}c@{}}Stockinette \\ (simulation)\end{tabular} & 0.182 & 0.684 & 0.453 & 0.202 \\
    \hline
    \begin{tabular}{@{}c@{}}Stockinette \\ (RS model)\end{tabular} & 0.210 & 0.528 & - & -\\
    \hline
    \begin{tabular}{@{}c@{}}Garter \\ (experiment)\end{tabular} & $0.223 \pm 0.021$ & $0.056 \pm 0.015$ & $0.481 \pm 0.004$ & $0.150 \pm 0.003$ \\
    \hline
    \begin{tabular}{@{}c@{}}Garter \\ (simulation)\end{tabular} & 0.200 &  0.030 & 0.504 & 0.407 \\
    \hline
    \begin{tabular}{@{}c@{}}Garter \\ (RS model)\end{tabular} & 0.275 & 0.106 & -& -\\
    \hline
    \begin{tabular}{@{}c@{}}Rib \\ (experiment)\end{tabular} & $0.010 \pm 0.006$ & $0.119 \pm 0.012$ & $0.208 \pm 0.001$ & $0.294 \pm 0.006$ \\
    \hline
    \begin{tabular}{@{}c@{}}Rib \\ (simulation)\end{tabular} & 0.022 & 0.129 & 0.200 & 0.461 \\
    \hline
    \begin{tabular}{@{}c@{}}Rib \\ (RS model)\end{tabular} & 0.024 & 0.109 &- &- \\
    \hline
    \begin{tabular}{@{}c@{}}Seed \\ (experiment)\end{tabular} & $0.070 \pm 0.007$ & $0.019 \pm 0.001$ & $0.471 \pm 0.004$ & $0.138 \pm 0.002$ \\
    \hline
    \begin{tabular}{@{}c@{}}Seed \\ (simulation)\end{tabular} & 0.103 & 0.046 & 0.373 & 0.515 \\
    \hline
    \begin{tabular}{@{}c@{}}Seed \\ (RS model)\end{tabular} & 0.077 & 0.029 &- &- \\
    \hline
\end{tabular}
}
\caption{\label{table:acrylic_youngpoisson_RS} List of parameters obtained by fitting the Young's moduli and Poisson ratios to experimental and simulation data representing fabric made from the acrylic yarn. Also included: estimates of stitch rigidity from the reduced-symmetry (RS) elastica model using geometric parameters obtained from simulations of relaxed stitches.}
\end{table}

\begin{table}[h!]
\centering
{
\setlength{\tabcolsep}{2mm}
\begin{tabular}{|c|c|c|c|c|}
    \hline
     & \begin{tabular}{@{}c@{}}$Y_x$ \\ (N/mm)\end{tabular} & \begin{tabular}{@{}c@{}}$Y_y$ \\ (N/mm)\end{tabular} & $\nu_{yx}$ & $\nu_{xy}$ \\
    \hline\hline 
    \begin{tabular}{@{}c@{}}Stockinette \\ (experiment)\end{tabular} & $0.122 \pm 0.018$ & $0.545 \pm 0.038$ & $0.420 \pm 0.003$ & $0.412 \pm 0.020$ \\
    \hline
    \begin{tabular}{@{}c@{}}Stockinette \\ (simulation)\end{tabular} & 0.298 & 0.536 & 0.441 & 0.359 \\
    \hline
    \begin{tabular}{@{}c@{}}Stockinette \\ (RS model)\end{tabular} & 0.312 & 0.602 & - & - \\
    \hline
    \begin{tabular}{@{}c@{}}Garter \\ (experiment)\end{tabular} & $0.051 \pm 0.006$ & $0.028 \pm 0.002$ & $0.561 \pm 0.006$  & $0.177 \pm 0.002$ \\
    \hline
    \begin{tabular}{@{}c@{}}Garter \\ (simulation)\end{tabular} & 0.188 & 0.047 & 0.459 & 0.210 \\
    \hline
    \begin{tabular}{@{}c@{}}Garter \\ (RS model)\end{tabular} & 0.099 & 0.052 & - & - \\
    \hline
    \begin{tabular}{@{}c@{}}Rib \\ (experiment)\end{tabular} & $0.003 \pm 0.005$ & $0.046 \pm 0.006$ & $0.205 \pm 0.003$ & $0.289 \pm 0.006$ \\
    \hline
    \begin{tabular}{@{}c@{}}Rib \\ (simulation)\end{tabular} & 0.008 & 0.026 & 0.195 & 0.439 \\
    \hline
    \begin{tabular}{@{}c@{}}Rib \\ (RS model)\end{tabular} & 0.029 & 0.099 & - & - \\
    \hline
    \begin{tabular}{@{}c@{}}Seed \\ (experiment)\end{tabular} & $0.035 \pm 0.005$ & $0.041 \pm 0.003$ & $0.435 \pm 0.006$ & $0.165 \pm 0.008$ \\
    \hline
    \begin{tabular}{@{}c@{}}Seed \\ (simulation)\end{tabular} & 0.107 & 0.180 & 0.359 & 0.168 \\
    \hline
    \begin{tabular}{@{}c@{}}Seed \\ (RS model)\end{tabular} & 0.079 & 0.033 & - & - \\
    \hline
\end{tabular}
}
\caption{\label{table:cotton_youngpoisson_RS} List of parameters obtained by fitting Young's moduli and Poisson ratios to experimental and simulation data representing fabric made from the cotton yarn. Also included: estimates of stitch rigidity from the reduced-symmetry (RS) elastica model using geometric parameters obtained from simulations of relaxed stitches.}
\end{table}

\subsubsection{The high-tension limit}

As the endpoints $\pm \mathbf{r}_0$ are brought further apart, the distribution of curvature in the elastica shifts and becomes increasingly concentrated at the endpoints, as the central segment straightens out.
Since the curve must maintain a fixed length, the force required to move these points apart increases until the distance between the points is equal to the total length of the curve.
In this pathological limit, the radius of curvature at the endpoints approaches zero and the bending energy diverges; the tension $T$ required to maintain the fixed-length constraint likewise diverges.
The nature of this strain-stiffening response can therefore be obtained in the asymptotic $T \to \infty$ limit of the elastica model.
This is equivalent to taking the large-$q$ limit in the evaluation of the energies \ref{eq:even_energy} and \ref{eq:odd_energy}.
In this limit, the integrands of the energy functionals for both even and odd connecting yarn segments have the same $q$-dependence.
Minimizing each energy with respect to $q$, we find an identical asymptotic form to the energy,
\begin{equation}
E_{\rm high-tension}(\hat{v}_\zeta, \lambda) \simeq \frac{B}{4L}\frac{\hat{v}_\zeta^4}{1 - \frac{\lambda}{L}} \, .
\end{equation}
Thus, we recover both the expected strain-stiffening response $f(\lambda) \sim (1 - \lambda/L)^{-2}$ and the observed universal strain-stiffening behavior, independent of stitch symmetry.

\subsection{Applying the constitutive model to simulate uniaxial stretching in fabrics of finite size}

To simulate the deformed shape of sample of fabric of finite extent using our constitutive model (\ref{eq:constitutive_model}), we turn to continuum elasticity theory.

\krish{There is considerable prior work on numerical homogenization of yarn level simulations that use micromechanical simulations to predict the bulk level elastic response that is then implemented in FEA \cite{Sperl2020,Wadekar2020}. 
We emphasize here that our FEA calculation is a proof of concept demonstration that the nonlinear constitutive model captures realistic fabric-level deformations. 
Rather than the micromechanical model being tied directly into the FEA calculation, we use the elastica model to derive the nonlinear form of the constitutive model, based on stitch microstructure.
Here, we use constitutive model coefficients that are found from fitting experimental (rather than simulation) data. }

We will denote points inside the undeformed fabric as $\mathbf{r} = (x,y)$, where $x \in [-W_0/2,W_0/2]$ and $y \in [-L_0/2,L_0/2]$.
The dimensions of the undeformed garter fabric made of acrylic yarn, measured prior to the stretching experiment, are $W_0 = 127\, {\rm mm}$ and $L_0 = 40\, {\rm mm}$.
Under applied uniaxial displacement of the boundaries, fabric points $\mathbf{r}$ are displaced by a vector field $\mathbf{u}(\mathbf{r})$ to new points $\mathbf{R}(\mathbf{r}) = \mathbf{r} + \mathbf{u}(\mathbf{r})$, resulting in a linearized strain tensor
\begin{equation}
    \varepsilon_{ij}(\mathbf{r}) = \frac{1}{2}\left(\partial_i u_j + \partial_j u_i\right) \, .
\end{equation}
This strain corresponds to an internal stress field $\sigma_{ij}(\mathbf{r})$ via our constitutive model (\ref{eq:constitutive_model}).
 \seg{The points $\mathbf{r}$ and $\mathbf{R}$ lie on two-dimensional triangular meshes with the same topology (no re-meshing is performed during the calculation).}
The final shape adopted by the fabric under set displacements of the $y$-boundary, $\mathbf{u}(x,\pm L_0/2) = \pm U \hat{\mathbf{y}}$, is determined by solving the continuum elasticity equilibrium equations,
\begin{equation}
    \partial_{j}\sigma_{ij}(\mathbf{r}) = 0 \, ,
\end{equation}
with boundary conditions $\sigma_{xj}(\pm W_0/2, y) = 0$ on the $x$-boundary.

To solve this boundary value problem, we turn to Finite Element Analysis (FEA), as implemented by FEniCS, an open-source finite element solver (see \url{https://fenicsproject.org/}).
Rather than directly solving the stress balance partial differential equation, the problem is cast in its ``weak form,'' derived from the energy functional
\begin{equation}
    E[\varepsilon] = \int_{-W_0/2}^{W_0/2}\int_{-L_0/2}^{L_0/2} {\rm d}x\,{\rm d}y\left\{\frac{1}{2}C^0_{ijkl}\varepsilon_{ij}\varepsilon_{kl} + \frac{\beta_{xx}\alpha_{xx}^2\varepsilon_{xx}^3}{1 - \alpha_{xx}\varepsilon_{xx}} + \frac{\beta_{yy}\alpha_{yy}^2\varepsilon_{yy}^3}{1 - \alpha_{yy}\varepsilon_{yy}}\right\} \, .
\end{equation}
Note that this energy functional requires the symmetry $C^0_{xxyy} = C^0_{yyxx}$, whereas our fits show significant asymmetry between these components.
To continue using this form of the energy functional, we use the average of the measured values of $C^0_{xxyy}$ and $C^0_{yyxx}$.
Following the standard FEA procedure, we create a meshed representation of the undeformed fabric.
The mesh elements at the top and bottom boundaries of the fabric are displaced by the fixed boundary displacement $U$.
Next, the program calculates the variations in the total energy $\delta E$ with respect to displacements of the mesh vertices $\delta u_k$.
Finally, the program iteratively searches for the root $\delta E(\mathbf{u}) = 0$.
To avoid numerical issues due to the singular form of the nonlinear part of the elastic energy functional, we approximate the nonlinear part by its series expansion, truncating at quartic order (dropping terms $\mathcal{O}(\varepsilon^5)$ and higher).
While we did not calculate the elastic constant $C_{xyxy}$ in experiments or simulations, for this demonstration, we chose $C_{xyxy} = 0.01$ N mm\textsuperscript{-1}, which is on a comparable scale as the other elastic constants. 
Since the majority of the uniaxial deformation involves $\varepsilon_{xx}$ and $\varepsilon_{yy}$ components of strain, the simulated deformations are relatively insensitive to this one elastic constant.

\subsection{Therapeutic Glove Prototype}

\begin{figure}[ht!]
\centering
\includegraphics[width=.9\textwidth]{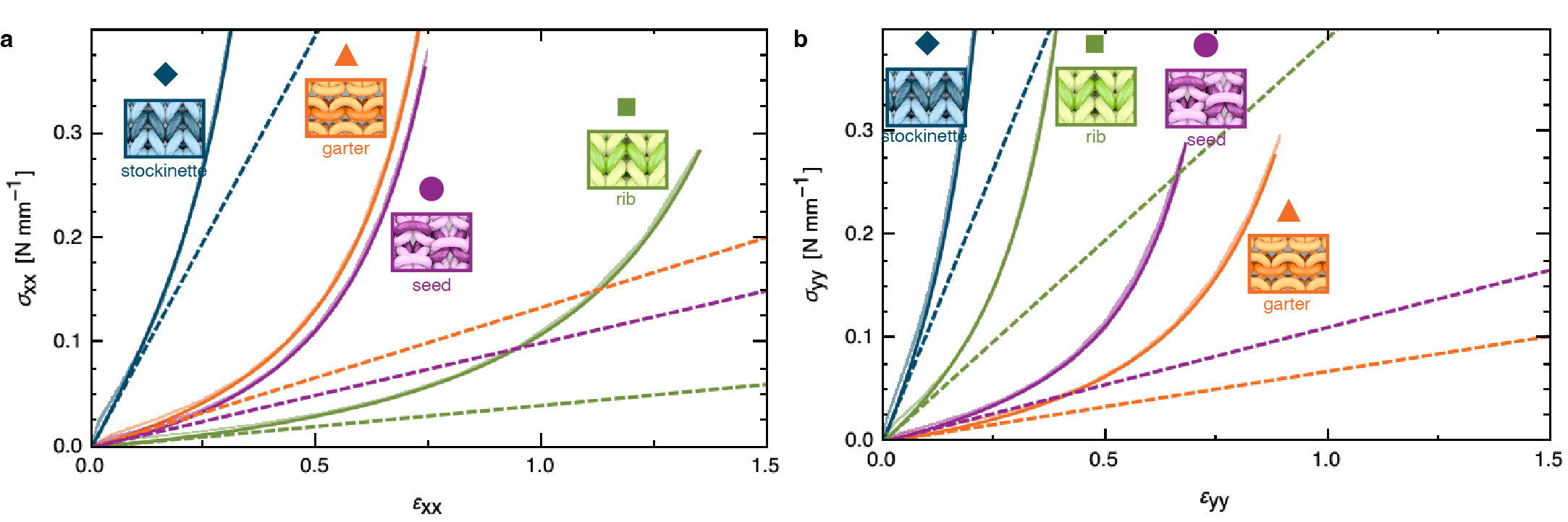}
\caption{\label{fig:stressstrain} The experimental stress-versus-strain relations for the four fabrics made for the therapeutic glove prototype in the (\textbf{a}) $x$- and (\textbf{b}) $y$-directions. All of the data for each type of fabric is displayed by a different color: stockinette in blue, garter in orange, rib in green, and seed in purple.
The experimental data is shown in the translucent regions where the width of the region is one standard deviation of the data.
The solid curves are fits to the constitutive relations. 
Dashed lines depict the linear response at zero stress. 
}
\label{fig:babystress}
\end{figure}

We fabricated the samples of different fabrics by hand using 2.75 mm and 2.0 mm needles.
We then measured the stiffness of each test sample using the uniaxial stretching experiment protocol, described in \ref{sec:uniaxial}.
The experiment results are shown in \ref{fig:stressstrain}, constitutive model fits are given in \ref{table:baby_constitutive}, and the Young's moduli and Poisson ratios are given in \ref{table:baby_youngpoisson}. \krish{The yarns used for the glove were chosen because visually they would help the reader discern which stitch patterns were implemented and where. These yarns were fairly similar to the acrylic yarn, which was unavailable in a large enough color variety for this experiment.}

We took measurements of the hand, in particular the relative location of joints and other parts of the anatomy.
We used those measurements to determine where in the glove we needed rigidity and where we needed flexibility to support natural hand motion for the specific glove wearer.
This determined which type of fabric was needed in each region of the glove.
We used the stitch gauge -- how many stitches comprise five centimeters of fabric\seg{, also known as the stitch density} -- from the test samples to determine of the number of stitches needed in each part of the glove to match the hand dimensions.
\krish{We targeted 20 - 30 mm Hg for the hand of the specific wearer but it was calculated to be approximately 3.75 kPa or 28 mm Hg. \seg{This pressure was calculated by measuring the rest, flat position of the circumference of the glove, then measuring the circumference of the glove on the the hand. From these measurements, we calculate the linear strain of the glove as it is being worn. Using \ref{fig:babystress}, we use this strain to find a correlated stress. Multiplying the stress by the width of the wrist support segment (the stockinette region around the wrist) gives a force, which is then divided by the area of the wrist support segment to estimate a pressure. The pressure provided by the theraputic glove prototype} is comparable to the pressure that compression stockings are, between 30-40 mm Hg \cite{Lim2014}. We have achieved this comparable pressure without the use of elastane. The placement of the other stitch patterns was chosen \textit{ad hoc} to illustrate the anisotropic behavior of the fabric to enable unrestricted motion of the human hand.} 

\begin{figure}[ht!]
\centering
\includegraphics[width=.66\textwidth]{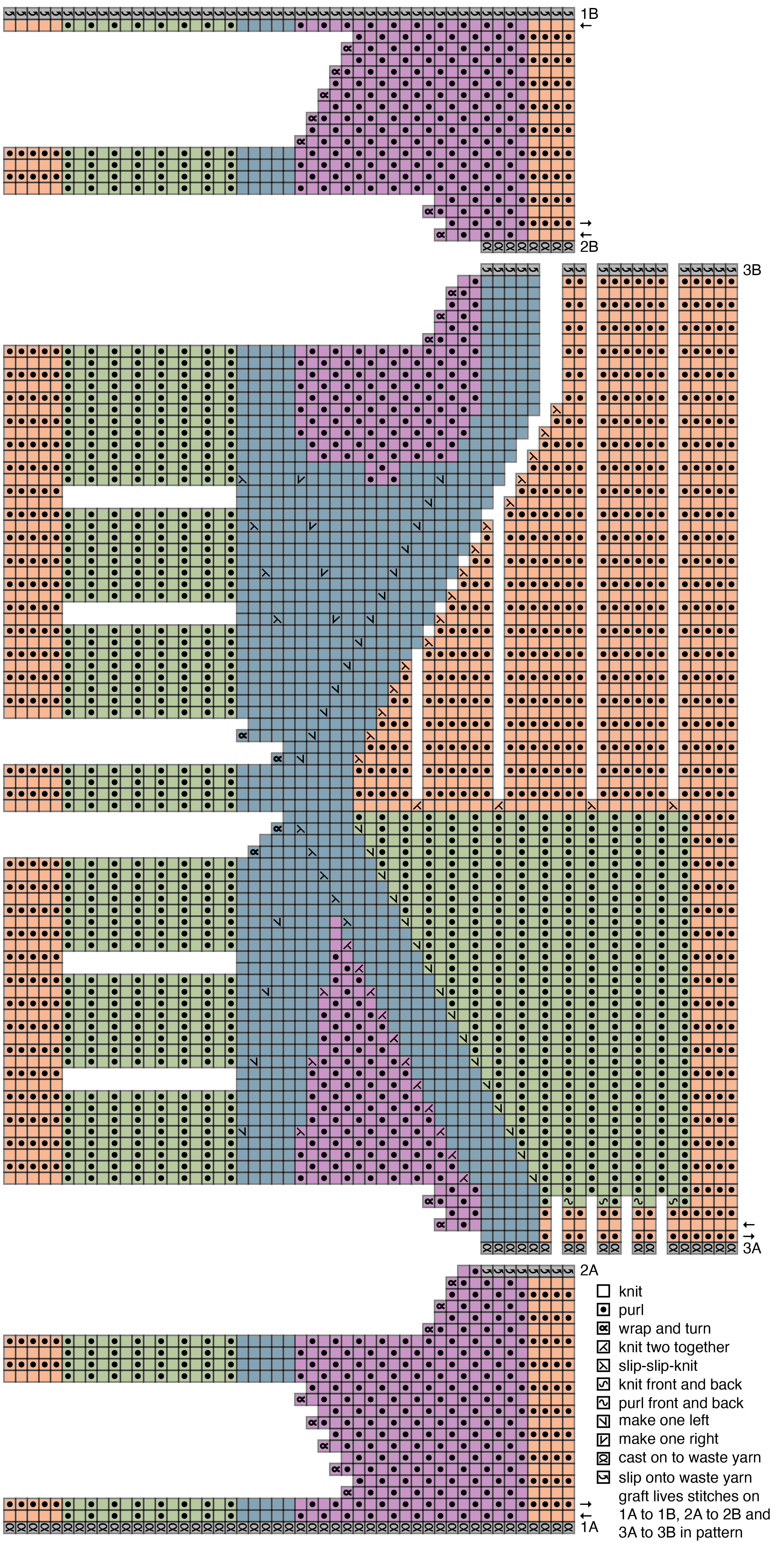}
\caption{\label{fig:gloveDesign} Pattern for the therapeutic glove. Arrows indicate direction of knitting.
}
\end{figure}

To create a seamless pattern which augments the stiffness of the stockinette fabric that supports the radiocarpal and intercarpal joints, we chose to knit the pattern as a single flat piece, starting from the thumb, wrapping the hand from the back to the front, and finally grafting the start and end of the fabric together into a glove.
\seg{Knitting this horizontally as a flat piece rather than as a tubular knit enabled us to exploit the stiffest direction of stockinette fabric to provide pressure to the wrist.}
For the illustration in this paper, we chose to highlight the different types of fabric with different-colored yarn.
These were knitted together \emph{in situ} using a seamless joining technique known as ``intarsia.''
The pattern shown in \ref{fig:gloveDesign} can also be knit with a single color of yarn without using intarsia.

\newpage

\begin{table}[h!]
\centering
{
\setlength{\tabcolsep}{2mm}
\color{black}\begin{tabular}{|c|c|c|c|c|c|c|c|c|}
    \hline
     & \begin{tabular}{@{}c@{}}$C^0_{xxxx}$ \\ (N/mm)\end{tabular} & \begin{tabular}{@{}c@{}}$C^0_{yyyy}$ \\ (N/mm)\end{tabular} & \begin{tabular}{@{}c@{}}$C^0_{xxyy}$ \\ (N/mm)\end{tabular} & \begin{tabular}{@{}c@{}}$C^0_{yyxx}$ \\ (N/mm)\end{tabular} & $\alpha_{xx}$ & $\alpha_{yy}$ & \begin{tabular}{@{}c@{}}$\beta_{xx}$ \\ (N/mm)\end{tabular}  & \begin{tabular}{@{}c@{}}$\beta_{yy}$ \\ (N/mm)\end{tabular}\\
    \hline\hline 
    \multirow{2}{*}{Stockinette} & \seg{0.210} & \seg{0.590} & \seg{0.116} & \seg{0.267} & \seg{0.926} & \seg{1.777} & \seg{0.045} & \seg{0.047} \\
    \cline{2-9}
    & 1.225\seg{$^*$}  & 1.646\seg{$^*$} & 0.758\seg{$^*$} & 0.936\seg{$^*$} & 1.806\seg{$^*$} & 2.426\seg{$^*$} & 0.051\seg{$^*$} & 0.088\seg{$^*$}\\
    \hline
        Garter  & 0.149 & 0.076 & 0.035 & 0.032 & 0.928 & 0.732 & 0.043 & 0.039 \\
    \hline
    Rib  & 0.046 & 0.452 & 0.028 & 0.100 & 0.483 & 1.547 & 0.039 & 0.059 \\
    \hline
    Seed  & 0.108 & 0.120 & 0.225 & 0.044 & 0.889 & 1.006 & 0.047 & 0.028 \\
    \hline
\end{tabular}
}
\caption{\label{table:baby_constitutive} List of parameters obtained by fitting the constitutive model to experimental  data representing test samples made for the therapeutic glove. \seg{Data that is starred was made on 2.00 mm knitting needles (US size 0) and all remaining data was knit on 2.75 mm knitting needles (US size 2).}}
\end{table}

\begin{table}[h!]
\centering
{
\setlength{\tabcolsep}{2mm}
\color{black}\begin{tabular}{|c|c|c|c|c|}
    \hline
     & \begin{tabular}{@{}c@{}}$Y_x$ \\ (N/mm)\end{tabular} & \begin{tabular}{@{}c@{}}$Y_y$ \\ (N/mm)\end{tabular} & $\nu_{yx}$ & $\nu_{xy}$ \\
    \hline\hline 
    \multirow{2}{*}{Stockinette} & \seg{0.157} & \seg{0.442} & \seg{0.452} & \seg{0.554} \\
    \cline{2-5}
    & 0.793\seg{$^*$}  & 1.066\seg{$^*$} & 0.569\seg{$^*$} & 0.619\seg{$^*$} \\
    \hline
        Garter & 0.134 & 0.068 & 0.419 & 0.236 \\
    \hline
    Rib & 0.040 & 0.392 & 0.222 & 0.599 \\
    \hline
    Seed  & 0.100 & 0.111 & 0.362 & 0.208 \\
    \hline
\end{tabular}
}
\caption{\label{table:baby_youngpoisson} List of parameters obtained by fitting the Young's moduli and Poisson ratios to experimental data representing the test samples of the therapeutic glove. \seg{Data that is starred was made on 2.00 mm knitting needles (US size 0) and all remaining data was knit on 2.75 mm knitting needles (US size 2).}}
\end{table}

\subsection{Knitting Machine versus Hand Knitting}

\begin{figure}[ht!]
\centering
\includegraphics[width=.66\textwidth]{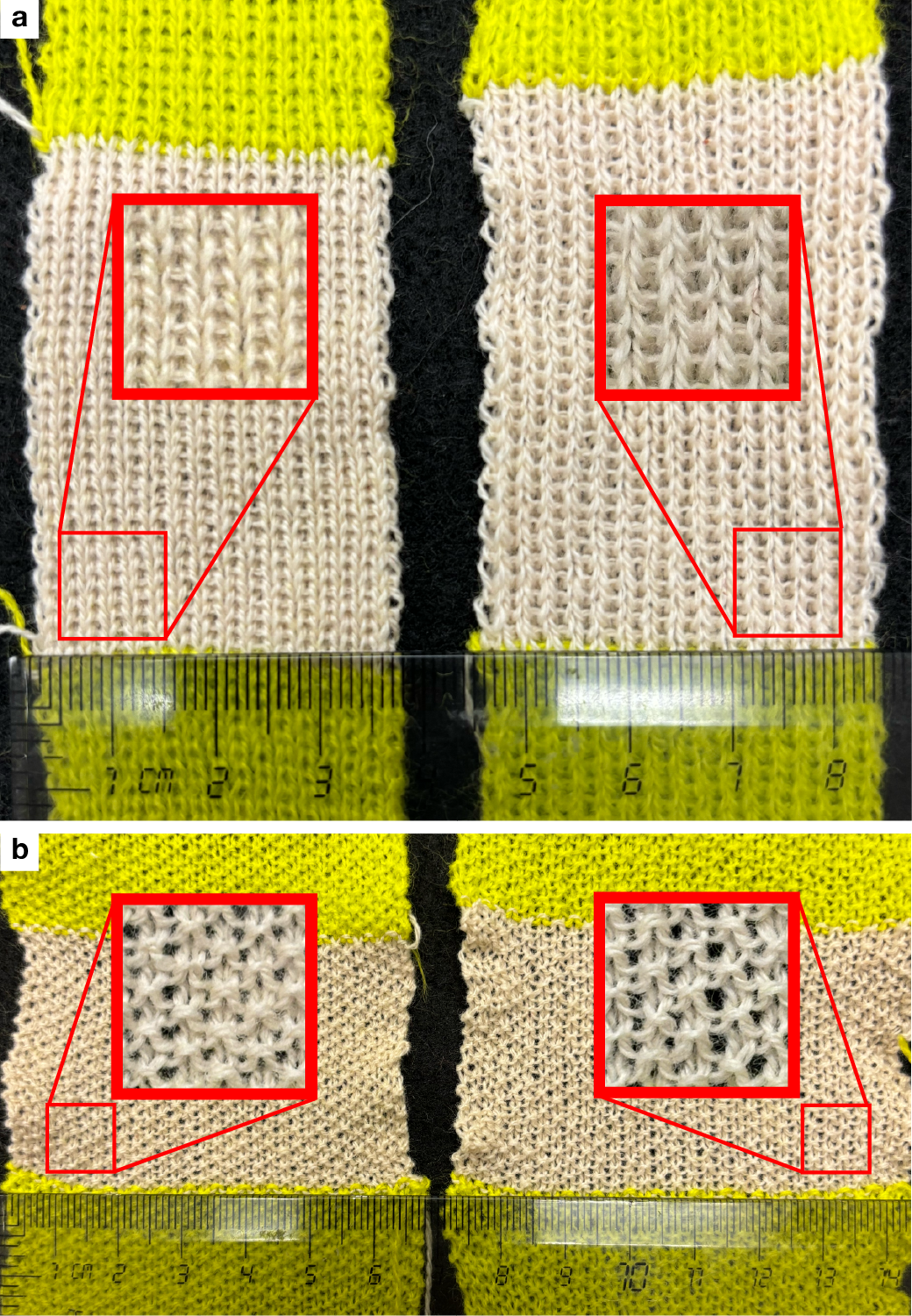}
\caption{\label{fig:stitchsizecomparison} \seg{Comparison of (\textbf{a}) rib and (\textbf{b}) seed fabrics made on the STOLL Industrial knitting machine at different assigned stitch sizes. On the left are fabrics made at size 11 whereas on the right they are made at size 12. The insets display a closeup on a 1 cm by 1 cm portion of the fabrics.}
}
\end{figure}

In general, ensuring uniform tension and uniform stitch size between types of fabrics is challenging for knitting. We were able to uniformly craft and replicate each fabric type with an equivalent number of rows and columns with a knitting machine. The knitting machine is ideal for ensuring uniform tension throughout the sample; however, it comes at the expense of not guaranteeing uniform stitch size between types of fabrics. \seg{For the lace weight samples made with the STOLL Industrial knitting machine, stockinette and garter were made with a stitch size setting of 12 while rib and seed were made at size 11. We find that if all four types of fabric are made at size 12, rib and seed are significantly more loose (\ref{fig:stitchsizecomparison}).} Hand knitting, in contrast, cannot guarantee uniform tension throughout the sample, but provides greater control of stitch size even while altering the pattern of the knit and purl stitches due to the fixed diameter of the knitting needle\seg{, as seen in \ref{table:acrylic_dims}}. Despite these differences, we do get consistent behavior of the four different types of fabric both machine made (Fig.~2 and \ref{fig:cotton}) and hand made (\ref{fig:stressstrain}). \ref{table:acrylic_dims} displays the differences in the yarn per stitch and the yarn diameter between machine-knit and hand-knit samples for the acrylic yarn.

\end{document}